\documentclass[aps,pra,twocolumn,reprint,showkeys,preprintnumbers,amsmath,amssymb,floatfix,a4paper,longbibliography,superscriptaddress]{revtex4-2}

\usepackage{amsmath}
\usepackage{amsfonts}
\usepackage{amssymb}
\usepackage{bm}
\usepackage{bbm}
\usepackage{color}
\usepackage{graphicx}

 \usepackage{hyperref}
\hypersetup{breaklinks=true}
\usepackage{dcolumn}
\usepackage{array}

\usepackage{orcidlink}

\newcommand{\figwidth}{1\columnwidth}

\newcommand{\nqubit}{N_{\text{q}}}
\newcommand{\Ntrotter}{N_{\rm ST}}

\newcommand{\Ntwoq}{\Lambda_{\text{2q}}}
\newcommand{\Natoms}{N}
\newcommand{\NHq}{L_{\text{q}}}
\newcommand{\Nbasis}{K}
\newcommand{\Nmodes}{N_{\text{m}}}

\newcommand{\dt}{\Delta t}
\newcommand{\decaytime}{\tau}

\DeclareMathOperator{\bin}{bin}
\DeclareMathOperator{\trace}{tr}
\newcommand{\quditd}{d}
\newcommand{\GellMann}{\lambda}
\newcommand{\erroroneq}{\epsilon_{\rm 1q}}
\newcommand{\errortwoq}{\epsilon_{\rm 2q}}
\newcommand{\errorn}[1]{\epsilon_{#1}}

\newcommand{\vmax}{v_{\rm max}}
\newcommand{\HO}{H_{\text{HO}}}
\newcommand{\HI}{H_{\text{AH}}}
\newcommand{\UST}{U}
\newcommand{\Id}{\mathbb{I}}
\newcommand{\Oo}{\Omega}
\newcommand{\GMindex}{j}

\begin{document}

\title{Simulation of vibrational dynamics using qubits and qudits}
\date{\today}
\author{Erik L\"otstedt\,\orcidlink{0000-0003-2555-8660}}
\email{loetstedte@riken.jp}
\affiliation{RIKEN iTHEMS, 2-1 Hirosawa, Wako, Saitama 351-0198, Japan}
\affiliation{Trapped Ion Quantum Computer Team, TRIP Headquarters, RIKEN, 2-1 Hirosawa, Wako, Saitama 351-0198, Japan}
\affiliation{Computational Condensed Matter Physics Laboratory, RIKEN Pioneering Research Institute (PRI), 2-1 Hirosawa, Wako, Saitama 351-0198, Japan}
\author{Kaoru Yamanouchi\,\orcidlink{0000-0003-3843-988X}}
\affiliation{Trapped Ion Quantum Computer Team, TRIP Headquarters, RIKEN,   2-1 Hirosawa, Wako, Saitama 351-0198, Japan}
\affiliation{Institute for Attosecond Laser Facility, The University of Tokyo, 7-3-1 Hongo, Bunkyo-ku, Tokyo 113-0033, Japan}

\begin{abstract}
We investigate the quantum computing of the vibrational dynamics of CO$_2$ and H$_2$O by constructing the   vibrational third-order anharmonic Hamiltonian 
 in qubit and qudit form  by two types of qubit encodings (binary and direct) and a qudit encoding. We simulate the time-dependent vibrational population transfer using the three different encodings, including the effect of noise and find that  the qudit encoding leads to the most accurate results both for  CO$_2$ and H$_2$O 
 because of the small number of terms in the qudit Hamiltonian as long as the same values of the entangling gate error rates are adopted.
Although the entangling gate errors of current qudit quantum computers are approximately one order of magnitude larger than the entangling gate errors of current qubit quantum computers, our results suggest that the simulation of vibrational dynamics is a promising application for future qudit quantum computers.
\end{abstract}


\maketitle
%
\newpage 

\section{Introduction}\label{Sec:Intro}
A quantum computer is usually constructed from qubits, quantum two-level systems having a ground state, $|0\rangle$, and an excited state, $|1\rangle$.  Currently, there are several types of qubit-based quantum computers in operation, including devices where the qubits are realized as superconducting circuits \cite{Bravyi2022}, trapped ions \cite{Moses2023_PRX,Chen2024}, and neutral atoms \cite{Chew2022,Bluvstein2023,Shao2024}. The number of qubits $\nqubit$ is at present around $\nqubit=150$ for superconducting-type quantum computers \cite{ibmqcloud}, $\nqubit=100$ for trapped-ion type quantum computers \cite{Ransford2026}, and about $\nqubit=300$ for neutral-atom type quantum computers \cite{Bluvstein2023}.  

Qubit-based quantum computers have been employed for pioneering applications in quantum chemistry  \cite{Caoetal_review2019,Baueretal2020,McArdleetal_review2020,Weidman2024}, 
where the electronic energies of small molecules such as as H$_2$O \cite{Eddinsetal2021} and F$_2$ \cite{Guo2024} were calculated. However, because of the noise present on current quantum computers, and because of the large number of terms in the qubit form of the electronic Hamiltonian \cite{Tranteretal_2018}, quantum computing of large molecules is still difficult. A possible way forward is the quantum-centric supercomputing approach \cite{Alexeev2024}, in which quantum computers are combined with supercomputers. A representative method is the quantum-selected configuration interaction method, in which the active configuration space is determined by sampling from a wave function prepared by a quantum computer and the electronic energy is obtained by the solution of the corresponding eigenvalue problem on a classical supercomputer 
\cite{Kanno2026,Nakagawa2024,RobledoMoreno2025}.

Another promising approach is the use of qudits as a basic unit of quantum computing \cite{Muthukrishnan2000,Vlasov2002,Wang2020}. A qudit is a 
$\quditd$-level quantum system with states
\begin{equation}
|\ell\rangle =|0\rangle, |1\rangle,\ldots,|\quditd-1\rangle.
\end{equation}
A quantum computer constructed from $\nqubit$ qudits is in a superposition of $\quditd^{\nqubit}$ states, which is exponentially  larger than the $2^{\nqubit}$ states available in a qubit-based quantum computer. Qudit-based quantum computers having up to around  ten qudits have been realized with 
superconducting circuits \cite{Bianchetti2010,Blok2021,Luo2023,Roy2023,Goss2024,WangParker2025}, 
trapped ions \cite{Randall2015,Senko2015,Ringbauer2022,Aksenov2023,Hrmo2023,Edmunds2025,Nikolaeva2024,Meth2025,Low2025,Zalivako2025,John2026},
photonic circuits \cite{Lanyon2008,Lanyon2009,Chi2022},
donor atoms \cite{FernandezdeFuentes2024},
neutral atoms \cite{Chaudhury2007,Omanakuttan2021,Lindon2023}, and
vacancy centers \cite{Soltamov2019,Adambukulam2024}.
Trapped-ion setups are particularly promising because of the long coherence time and the high-fidelity gate operations. So far, different values of $\quditd$ were adopted depending on the atomic ion species:  $\quditd=8$ in ${}^{40}$Ca$^{+}$ \cite{Ringbauer2022}, $\quditd=13$ in ${}^{137}$Ba$^{+}$ \cite{Low2025}, and $\quditd=4$ in ${}^{171}$Yb$^{+}$ \cite{Aksenov2023,Zalivako2025}.
The error rate of the two-qudit gate operations is around $\errortwoq \approx 1\times 10^{-2}$ \cite{Hrmo2023} for current qudit devices, which is about one order of 
magnitude larger than $\errortwoq \approx 1\times 10^{-3}$  \cite{Ransford2026} for current qubit-based quantum computers.

Several  advantages of using qudits for quantum computation have been proposed, including the simulation of fermions \cite{Chizzini2024}, chemical dynamics \cite{MacDonell2021}, lattice gauge theory \cite{Meth2025}, and spin dynamics \cite{Senko2015,Edmunds2025},
simplifications of multi-qubits gates \cite{Lanyon2009,Kiktenko2025} and the reduction of the circuit depth \cite{Nikolaeva2023,NikolaevaKiktenkoFedorov2024,Kiktenko2025}. Qudits may also be used in error correction \cite{Gottesman1999_qudit,Campbell2014,Keppens2025}. Indeed, a logical qudit ($\quditd=3$ and $4$) encoded using the Gottesman-Kitaev-Preskill bosonic code \cite{Gottesman2001} was demonstrated to have an effective lifetime longer than a physical qudit by a factor of about two \cite{Brock2025}.

In the present study, we show that qudits are advantageous for the simulation of molecular vibration.  Because a nonlinear $\Natoms$-atomic molecule has $3\Natoms-6$ vibrational modes, the calculation of the  vibrational energy levels   of a molecule is a hard computational problem. If we assume that each vibrational mode is described by $\Nbasis$ basis functions, the number of expansion coefficients in the vibrational wave function becomes $\Nbasis^{3\Natoms-6}$, which increases exponentially with increasing  $\Natoms$. Although quantum computing of molecular vibration  has attracted  less attention compared to the quantum computing of the electronic structure of molecules, theoretical rovibrational spectroscopy is recognized as one of the areas in which quantum computers are expected to be useful \cite{Baueretal2020}. 
Several attempts have been made to evaluate vibrational energies using qubit-based quantum computers \cite{Teplukhin2019,McArdleetal2019,SawayaHuh_2019,Loetstedt2021,Loetstedt2022_copy,Sawaya2021,Majland2023,Somasundarametal2025,LotstedtSzidarovszky2026,Asnaashari2026},
but demonstrations using quantum hardware have so far been limited to few-atomic molecules such as 
CO$_2$ \cite{Ollitraultetal2020,Loetstedt2021,Loetstedt2022_copy,Somasundarametal2025}, H$_2$O \cite{LotstedtSzidarovszky2026}, and NH$_3$ \cite{Somasundarametal2025}.

We demonstrate that the qudit representation of the vibrational Hamiltonian has much fewer terms than  the qubit representation, and consequently, a qudit quantum circuit  for the Suzuki-Trotter approximation of the time evolution operator contains much fewer two-qudit gates than the number of two-qubit gates in the corresponding qubit quantum circuit, resulting in much smaller error originating from the noise. We then assess quantitatively the qudit advantage over qubits for simulation of vibrational dynamics on noisy quantum computers by carrying out quantum computing of CO$_2$ and H$_2$O using a completely depolarizing noise model.
Because of the smaller number of gates,  we find that the simulations using qudits are more accurate than simulations using qubits, assuming that the same value of the two-qubit/qudit gate error rate is adopted. The error rates for qudit devices are currently about ten times larger than the qubit error rates, which means that there is no immediate advantage of using qudits. Our results suggest that the  simulation of vibrational dynamics is a promising application for  future qudit quantum computers.

%

\section{Theory}\label{Sec:Theory}

\subsection{Vibrational Hamiltonian}\label{Subsec:vibH}
We consider the vibrational Hamiltonian of a general nonlinear molecule,
\begin{equation}
H=\HO+\HI,\end{equation}
where
\begin{equation}\label{Eq:harmonicpot}
\HO=\sum_{k=1}^{\Nmodes}\HO^{(k)} =\sum_{k=1}^{\Nmodes}\frac{\omega_k}{2}\left(-\frac{\partial^2}{\partial q_k^2} + q_k^2 -1\right) 
\end{equation}
is the harmonic oscillator part defined in terms of the dimensionless normal mode coordinates $q_k$,
$\Nmodes$ is the number of modes, and 
\begin{equation}\label{Eq:anharmonicpot}
\HI= \sum_{\substack{j,k,l=1\\j\le k\le l}}^{\Nmodes} f_{jkl}q_jq_kq_l
\end{equation}
defines the  anharmonic potential in terms of the anharmonic coupling constants $f_{jkl}$ ($j,k,l=1,2,\ldots,\Nmodes$). In the current investigation, we consider anharmonic coupling up to third order in $q_k$, but in the general case, the anharmonic potential \eqref{Eq:anharmonicpot} may contain terms of the fourth and higher orders \cite{Csaszar2011}.

The time-dependent vibrational wave function is expanded according to 
\begin{equation}
|\psi(t)\rangle=\sum_{v_1,v_2,\ldots,v_{\Nmodes}=0}^{\vmax} c_{\bm{v}}(t)|\bm{v}\rangle,
\end{equation}
where $v_k$ is the vibrational quantum number of the $k$th mode, $\bm{v}=(v_1,v_2,\cdots, v_{\Nmodes})$ is a composite index, 
$\vmax$ is the maximum vibrational quantum number, and 
\begin{equation}
|\bm{v}\rangle =|v_1\rangle|v_2\rangle\cdots |v_{\Nmodes}\rangle
\end{equation}
is a direct product of $\Nmodes$ harmonic-oscillator eigenstates $|v_k\rangle$. The basis functions $|\bm{v}\rangle$ are eigenfunctions of $\HO$,  
\begin{equation}
\HO |\bm{v}\rangle = \left(\sum_{k=1}^{\Nmodes} v_k \omega_k\right)|\bm{v}\rangle, 
\end{equation}
but not of the total Hamiltonian $H$ because of the anharmonic coupling $\HI$.

\subsection{Qubit and qudit encodings}\label{Subsec:qencode}

In order to encode the basis states as qubit states, we consider two different encoding schemes, i.e., the binary encoding  \cite{Sawayaetal2020} and the direct encoding \cite{Sawayaetal2020}. In the binary encoding, we map a basis state to a qubit state according to
\begin{equation}
|\bm{v}\rangle = |\bin(v_1)\bin(v_2)\cdots \bin(v_{\Nmodes})\rangle_{\text{q}},
\end{equation}
where $\bin(v)$ is the binary representation of the integer $v$. The binary encoding requires $\nqubit=\Nmodes \lceil\log_2(\vmax+1)\rceil$ qubits, where $\lceil\cdot\rceil$ is the ceiling function.  For example, a two-mode basis state $|v_1=0,v_2=3\rangle$ is represented as the qubit state $|\bin(0)\bin(3)\rangle_{\text{q}}=|0011\rangle_{\text{q}}$. In the case when 
$\vmax+1$ is not equal to a power of 2, some qubit states do not represent vibrational basis states and are not used in the encoding.  

In the direct encoding (also referred to as the unary encoding \cite{Sawayaetal2020} or  one-hot encoding \cite{Hadfield2019}), we map a single-mode basis state $|v\rangle$ to  a qubit state 
where the $v$th qubit is in the excited state, and the rest of the qubits are in the ground state,
\begin{equation}
|v\rangle = X_{v}|\overline{0}\rangle_{\text{q}},
\end{equation}
where $X_v$ is a Pauli $\sigma_x$ operator acting on the $v$th qubit, and $|\overline{0}\rangle_{\text{q}}=|0\cdots 00\rangle_{\text{q}}$ is the all-zero qubit state. The multi-mode basis state 
$|\bm{v}\rangle$ is encoded accordingly as
\begin{equation}
|\bm{v}\rangle = \prod_{k=1}^{\Nmodes} X_{\mu(k)}|\overline{0}\rangle_{\text{q}},\quad \mu(k)=(k-1)(\vmax+1)+v_k.
\end{equation}
The direct encoding requires $\nqubit = \Nmodes(\vmax+1)$ qubits. Only a small fraction $[(\vmax+1)/2^{\vmax+1}]^{\Nmodes}$ of the total number of qubit states is employed.

In the case of qudits, we  map each harmonic oscillator basis function $|v\rangle$ to a qudit state $|\ell\rangle_{\text{q}}$, 
\begin{equation}
|\bm{v}\rangle = |\ell_1\ell_2\cdots \ell_{\Nmodes}\rangle_{\text{q}}, \quad \ell_k=v_k,
\end{equation} 
which requires $\Nmodes$  qudits having $\quditd=\vmax+1$ levels.

The qubit or qudit  Hamiltonian is expressed as
\begin{equation}\label{Eq:Hq}
H_{\text{q}}=\sum_{n=1}^{\NHq}h_n\Gamma_n,
\end{equation}
where $h_n$ is a numerical coefficient, and $\Gamma_n$ is defined as a direct product of generalized Gell-Mann matrices $\GellMann_j$ \cite{Luo2014},
\begin{equation}
	\Gamma_n = \GellMann_{\GMindex_{n(\nqubit-1)}}\GellMann_{\GMindex_{n(\nqubit-2)}}\cdots \GellMann_{\GMindex_{k0}},
\end{equation} 
where $\GMindex_{nm}\in\{0,\ldots,\quditd^2-1\}$ indicates the index of the Gell-Mann matrix operating on the $m$th qudit.
The generalized Gell-Mann matrices $\GellMann_j$  are a convenient basis for qudit operators \cite{Ringbauer2022}. They are of size $\quditd\times\quditd$, and have the properties $\trace(\GellMann_j) =0$ and $\trace(\GellMann_j\GellMann_k) = 2\delta_{jk}$. When $\quditd=2$, $\GellMann_j$ become the standard Pauli matrices: $\GellMann_0(d=2)=I$, $\GellMann_1(d=2)=X$, $\GellMann_2(d=2)=Y$, and $\GellMann_3(d=2)=Z$. 
The detailed definition of $\GellMann_j$ employed in this paper  can be found  in Appendix B of Ref.~\cite{Loetstedt2025}. Note however  that in this paper, we label the generalized Gell-Mann matrices starting from $j=0$ instead of from $j=1$ in \cite{Loetstedt2025}.

The numerical coefficients $h_n$ are obtained in two steps.  First, we derive the qubit (or qudit) representation of the  single-mode operators $\HO^{(k)}$ [see Eq.~\eqref{Eq:harmonicpot}] and $q^\kappa$ ($\kappa=1,2,3$) according to 
\begin{equation}\label{Eq:signlemodeopexp}
o^{(1)}=\sum_{n=1}^{\NHq^{(1)}}h_n^{(1)}\Gamma_n^{(1)}, \quad h_n^{(1)}  =2^{-\frac{\nqubit}{\Nmodes}}\trace (o^{(1)}\Gamma_n^{(1)}),
\end{equation}
where $o^{(1)}$ is $\HO^{(k)}$ or $q^\kappa$, and $\Gamma_n^{(1)}$ acts on  $\frac{\nqubit}{\Nmodes}$ qubits (or on one qudit). Second, the expansion of the complete $\Nmodes$-mode operator  is obtained as a product of the expansions in \eqref{Eq:signlemodeopexp}. For example, the qubit representation of the operator $q_1q_2q_3$ is 
\begin{equation}
(q_1q_2q_3)_{\text{q}}=\sum_{n_1,n_2,n_3=1}^{\NHq^{(1)}}h_{n_1}^{(1)}h_{n_2}^{(1)}h_{n_3}^{(1)}\Gamma_{n_1}^{(1)}\Gamma_{n_2}^{(1)}\Gamma_{n_3}^{(1)},
\end{equation}
where $h_n^{(1)}$ and $\Gamma_n^{(1)}$ are derived in \eqref{Eq:signlemodeopexp} with $o^{(1)}=q$. Third, the qubit representations of $\HO$ and $q_jq_kq_l$ are combined to generate the 
expansion \eqref{Eq:Hq} for the complete Hamiltonian $H_{\text{q}}$.

In order to compare the qubit Hamiltonians in the different encodings with the  qudit Hamiltonian, we define the operator order $\Oo(\Gamma_n)$ as  the number of qubits or qudits 
that
$\Gamma_n$ acts non-trivially on. We use $\NHq(\Oo)$ to denote the number of terms in \eqref{Eq:Hq} having an operator order  $\Oo$.

\subsection{Time evolution by the Suzuki-Trotter approximation}\label{Subsec:suzukitrotter}
Given the qubit or qudit representation \eqref{Eq:Hq}, we calculate a time-dependent  density matrix $\rho(t)$ by the Suzuki-Trotter approximation \cite{Trotter1959,Suzuki1976},
\begin{equation}\label{Eq:STapproxDef}
\rho(t)=\UST^{\Ntrotter}\rho_0 \UST^{\dagger\Ntrotter},
\end{equation}
where $\Ntrotter$ is the number of Suzuki-Trotter steps, $\rho_0$ is the initial state, and 
\begin{equation}\label{Eq:USTDef}
U=\prod_{n=1}^{\NHq}e^{-i\dt h_n \Gamma_n/\hbar}
\end{equation}
is the  first-order Suzuki-Trotter time evolution operator. We use a pure, uncorrelated initial state,
\begin{equation}
\rho_0=|\bm{v}_0\rangle\langle\bm{v}_0|,
\end{equation}
and aim to compute the time-dependent populations $p_{\bm{v}}(t)$, defined as the diagonal elements of $\rho(t)$, 
\begin{equation}
p_{\bm{v}}(t)=\langle\bm{v}|\rho(t)|\bm{v}\rangle.
\end{equation}
The exact populations obtained without the Suzuki-Trotter approximations at a certain value of $\vmax$ can be expressed as 
\begin{align}\label{Eq:exactpop}
p_{\bm{v}}^{\text{exact}}(t)&=\left |\langle \bm{v}|e^{-itH/\hbar}|\bm{v}_0\rangle\right|^2 
\nonumber 
\\
&=\sum_{m,n}e^{-it(E_{m}-E_{n})/\hbar} \alpha_{mn}^{\bm{v}\bm{v}_0},
\end{align}
where the coefficient $\alpha_{mn}^{\bm{v}\bm{v}_0}$ is defined as 
\begin{equation}\label{Eq:alphacoeffDef}
\alpha_{mn}^{\bm{v}\bm{v}_0} = \langle \psi_{m}|\bm{v}_0\rangle \langle \bm{v}_0| \psi_{n}\rangle   \langle \psi_{n}|\bm{v}\rangle \langle \bm{v}| \psi_{m}\rangle,
\end{equation}
 $|\psi_n\rangle$ is an approximate eigenfunction of $H$ defined as 
 \begin{equation}\label{Eq:Happroxpsi}
 |\psi_n\rangle = \sum_{\bm{v}} \psi_{\bm{v}n}|\bm{v}\rangle,
\end{equation}
and  $\psi_{\bm{v}n}$ is an eigenfunction of the matrix representation of $H$ satisfying
\begin{equation}
E_n\psi_{\bm{v}n}=\sum_{v'_1,\ldots,v'_{\Nmodes}=0}^{\vmax}\langle \bm{v}|H|\bm{v}'\rangle \psi_{\bm{v}'n}.
\end{equation}
 
In order to simulate the noise due to the decoherence, present in all currently available quantum computers, we employ the completely depolarizing channel \cite{NielsenChuang}. After each application of the unitary operation $e^{-i\dt h_n \Gamma_n/\hbar}$, we evolve $\rho(t)$ according to 
\begin{equation}\label{Eq:DepolNoisemodel}
\rho(t) \to \errorn{n}\frac{\Id}{d^{\nqubit}} + (1-\errorn{n})\rho(t),
\end{equation}   
where $\Id$ is the $d^{\nqubit}\times d^{\nqubit}$ identity matrix ($\quditd=2$ for qubits and $\quditd>2$ for qudits), and $\errorn{n}$ is the error of the $n$th gate in one Suzuki-Trotter step. We use a simple error model where the gate error $\errorn{n}$ is defined as  
\begin{equation}\label{Eq:errornDef}
\errorn{n} =
\begin{cases}
0 & \text{if } \Oo(\Gamma_n)=1, \\
[2\Oo(\Gamma_n)-3]\errortwoq
 & \text{if } \Oo(\Gamma_n)>1,
\end{cases}
\end{equation}
where $\errortwoq$ is the two-qubit gate error.  
The gate error $\errorn{n}$ is set to zero when the operator order $\Oo(\Gamma_n)=1$ because the single-qubit gate error is typically  more than one order of magnitude smaller than the two-qubit gate error. For example, on Quantinuum's recent 98-qubit Helios trapped-ion device \cite{Ransford2026}, the single-qubit error is $\erroroneq \approx 3\times 10^{-5}$, and $\errortwoq\approx 8\times 10^{-4}$. 
For trapped-ion qudits, a recent study \cite{John2026} reports $\erroroneq \approx 4\times 10^{-4}$ and $\errortwoq\approx 8\times 10^{-3}$.  
The error for a multi-qubit or qudit gate is set to $[2\Oo(\Gamma_n)-3]\errortwoq$ because an $\Oo$-qubit ($\Oo\ge 2$) Pauli rotation gate can be decomposed into $2\Oo-3$ two-qubit gates \cite{Sriluckshmy2023}. On existing quantum computers, only one- and two-qubit gates are implemented. 
The model \eqref{Eq:errornDef} accounts for the decomposition of  a multi-qubit gate into single- and two-qubit gates before executing the quantum circuit.
In the qudit simulations, we consider in Sec.~\ref{Sec:Results} models where only one-qudit  and two-qudit gates are required.

Even in the absence of noise ($\errortwoq=0$), the Suzuki-Trotter approximation  defined in Eqs.~\eqref{Eq:STapproxDef} and \eqref{Eq:USTDef} implies an algorithmic error 
\begin{equation}\label{Eq:STerror}
|e^{-i\dt H_{\text{q}}/\hbar}-U|\approx \varepsilon_{\text{ST}}= \frac{\dt^2}{2\hbar^2}\left|\sum_{n}^{\NHq}\sum_{m=n+1}^{\NHq}h_nh_m[\Gamma_n,\Gamma_m]\right|,
\end{equation}
where $[\cdot,\cdot]$ is the commutator and we assume that $\dt$ is small. Because the signs of the different terms in the sum on the right hand side in Eq.~\eqref{Eq:STerror} can be changed by changing the order of the terms in the expansion \eqref{Eq:Hq}, the Suzuki-Trotter error depends on the order of the terms in  \eqref{Eq:Hq}. We have applied the following simplified method for optimizing the order of the terms in the Suzuki-Trotter expansion. First, we compute the commutator score 
\begin{equation}\label{Eq:CommScore}
s_n=\sum_{m=1}^{\NHq}\left|h_nh_m[\Gamma_n,\Gamma_m]\right|,
\end{equation}
where $|\cdot |$ refers to the Frobenius norm. We then define an initial ordering by sorting the terms $h_n\Gamma_n$ in order of decreasing $s_n$. In order to minimize the error $\varepsilon_{\text{ST}}$, we 
sequentially check the  terms $h_n\Gamma_n$ from $n=1$ to $n=\NHq-1$ and change the order if a local swap $h_n\Gamma_n\leftrightarrow h_{n+1}\Gamma_{n+1}$ lowers the value of $\varepsilon_{\text{ST}}$. We repeat the local swapping until $\varepsilon_{\text{ST}}$ is not further lowered, and use this final ordering in the Suzuki-Trotter simulations. Although this simple optimization method does not guarantee the best ordering, it results in a small  Suzuki-Trotter error where the populations obtained using the Suzuki-Trotter approximation differ from the exact populations by less than 0.1 (see Sec.~\ref{Sec:Results}), and ensures that the first-order Suzuki-Trotter approximation \eqref{Eq:USTDef} is of similar accuracy  for all three encoding schemes (binary, direct, and qudit). 

We  compare the ``local swap'' method introduced above with two other ordering schemes, the ``magnitude'' scheme where the terms are ordered according to the value of 
$|h_n|$ and the ``commuting subset'' scheme, where the terms are grouped into mutually commuting subsets \cite{Tranter2019}.
  A comparison of the three different ordering schemes is summarized in Appendix~\ref{Sec:STOrdering}, where we conclude that the ``local swap'' method employed here gives good results for both qubit and qudit encodings.
An extended  discussion of the operator ordering in the Suzuki-Trotter approximation can be found in 
 \cite{Hastingsetal2015,Poulinetal2015,Tranter2019}.

\section{Results}\label{Sec:Results}

\subsection{Two-mode model of CO$_2$}\label{Subsec:CO2}
In this section, we compare the three different encoding schemes, qubit binary, qubit direct, and qudit by carrying out time-dependent Suzuki-Trotter simulations of population dynamics in 
a two-mode ($\Nmodes = 2$) model of CO$_2$ with the symmetric stretch ($\nu_1$) and bending ($\nu_2$) modes. This model is the same as that we employed in our previous investigations \cite{Loetstedt2021,Loetstedt2022_copy}. The numerical values of the harmonic frequencies $\omega_k$ and the anharmonic constants $f_{jkl}$  taken from \cite{Suzuki1968} are $\omega_1=1354.31$ cm$^{-1}$, $\omega_2=672.85 \text{ cm}^{-1}$,
$f_{111}=-45.78 \text{ cm}^{-1}$, and
$f_{122}=74.72 \text{ cm}^{-1}$ (all other $f_{jkl}=0$).
The programs employed for the simulations described in this study  are written using {\sc matlab} \cite{MATLAB_2024b}.

CO$_2$ features a well-known anharmonic resonance, referred to as a Fermi resonance \cite{Fermi1931,RodriguezGarciaetal2007}, where the singly excited symmetric stretching mode is strongly coupled to the doubly excited bending mode. The strong coupling arises from the non-zero value of the anharmonic coupling $f_{122}$ ($=74.72$ cm$^{-1}$ in our model) and because 
$2\omega_2\approx \omega_1$. The Fermi doublet refers to the pair of states having approximate wave functions $|\psi_{\text{Fermi}\pm}\rangle\approx (|10\rangle\pm |02\rangle)/\sqrt{2}$, separated in energy by approximately  $f_{122}$.

\subsubsection{Qubit and qudit Hamiltonians}\label{Subsubsec:CO2Hamiltonian}

\begin{figure}
	\includegraphics[width=\figwidth]{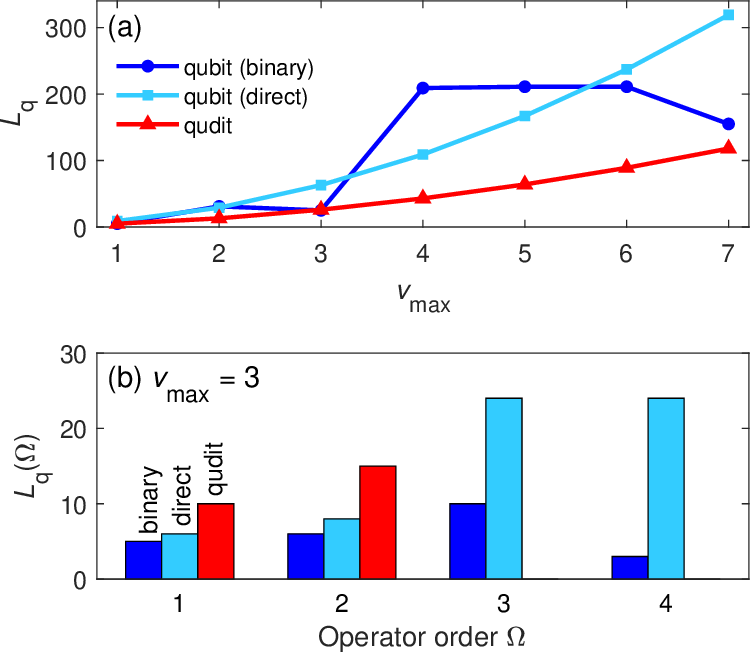}
	\caption{\label{Fig1} (a) Number of terms $\NHq$ in the qubit/qudit Hamiltonian for the two-mode CO$_2$ model.  The binary, direct, and qudit encodings are compared. (b) Breakdown of $\NHq$ into the number of terms of different operator order $\Oo$, for a maximum vibrational quantum number $\vmax=3$. $\Oo$ is defined as the number of qubits or qudits an operator acts non-trivially on. We have $\NHq(3)=\NHq(4)=0$ for the qudit encoding because the number of qudits is $\nqubit=2$.
	}
\end{figure}

In Fig.~\ref{Fig1}(a), we show the number of terms $\NHq$ in the  expansion \eqref{Eq:Hq} of the qubit and qudit Hamiltonians for CO$_2$. We can see Fig.~\ref{Fig1} that $\NHq$ increases approximately as a quadratic function of $\vmax$ for the qubit direct and qudit encodings. We find that in the range $\vmax\le 7$ shown in Fig.~\ref{Fig1}, $\NHq$ for the direct and qudit encodings can be  fitted approximately to $\NHq\approx a\vmax^2$ with 
with $a \approx 6.59$ for the direct encoding and $a \approx 2.47$ for the qudit encoding. The reason for the approximately quadratic increase of $\NHq$ with increasing $\vmax$ is that we have two modes, and the number of non-zero matrix elements  of single mode operators like $q$, $q^2$, and $q^3$ increases linearly with increasing $\vmax$. 
On the other hand, the number of terms $\NHq$ in the binary encoding displays a non-monotonic $\vmax$-dependence. $\NHq$ takes a small value when $\vmax +1$ can be expressed as a power of 2, that is, at $\vmax = 3$ and 7, something which is an intrinsic feature of the binary encoding \cite{Sawayaetal2020}. When $\vmax +1=2^\eta$ with an integer $\eta$, all qubit states represent vibrational basis states.

In Fig.~\ref{Fig1}(b), we show the number of terms in the qubit/qudit Hamiltonians at each operator order $\Oo$, for $\vmax = 3$. 
While the qudit Hamiltonian contains only single-and two-qudit operators ($\Oo=1$ and 2), both the binary and direct qubit Hamiltonians contain terms with operator orders three and four. Because more than one qubit is used to represent one vibrational mode in the binary and direct encodings, many-qubit operators are required to represent both the anharmonicity of a single mode (terms like $q_1^3$ in $\HI$) as well as the anharmonic coupling between the two modes. In the case of the qudit encoding, where one mode is encoded in one qudit, the intra-mode anharmonic coupling is implemented by single-qudit operators, and the inter-mode couplings are implemented by two-qudit operators. 

Because the major source of noise in currently available quantum computers is the two-qubit gates, a smaller number of two-qubit or two-qudit gates results in more accurate (less noise-prone) simulations. Assuming that a Pauli rotation gate $\exp(-i \dt h_n \Gamma_n/\hbar)$ can be decomposed into  
$2\Oo(\Gamma_n)-3$ two-qubit gates as discussed below Eq.~\eqref{Eq:errornDef}, we obtain at $\vmax =3$ 
the  number of two-qubit gates in the binary encoding, the  number of two-qubit gates in the direct encoding, and the  number of two-qudit gates in one Suzuki-Trotter step as
\begin{equation}\label{Eq:N2qcompare}
\begin{aligned}
&\Ntwoq(\text{binary})&={}&  51, 
\\ 
&\Ntwoq(\text{direct}) &={}& 200, \text{ and}   
\\ 
&\Ntwoq(\text{qudit}) &={}& 15.
\end{aligned}
\end{equation}
 Note that we do not account for circuit transpilation, which may change the two-qubit gate count depending on the  qubit layout on a particular quantum device \cite{Sivarajah2020, Kremer2024}.
As is clear from Eq.~\eqref{Eq:N2qcompare}, the qudit encoding results in much fewer two-qudit gates than the binary and direct encodings, and we therefore expect that the qudit encoding leads to more accurate population dynamics.

\subsubsection{Time-dependent vibrational dynamics}\label{Subsubsec:CO2TimeDepVibDyn}

\begin{figure}
	\includegraphics[width=\figwidth]{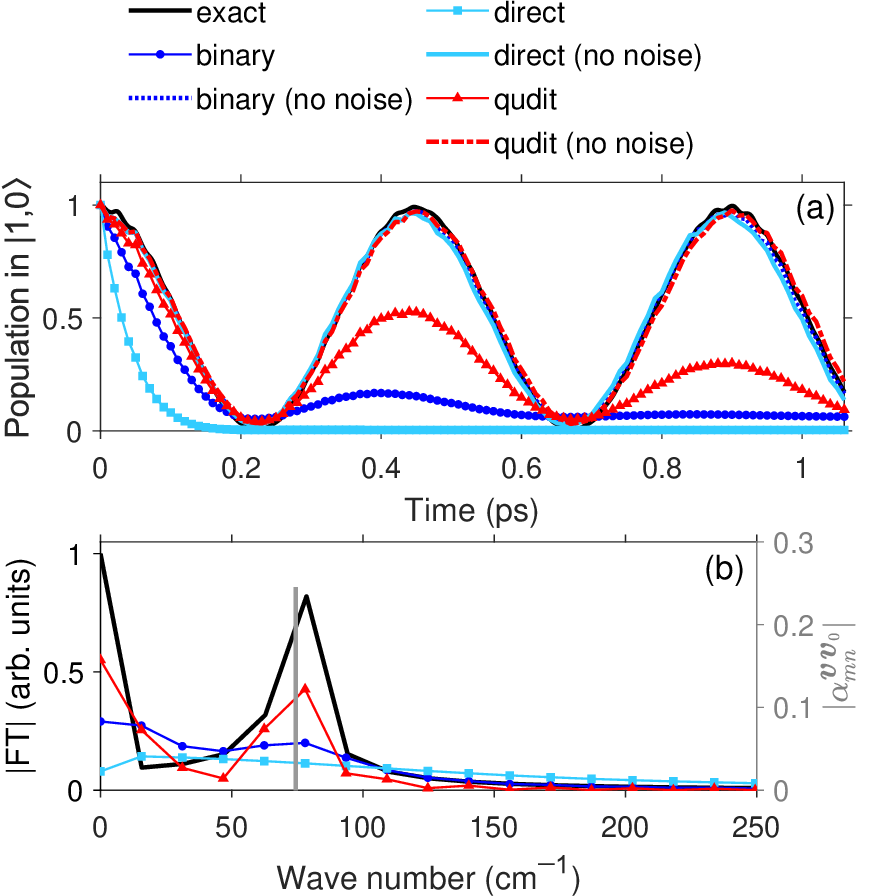}
	\caption{\label{Fig2} (a) Time-dependent population $p_{10}(t)$ in the $|\bm{v}\rangle = |1,0\rangle$ state in the two-mode CO$_2$ model with $\vmax = 3$. The initial state is $|\bm{v}_0\rangle = |1,0\rangle$ and the Suzuki-Trotter step size is $\dt=0.01$ ps.
	(b) Normalized absolute value of the Fourier transform of the time-dependent populations in (a). The Fourier transform of the noiseless Suzuki-Trotter simulations are not shown because they are almost the same as as the exact curve. The vertical gray stick indicates the Fermi resonance energy difference $\Delta E_{\text{Fermi}} = 74.4$ cm$^{-1}$ obtained by matrix diagonalization of $H_{\text{q}}$. The length of the gray stick (numerical value indicated on the right-hand vertical axis) is given by the absolute value of the expansion coefficient $\alpha_{mn}^{\bm{v}\bm{v}_0}$ [see Eq.~\eqref{Eq:alphacoeffDef}] in the exact population.
	}
\end{figure}

In Fig.~\ref{Fig2}, we show an example of the time-dependent population $p_{v_1v_2=10}(t)$ in the two-mode CO$_2$ model at $\vmax=3$, for which the number of terms in the binary qubit Hamiltonian, $\NHq(\text{binary})=25$, and the qudit Hamiltonian, $\NHq(\text{qudit})=26$,  are almost equal. 
 The total number of qubits required for $\vmax=3$ is $\nqubit(\text{binary})=4$ and $\nqubit(\text{direct})=8$.  For the qudit simulation, we require two qudits having $\quditd=4$. 
The numerical values of $h_n$ and the ordering of the terms in the Suzuki-Trotter approximation \eqref{Eq:USTDef} are available at \cite{Lotstedt_qubitquditDataset}. We employ the same value for the two-qubit gate error and the two-qudit gate error, $\errortwoq=10^{-3}$. This value is similar to what is achieved in currently available qubit-based trapped-ion quantum computers  \cite{Moses2023_PRX,Ransford2026}. For qudit-based trapped-ion quantum computers, two-qudit gate errors ranging from about $\errortwoq = 5\times 10^{-2}$
\cite{Ringbauer2022,Zalivako2025} to $\errortwoq \approx 1\times 10^{-2}$ \cite{Hrmo2023,John2026} have been demonstrated. 

The initial state is taken to be $|\bm{v}_0\rangle = |1,0\rangle$, corresponding to one of the basis states involved in the Fermi resonance. Because of the  coupling between the $|1,0\rangle$ and  $|0,2\rangle$ states, the population in $|1,0\rangle$ is transferred to $|0,2\rangle$ on a time scale $2\pi \hbar/\Delta E_{\text{Fermi}}$, where $\Delta E_{\text{Fermi}}$ is the Fermi doublet energy splitting.

We can see in Fig.~\ref{Fig2}(a) that the exact population $p_{10}^{\text{exact}}(t)$  (black solid curve) obtained without using the Suzuki-Trotter approximation and the populations obtained using the Suzuki-Trotter approximation with zero noise (dotted, solid, and dash-dotted curves) almost overlap, showing that the algorithmic error in the Suzuki-Trotter approximation is sufficiently small. We have $|p_{10}^{\text{no noise}}(t)-p_{10}^{\text{exact}}(t)|<0.08$ during the time range $0\le t \le 1$ ps for both the qubit and qudit encodings. The populations simulated using the completely depolarizing noise model decay because of the noise. According to the noise model \eqref{Eq:DepolNoisemodel}, the noisy populations are approximately given by 
\begin{align}\label{Eq:pnoisyapprox}
p_{\bm{v}}^{\text{noisy}}(t)&\approx(1-\errortwoq)^{\Ntrotter \Ntwoq}p_{\bm{v}}^{\text{no noise}}(t)
\nonumber
\\
&\approx e^{-\frac{t}{\decaytime}}p_{\bm{v}}^{\text{no noise}}(t),
\end{align}
where the decay time $\decaytime$ is given by 
\begin{equation}\label{Eq:decaytimeDef}
\decaytime=\frac{\dt}{\Ntwoq \errortwoq},
\end{equation}
and $\Ntwoq$ is the number of two-qubit/qudit gates in  one Suzuki-Trotter step. We obtain $\decaytime_{\text{binary}} \approx 0.20$ ps,  $\decaytime_{\text{direct}} \approx 0.050$ ps, and $\decaytime_{\text{qudit}} \approx 0.67$ ps for the time-dependent populations shown in Fig.~\ref{Fig2}.

In order to corroborate the assumption of a vanishing single-qubit/qudit gate error  in the noise model [see Eq.~\eqref{Eq:errornDef}],
  we carry  out simulations under the same conditions as those in the simulation shown in Fig.~\ref{Fig2} and add a  depolarization channel 
  [see Eq.~\eqref{Eq:DepolNoisemodel}] also for each single-qubit/qudit gate. We consider two  values of the single-qubit/qudit gate error, $\erroroneq=10^{-5}$ and $\erroroneq=10^{-4}$. We find that for the qubit simulations, the  deviation  from the populations obtained using  $\erroroneq=0$ during  $0\le t\le1$ ps is as small as 
  $|p_{10}^{\erroroneq=0}(t)-p_{10}^{\erroroneq=10^{-5}}(t)|< 2.2\times 10^{-4}$ and
    $|p_{10}^{\erroroneq=0}(t)-p_{10}^{\erroroneq=10^{-4}}(t)|< 2.2\times 10^{-3}$ for the binary encoding and 
  $|p_{10}^{\erroroneq=0}(t)-p_{10}^{\erroroneq=10^{-5}}(t)|< 1\times 10^{-4}$ and
    $|p_{10}^{\erroroneq=0}(t)-p_{10}^{\erroroneq=10^{-4}}(t)| < 1\times 10^{-3}$ for the direct encoding. For the qudit encoding, we obtain 
  $|p_{10}^{\erroroneq=0}(t)-p_{10}^{\erroroneq=10^{-5}}(t)|< 2.2\times 10^{-3}$ and
    $|p_{10}^{\erroroneq=0}(t)-p_{10}^{\erroroneq=10^{-4}}(t)| < 2.1\times 10^{-2}$, which are about ten times  larger than for the qubit encoding, but still small enough to validate the $\erroroneq=0$ approximation. We conclude that as long as the single-qubit/qudit gate error is at least one order of magnitude smaller than the two-qubit gate error, we can neglect 
    the error originated from the single-qubit/qudit gates.
    
   In order to check  how the results shown in Fig.~\ref{Fig2}(a) depend on  the noise model,  we also investigate a local depolarizing noise model, where only the qubits/qudits acted on by a gate are depolarized. As shown in detail in Appendix~\ref{Sec:CO2localdepolnoise}, we obtain very similar results for both the global depolarization channel defined in 
    Eq.~\eqref{Eq:DepolNoisemodel} and the local depolarization channel defined in Eq.~\eqref{Eq:DepolNoisemodelLocal}, confirming that 
   the global depolarizing channel is an adequate noise  model for analyzing the difference of the qudit and qubit encodings.

In  Fig.~\ref{Fig2}(b), we show the Fourier transform of the time-dependent population in Fig.~\ref{Fig2}(a). The exact and qudit Fourier spectra show a clear peak at $74$ cm$^{-1}$ (the Fermi resonance gap). The simulation performed using the qubit binary encoding results in a broad peak around $74$ cm$^{-1}$, while no peak can be seen in the curve obtained using the qubit direct encoding, reflecting the fast decay time. The Fourier spectra in Fig.~\ref{Fig2}(b) suggest that  we can retrieve energy differences by recording a time-dependent  observable (a population, for example) and computing the Fourier transform, as is also clear from Eq.~\eqref{Eq:exactpop}. This Fourier transform approach has been successfully implemented experimentally for the high-precision measurements of atomic and molecular transition energies \cite{Ando2018,Ando2025}. However, in order to derive  vibrational energy differences with a resolution  of 
$\Delta\tilde{\nu}= 1 \text{ cm}^{-1}$  from the Fourier transform of a signal calculated on a quantum computer, we would need to continue the simulation until $t\sim 2\pi/c\Delta\tilde{\nu}\approx 200$ ps, which is difficult at the currently realized error rate  of $\errortwoq \approx 10^{-3}$. 

While we have used the same value ($\errortwoq=10^{-3}$) of the two-qubit/qudit gate error for both qubit and qudit simulations in Fig.~\ref{Fig2}, 
resulting in a longer decay time $\decaytime$ for qudits, we can  use Eq.~\eqref{Eq:decaytimeDef} to estimate the two-qudit gate error $\errortwoq^{\text{qudit}}$ 
for which the decay time becomes the same for qubit and qudit quantum computers.  Assuming the same value of $\dt$ for qubits and qudits, we should have 
\begin{equation}
\errortwoq^{\text{qudit}}=\frac{\Ntwoq^{\text{qubit}} }{\Ntwoq^{\text{qudit}}}\errortwoq^{\text{qubit}}
\end{equation}
for obtaining $\decaytime_{\text{qudit}}=\decaytime_{\text{qubit}}$.
Using   $\Ntwoq^{\text{qubit}}=\Ntwoq(\text{binary})$ from Eq.~\eqref{Eq:N2qcompare}, and assuming the currently realized value $\errortwoq^{\text{qubit}}=10^{-3}$ for the two-qubit gate error,  we obtain $\errortwoq^{\text{qudit}}=3\times 10^{-3}$. This value is roughly one order of magnitude smaller than the two-qudit gate errors reported for qudit-based trapped-ion quantum computers \cite{Ringbauer2022,Hrmo2023,Zalivako2025,John2026}.
Theoretically, it has been suggested that two-qudit gate errors smaller than $10^{-2}$ can in principle be achieved by trapped-ion quantum computers \cite{Low2020}.

\subsection{Three-mode model of H$_2$O}\label{Subsec:H2O}
In this section, we compare the qubit and qudit  encodings for a  model of H$_2$O, including three vibrational modes: symmetric stretch ($\nu_1$), bending ($\nu_2$), and anti-symmetric  stretch ($\nu_3$). We employ the following values of the harmonic frequencies $\omega_k$ and the anharmonic constants $f_{jkl}$, taken  from \cite{Csaszar1997}:
\begin{equation}\label{Eq:H2Oomega}
\begin{array}{@{} r @{} >{{}}l<{{}} @{} r @{}}
\omega_1 & =& 3843.74 \text{ cm}^{-1},  
\\
\omega_2 & =&1641.18 \text{ cm}^{-1},  
\\
\omega_3&=&3948.48 \text{ cm}^{-1},  
\end{array}
\end{equation}
and
\begin{equation}\label{Eq:H2Ofanharm}
\begin{array}{@{} r @{} >{{}}l<{{}} @{} r @{}}
f_{ 111 }  & =&  303.64 \text{ cm}^{-1},  
\\
f_{ 112 }  & = & 39.02 \text{ cm}^{-1},  
\\
f_{ 122 }  & = & -162.13 \text{ cm}^{-1},  
\\
f_{ 222 }  & = & -43.96 \text{ cm}^{-1},  
\\
f_{ 133 }  & = & 911.05 \text{ cm}^{-1},  
\\
f_{ 233 }  & = & 134.59 \text{ cm}^{-1}. 
\end{array}
\end{equation}
All other $f_{jkl}$ equal 0.

\subsubsection{Qubit and qudit Hamiltonians}\label{Subsubsec:H2OHamiltonian}

\begin{figure}
	\includegraphics[width=\figwidth]{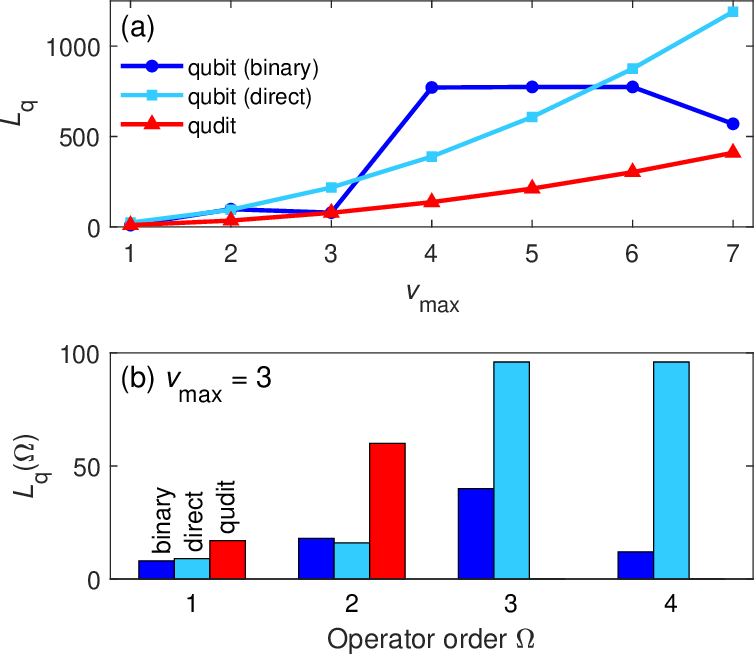}
	\caption{\label{Fig3} (a) Number of terms $\NHq$ in the qubit/qudit Hamiltonian for the three-mode model of H$_2$O. (b) Number of terms $\NHq(\Oo)$ for each operator order $\Oo$, at $\vmax=3$. 
	}
\end{figure}

In Fig.~\ref{Fig3}, we show the number of terms $\NHq$ in the qubit and qudit  H$_2$O Hamiltonians. Because of the larger number of modes, $\NHq$ is larger than for the two-mode CO$_2$ model, but shows the same behavior as a function of $\vmax$ as in Fig.~\ref{Fig1}(a). The distribution of the number of terms as a function of the operator order shown in Fig.~\ref{Fig3}(b) is also similar to that of CO$_2$: The direct and binary qubit encodings result in  qubit Hamiltonians containing  three-and four-qubit operators, while the qudit Hamiltonian only contains one- and two-qudit  operators. 
At $\vmax=3$, the binary qubit encoding requires six qubits, the direct qubit encoding requires 12 qubits, and three qudits are used in the qudit encoding. The reason for the absence of  three-qudit operators in the qudit Hamiltonian is that there are no three-mode interaction terms in the vibrational Hamiltonian [see Eq.~\eqref{Eq:H2Ofanharm}].

\subsubsection{Time-dependent vibrational dynamics}\label{Subsubsec:H2OTimeDepVibDyn}

\begin{figure}
	\includegraphics[width=\figwidth]{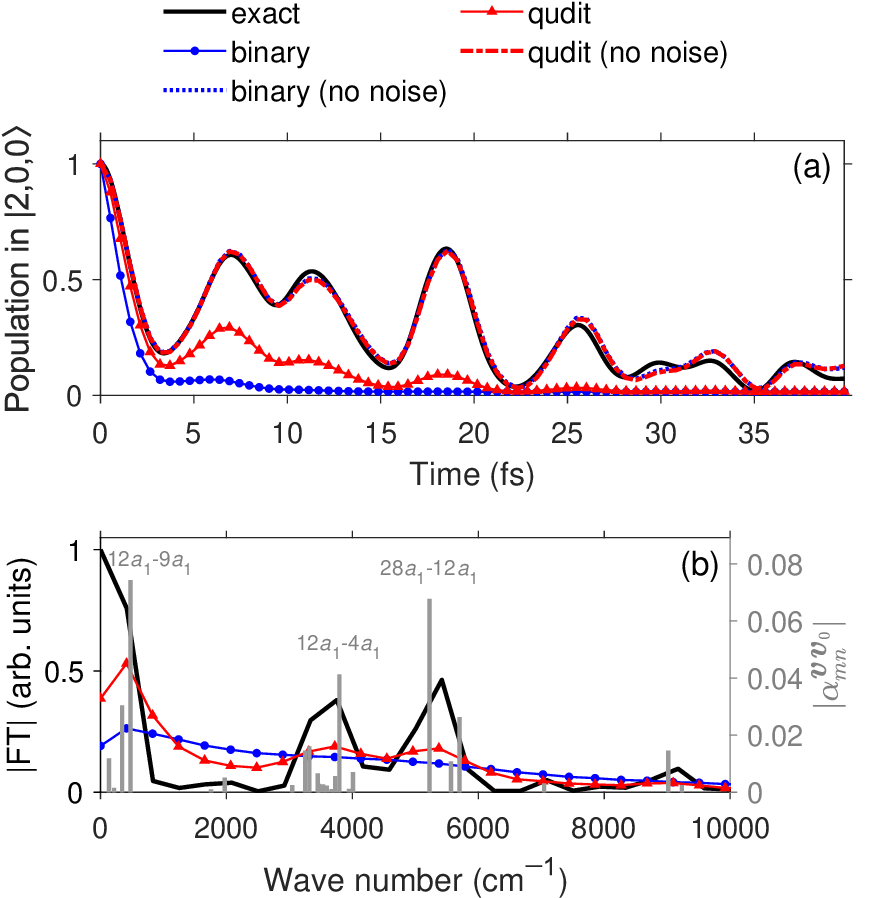}
	\caption{\label{Fig4} (a) Time-dependent population $p_{200}(t)$ in the $|\bm{v}\rangle = |2,0,0\rangle$ state for H$_2$O with $\vmax = 3$. The initial state is $|\bm{v}_0\rangle = |2,0,0\rangle$, the Suzuki-Trotter step size is $\dt=0.53$ fs, and the two-qubit/qudit gate error is $\errortwoq = 10^{-3}$.
The ``binary (no noise)'' and ``qudit (no noise)'' curves overlap almost completely.
	(b) Normalized absolute value of the Fourier transform of the time-dependent populations in (a). The Fourier transform of the noiseless Suzuki-Trotter simulations are not shown. 
	The horizontal positions of the gray sticks indicate the energy differences $\Delta E_{mn}= E_m-E_n$
	 obtained by matrix diagonalization of $H_{\text{q}}$.  The length of each gray stick (right-hand vertical axis) is determined by the absolute value of the expansion coefficients $\alpha_{mn}^{\bm{v}\bm{v}_0}$ [see Eq.~\eqref{Eq:alphacoeffDef}]. The state labels of the dominant transitions (see Table~\ref{Table1} in Appendix~\ref{Sec:VibEnergiesH2O}) are indicated above the respective stick.
	}
\end{figure}

In Fig.~\ref{Fig4}, we show time-dependent populations in H$_2$O for $\vmax=3$, assuming that the initial state is $|v_1,v_2,v_3\rangle = |2,0,0\rangle$. Because the direct qubit encoding results in a large number of terms in the qubit Hamiltonian ($\NHq=218$), we only compare the qubit binary encoding ($\NHq=79$) and the qudit $(\quditd=4)$ encoding ($\NHq=78$). We use a value of $\errortwoq=10^{-3}$ in the qubit as well as in the qudit simulations.  The number of two-qubit gates in the circuit for one Suzuki-Trotter step at  $\vmax=3$ becomes $\Ntwoq(\text{binary})= 198$ compared to $\Ntwoq(\text{qudit})= 60$ two-qudit gates in the qudit circuit. Even though $\NHq(\text{qudit})\approx \NHq(\text{binary})$, $\Ntwoq(\text{qudit})$ is much smaller than $\Ntwoq(\text{binary})$ because of the absence of three- and four qudit terms in the qudit Hamiltonian.

The error arising from the Suzuki-Trotter approximation is small, as can be seen by comparing the ``no noise'' and ``exact'' curves in Fig.~\ref{Fig4}(a). We have $|p_{200}^{\text{no noise}}(t)-p_{200}^{\text{exact}}(t)|<0.06$ for both qubit and qudit encodings in the time range $0\le t\le 40$ fs shown in Fig.~\ref{Fig4}(a). The curves obtained by the simulation including the noise model are strongly damped. We obtain
$\decaytime_{\text{binary}} \approx 2.7$ fs  and $\decaytime_{\text{qudit}} \approx 8.8$ fs for the decay times as defined in Eq.~\eqref{Eq:decaytimeDef}.

We comment that the results shown in Figs.~\ref{Fig2} and \ref{Fig4} could be improved by applying error mitigation \cite{EndoCaiBenjaminYuan2021,Cai2023} such as 
 zero-noise extrapolation \cite{TemmeBravyiGambetta2017,LiBenjamin2017}, or 
probabilistic error cancellation \cite{TemmeBravyiGambetta2017,Berg2023}. While error mitigation methods specific to qudit systems have not been widely studied, 
error mitigation methods originally developed for qubits  have been experimentally demonstrated on superconducting transmon qutrits \cite{Goss2024,Spagnoli2025}. 
We also note that if the noise can be modeled as a completely depolarizing channel and that if the gate error can be inferred from measurements of auxiliary circuits or by curve fitting, the error can be mitigated by 
applying Eq.~\eqref{Eq:pnoisyapprox} to estimate the noiseless population \cite{Urbaneketal2021,Shinjo2026}.

The Fourier spectrum of the time-dependent populations is plotted in Fig.~\ref{Fig4}(b). There are many frequency components contributing to the time-dependent populations, as can be seen by the large number of gray vertical sticks in Fig.~\ref{Fig4}(b). For reference, we list in Appendix~\ref{Sec:VibEnergiesH2O} the eigenenergies of our  H$_2$O model ($\vmax = 3$) 
obtained by matrix diagonalization. Broad peaks can be seen around the main three groups of transitions at $500$ cm$^{-1}$, $3800$ cm$^{-1}$, and  $5200$ cm$^{-1}$ in the case of the exact and qudit simulations. No clear peaks can be seen in the Fourier spectrum obtained in the qubit simulation. 

In order to assess the advantage of using qudits for simulations of molecules modeled by an anharmonic potential containing terms of higher order than three, we conduct simulations using a fourth-order anharmonic potential. In this case, the qudit Hamiltonian contains three-qudit terms for which three-qudit gates are required. The details of the simulation are described in Appendix~\ref{Sec:H2OfourthorderAH}. We find that  even when a fourth-order anharmonic potential is adopted, the qudit encoding results in slightly more accurate time-dependent populations if we assume that the  error of the three-qudit gates is the same as for the three-qubit gates.

\subsubsection{Time-dependent vibrational dynamics at small error rates}\label{Subsubsec:H2OTimeDepVibDynFTQC}

\begin{figure}
	\includegraphics[width=\figwidth]{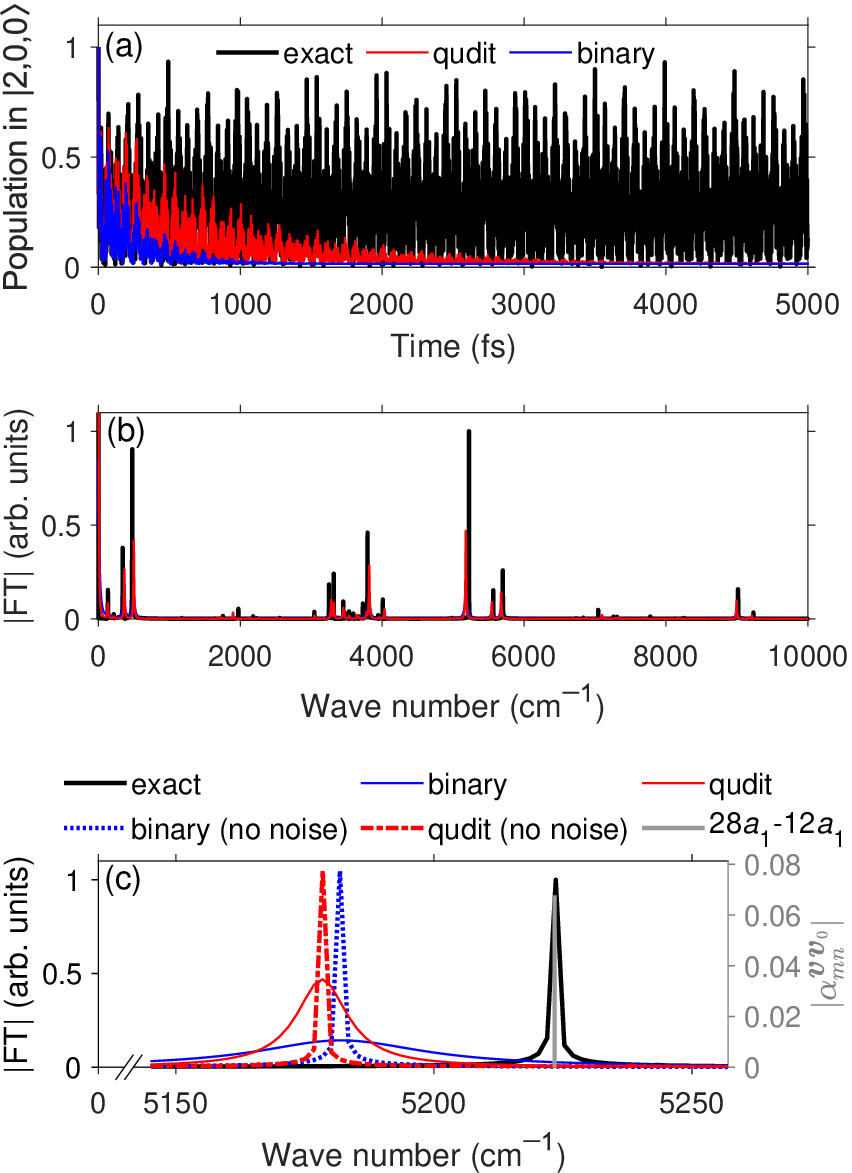}
	\caption{\label{Fig5} (a) Time-dependent population $p_{200}(t)$ in the $|\bm{v}\rangle = |2,0,0\rangle$ state of H$_2$O. All parameters are the same as those employed in Fig.~\ref{Fig4} ($\vmax = 3$, $|\bm{v}_0\rangle = |2,0,0\rangle$, $\dt=0.53$ fs), except   the two-qubit/qudit gate error which is $\errortwoq = 10^{-5}$.
	(b) Normalized absolute value of the Fourier transform of the time-dependent populations in (a). In panels (a) and (b), the no-noise curves are not shown to avoid cluttering the plot.
	(c) Zoom-in of the Fourier spectra in (b) in the wave number range $5150 \text{ cm}^{-1}\le\tilde{\nu}\le 5250 \text{ cm}^{-1}$. The curves obtained without noise (but with the Suzuki-Trotter approximation) are also shown.
	The gray vertical stick indicates the exact energy difference $E_{28a_1}-E_{12a_1}=5223.4$ cm$^{-1}$ obtained by matrix diagonalization of $H_{\text{q}}$. 
	The length of the gray stick is given by $|\alpha_{mn}^{\bm{v}\bm{v}_0}|$ [see Eq.~\eqref{Eq:alphacoeffDef}], with the numerical value shown on the right-hand vertical axis. The ``binary'' and ``qudit'' curves obtained with a noise model in  (b) and (c) are multiplied by a factor of five  to make the comparison with the other curves easier.
	}
\end{figure}

In view of the prospects of fault-tolerant quantum computing, it is interesting to consider smaller gate errors $\errortwoq$ than what is realized in current quantum hardware. For example, in
Quantinuum's roadmap \cite{QuantinuumRoadmap}, a logical gate error between $10^{-5}$ and $10^{-10}$ is expected to be realized in 2029. In Fig.~\ref{Fig5}, we show a simulation of the time-dependent vibrational dynamics employing the same parameters as in Fig.~\ref{Fig4}, except for a smaller two-qubit/qudit gate error of $\errortwoq = 10^{-5}$. The decay times are two orders of magnitude longer, $\decaytime_{\text{binary}} \approx 270$ fs  and $\decaytime_{\text{qudit}} \approx 880$ fs. As can be seen in Fig.~\ref{Fig5}(b), because of the increased resolution due to the long decay time, most of the peaks in the Fourier spectrum obtained by the noisy qudit simulation can now be resolved. 
In Figs.~\ref{Fig5}(a) and (b), the curves obtained using the Suzuki-Trotter approximation without noise are not shown to avoid cluttering the plots. As
 illustrated in the enlarged view in Fig.~\ref{Fig5}(c), the positions of the peaks obtained in the Suzuki-Trotter simulations are different from those obtained in the exact simulation because of the algorithmic error in the Suzuki-Trotter approximation [see Eq.~\eqref{Eq:STerror}]. In order to increase the accuracy, a higher-order Suzuki-Trotter formula \cite{Ostmeyer2023} needs to be used.

\section{Summary}\label{Sec:Summary}
We have simulated vibrational dynamics of two molecules, CO$_2$ and H$_2$O, using qubit and qudit encodings, including the effect of noise. We showed that for both molecules, the qudit representation of the Hamiltonian is more compact than the qubit representation, and therefore, the  simulations using the  qudit encoding are more accurate than those using the qubit encodings resulting in the time-dependent populations  closer to the populations obtained in the absence of noise. Our results show that simulation of vibrational dynamics is an interesting  application of existing and future qudit quantum computers. 
We mention that similarly to the damped spin dynamics considered in \cite{Rost2020,Leppaekangas2023},  damped vibrational  dynamics  simulated on noisy quantum computers  may serve as a model for the simulation of vibrational energy relaxation in liquids \cite{Owrutsky1994,Egorov1997}. 
Once a gate error of $\errortwoq \sim 10^{-5}$ or smaller is achieved, more accurate quantum computing of vibrational dynamics of polyatomic molecules becomes possible so that vibrational energy transfer and vibrational energy relaxation can be simulated.

\section{Data availability}\label{Sec:Dataavailability}
The data supporting the findings of this article are
openly available at \cite{Lotstedt_qubitquditDataset}.

\acknowledgments
We thank T.~Nishi (RIKEN Center for Quantum Computing)  for helpful comments. 
We are supported by the JSPS (Kakenhi no.~JP24K08336), the RIKEN TRIP initiative (RIKEN Quantum), and JST-CREST Quantum Frontiers (grant no.~JPMJCR23I7).  We are grateful to the DIC Corporation for their support through the Applied Quantum Chemistry by Qubits
(AQUABIT) project under the UTokyo Quantum Initiative.

\appendix

\section{Comparison of orderings schemes in the Suzuki-Trotter approximation}\label{Sec:STOrdering}

We compare three different ordering schemes for determining the optimal order of the operators $h_n\Gamma_n$ in the Suzuki-Trotter approximation \eqref{Eq:USTDef}.
The first, denoted as the ``Local swap'' scheme, is the scheme described below Eq.~\eqref{Eq:CommScore} and is used in the simulations whose results are shown in Figs.~\ref{Fig2}, \ref{Fig4}, and \ref{Fig5} in the main text. 

\begin{figure}
	\includegraphics[width=\figwidth]{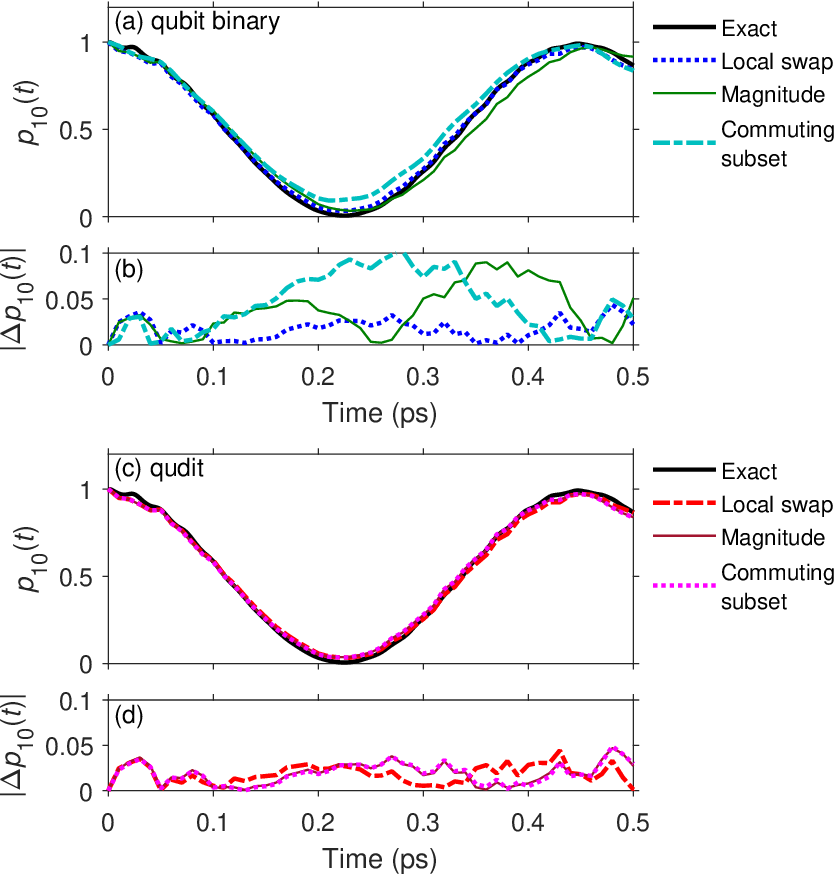}
	\caption{\label{Fig0_appendix}
	Time-dependent population $p_{10}(t)$ in the $|\bm{v}\rangle = |1,0\rangle$ state in the two-mode CO$_2$ model with $\vmax = 3$, simulated using (a) the binary qubit encoding and (c) the qubit encoding. The initial state is $|\bm{v}_0\rangle = |1,0\rangle$ and the Suzuki-Trotter step size is $\dt=0.01$ ps. Noise is not included in the simulations ($\errortwoq=0$). Three ordering schemes,  ``Local swap'', ``Magnitude'', and ``Commuting subset'', are compared with the exact population obtained without using the Suzuki-Trotter approximation. In (b) and (d), we show the absolute value of the difference of the populations obtained using the Suzuki-Trotter approximation and the exact population, as defined in Eq.~\eqref{Eq:DeltapST}. 
	In (c) and (d), the  ``Magnitude'' and ``Commuting subset'' curves overlap almost completely.
	}
\end{figure}

The second scheme, called the ``Magnitude'' scheme  \cite{Tranter2019}, consists in the  ordering of the coefficients $h_n$ in the expansion of the Hamiltonian \eqref{Eq:Hq} in decreasing order and in operating with the largest term first, that is,
\begin{equation}
U=e^{-i\dt h_{\NHq} \Gamma_{\NHq}/\hbar}\cdots e^{-i\dt h_2 \Gamma_2/\hbar}e^{-i\dt h_1 \Gamma_1/\hbar}, 
\end{equation}
with $|h_1|\ge|h_2|\ge\ldots$.

In the third scheme, taken from \cite{Tranter2019} and called here the ``Commuting subset'' scheme, we first group the terms $h_n\Gamma_n$ in the qubit/qudit Hamiltonian  in mutually commuting subgroups according to a greedy algorithm \cite{Tranter2019,Mukhopadhyay2023}. The initial ordering of the terms is determined by  $|h_n|$ with the term having largest $|h_n|$ placed first.  We then go through the mutually commuting subgroups sequentially and pick up  the  term having the largest value of $|h_n|$ as the next term in the qubit/qudit Hamiltonian, so that the term with the   largest $|h_n|$ in group 1 is the first term, the term with the   largest $|h_n|$ in group 2 is the second term, and so on. 

We remark that neither of the three schemes above is guaranteed to produce the optimal Suzuki-Trotter ordering, but that all three schemes are simple to be implemented and scalable to a large number of qubits/qudits.

In Fig.~\ref{Fig0_appendix}, we show the comparison of the three ordering  schemes applied to the simulation of the time-dependent vibrational population in the $|1,0\rangle$ state of  CO$_2$. In order to compare the three schemes quantitatively, we show in  Figs.~\ref{Fig0_appendix}(b) and (d) the difference $\Delta p_{10}(t)$ of the population obtained using the Suzuki-Trotter approximation and the exact population, defined as 
\begin{equation}\label{Eq:DeltapST}
\Delta p_{\bm{v}}(t)=p^{\text{ST}}_{\bm{v}}(t) - p^{\text{exact}}_{\bm{v}}(t).
\end{equation}
We can see in Fig.~\ref{Fig0_appendix} that all three methods result in accurate time-dependent populations with $|\Delta p_{10}(t)|<0.1$.   In the  qubit simulation, 
Fig.~\ref{Fig0_appendix}(b) reveals that  the ``Local swap'' method  gives a slightly smaller population difference than the  other two methods. In the qudit simulations, all methods exhibit a similar performance with $|\Delta p_{10}(t)|<0.05$. For the qudit encoding, the ``Local swap'' and ``Commuting subset'' methods lead to very similar orderings with populations differing by less than $0.01$,  
$|p^{\text{Magnitude}}_{10}(t) - p^{\text{Comm. subset}}_{10}(t)|<0.01$. 

We conclude that the ``Local swap'' method for determining the  order of the operators in the Suzuki-Trotter approximation is an efficient and easily applicable method for time-dependent vibrational dynamics, giving good results for both qubit and qudit encodings.
\section{CO$_2$ simulations using a local depolarizing noise model}\label{Sec:CO2localdepolnoise}
We compare the time-dependent vibrational populations of CO$_2$ obtained by simulations using the global depolariziation noise model defined in Eq.~\eqref{Eq:DepolNoisemodel}, and a local depolarization noise model. In the local depolarization noise model, after each multi-qubit gate $e^{-i\dt h_n \Gamma_n/\hbar}$, the density matrix is transformed as 
\begin{align}\label{Eq:DepolNoisemodelLocal}
\rho(t)\to{} & \frac{\errorn{n}}{4^{\Oo_n}} \sum_{j_1,j_2,\ldots,j_{\Omega_n}=0}^3  \lambda_{j_1}\lambda_{j_2}\cdots \lambda_{j_{\Oo_n}} \rho(t) \lambda_{j_1}\lambda_{j_2}\cdots \lambda_{j_{\Oo_n}}
 \nonumber
 \\
   &+ (1-\errorn{n})\rho(t),
\end{align}
where $\Oo_n=\Oo(\Gamma_n)$ is the operator order of $\Gamma_n$, $\errorn{n}$ is defined in Eq.~\eqref{Eq:errornDef}, and the Pauli matrices $\lambda_j$ ($\lambda_0=I$, $\lambda_1=X$, $\lambda_2=Y$, and $\lambda_3=Z$) operate on the same qubits as $\Gamma_n$. 
Whereas all qubits are affected by the noise in the global  depolarization noise channel defined in   Eq.~\eqref{Eq:DepolNoisemodel}, only the qubits acted on by the gate  $e^{-i\dt h_n \Gamma_n/\hbar}$ are depolarized in the local depolarization  model defined in Eq.~\eqref{Eq:DepolNoisemodelLocal}. Therefore, the local depolarization noise channel is a  better noise model, however, the practical implementation is more demanding. When the total number of qubits is the same as the number of qubits acted on by the gate, the two noise models give identical results.

\begin{figure}
	\includegraphics[width=\figwidth]{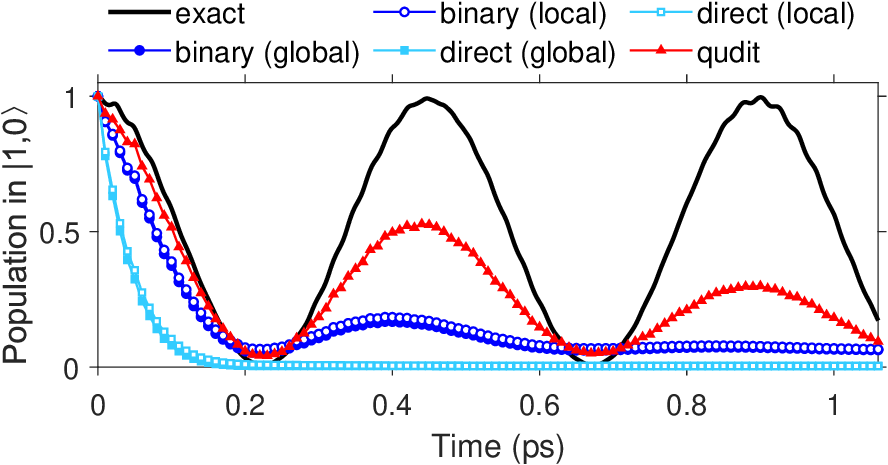}
	\caption{\label{Fig1_appendix}
	Time-dependent population $p_{10}(t)$ in the $|\bm{v}\rangle = |1,0\rangle$ state in the two-mode CO$_2$ model with $\vmax = 3$. The initial state is $|\bm{v}_0\rangle = |1,0\rangle$ and the Suzuki-Trotter step size is $\dt=0.01$ ps.
	Qubit results obtained using the global depolarization model [Eq.~\eqref{Eq:DepolNoisemodel}] are shown by filled symbols, and qubit results obtained using  the local depolarization noise model [Eq.~\eqref{Eq:DepolNoisemodelLocal}] are shown using open symbols.
	The ``exact,'' ``qudit,'', ``binary (global),'', and ``direct (global),'' are the same data as shown in Fig.~\ref{Fig2}(a).}
	
\end{figure}

In Fig.~\ref{Fig1_appendix}, we show the comparison of the time-dependent vibrational population in the $|1,0\rangle$ state of CO$_2$ obtained by the local and global depolarization noise models using $\errortwoq=10^{-3}$. For reference, we also show the exact population and the population obtained using the qudit encoding. In the qudit simulation, the global and local noise models are equivalent because there are only two qudits and the single-qudit gate error is assumed to be zero.  We can see that the global and local noise models result in very similar time-dependent populations, and that the local noise model results in slightly larger values of $p_{01}(t)$. We have for the time interval $0\le t\le 1$ ps shown in Fig.~\ref{Fig1_appendix} $p_{10}^{\text{local}}(t)-p_{10}^{\text{global}}(t)<0.02$ for the binary encoding, and $p_{10}^{\text{local}}(t)-p_{10}^{\text{global}}(t)<0.04$ for the direct encoding.

\section{H$_2$O simulations using a fourth-order anharmonic potential}\label{Sec:H2OfourthorderAH}
In this appendix, we consider an anharmonic potential of H$_2$O of the form
\begin{equation}\label{Eq:anharmonicpotfourthorder}
\HI'= \sum_{\substack{j,k,l=1\\j\le k\le l}}^{\Nmodes} f_{jkl}q_jq_kq_l + \sum_{\substack{j,k,l,m=1\\j\le k\le l\le m}}^{\Nmodes} g_{jklm}q_jq_kq_lq_m,
\end{equation}
where $f_{jkl}$ is given in Eq.~\eqref{Eq:H2Ofanharm}, and $g_{jklm}$ (from  \cite{Csaszar1997}) take the numerical values
\begin{equation}\label{Eq:AHpotfourthorder}
\begin{array}{@{} r @{} >{{}}l<{{}} @{} r @{}}
g_{ 1111 }  & =  31.59 \text{ cm}^{-1} ,
\nonumber \\ 
g_{ 1112 }  & =  10.35 \text{ cm}^{-1} ,
\nonumber \\ 
g_{ 1122 }  & =  -77.23 \text{ cm}^{-1}, 
\nonumber \\ 
g_{ 1222 }  & =  -27.10 \text{ cm}^{-1} ,
\nonumber \\ 
g_{ 2222 }  & =  -0.44 \text{ cm}^{-1} ,
\nonumber \\ 
g_{ 1133 }  & =  190.49 \text{ cm}^{-1} ,
\nonumber \\ 
g_{ 1233 }  & =  59.77 \text{ cm}^{-1} ,
\nonumber \\ 
g_{ 2233 }  & =  -92.80 \text{ cm}^{-1} ,
\nonumber \\ 
g_{ 3333 }  & =  31.89 \text{ cm}^{-1},
\end{array}
\end{equation}
and all other $g_{jklm}=0$.

\begin{figure}
	\includegraphics[width=\figwidth]{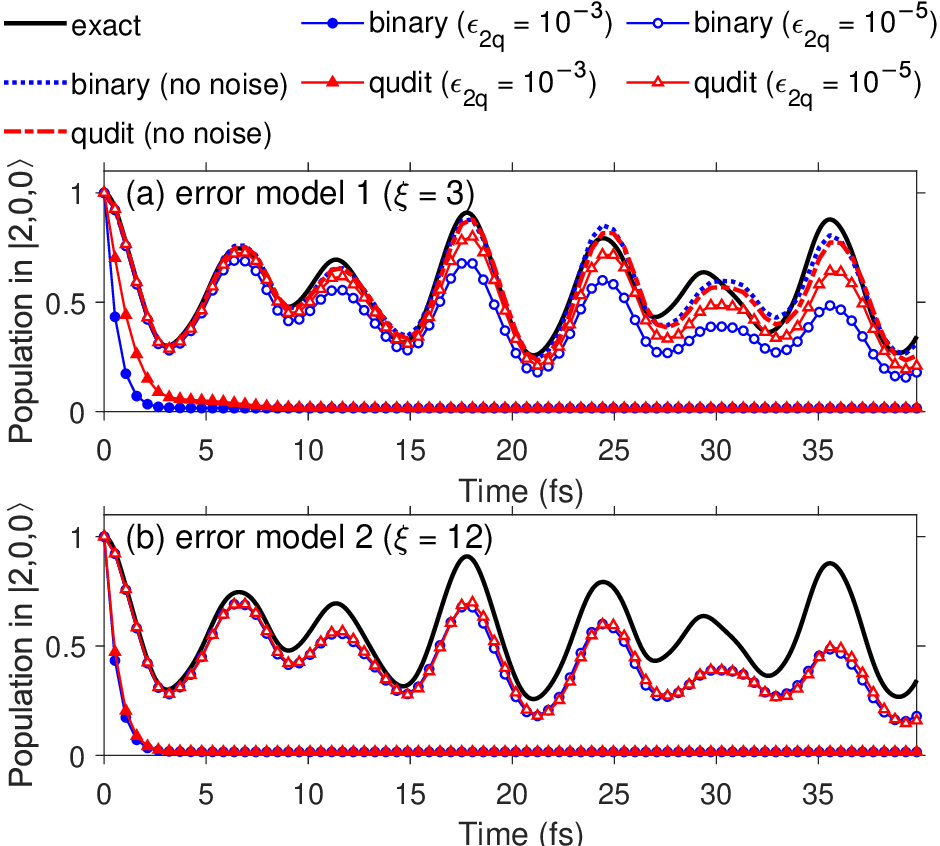}
	\caption{\label{Fig2_appendix}
	Time-dependent population $p_{200}(t)$ in the $|\bm{v}\rangle = |2,0,0\rangle$ state in the three-mode H$_2$O model with $\vmax = 3$, and the  anharmonic potential defined in Eq.~\eqref{Eq:anharmonicpotfourthorder}. The initial state is $|\bm{v}_0\rangle = |2,0,0\rangle$ and the Suzuki-Trotter step size is $\dt=0.53$ fs. 
	In (a) and (b),  $\xi=3$  and $\xi=12$ are employed in the error model \eqref{Eq:errornDefQudit}, respectively.  Two values of the  two-qubit/qudit gate error, 
	$\errortwoq=10^{-3}$ (filled symbols) and $\errortwoq=10^{-5}$ (open symbols), are employed.  Populations obtained without noise (dotted and dash-dotted curves) are shown only in (a).
	}
\end{figure}

The six-qubit Hamiltonian corresponding to $H=\HO+\HI'$   using the binary encoding contains $\NHq=215$ terms [$\NHq(\Oo=1)=11$, $\NHq(2)=52$, $\NHq(3)=67$, $\NHq(4)=43$, $\NHq(5)=33$, and $\NHq(6)=9$] and the three-qudit ($\quditd=4$) Hamiltonian contains   $\NHq=218$ terms [$\NHq(1)=23$, $\NHq(2)=150$, and $\NHq(3)=45$]. In order to account for the noise of the three-qudit gate, we take the error rate of the $n$th qudit gate in one Suzuki-Trotter step to be 
\begin{equation}\label{Eq:errornDefQudit}
\errorn{n}^{\text{qudit}} =
\begin{cases}
0 & \text{if } \Oo(\Gamma_n)=1, \\
\errortwoq  & \text{if } \Oo(\Gamma_n)=2,\\
\xi\errortwoq  & \text{if } \Oo(\Gamma_n)=3,
\end{cases}
\end{equation}
where $\xi$ is a numerical constant.  
The optimal decomposition of 
multi-qudit gates of the form $e^{i\theta \lambda_{j_1}\lambda_{j_2}\cdots \lambda_{j_{\Oo}} }$ 
appearing in the Suzuki-Trotter approximation of a time evolution operator depends on the 
available two-qudit gates \cite{Di2013, Di2015,Trisciani2025} and there is no general formula for the number of  two-qudit gates required for the decomposition of a three-qudit gates. We 
investigate two error models by adopting two different values of $\xi$, $\xi=3$ and $\xi =12$, in Eq.~\eqref{Eq:errornDefQudit},
where $\xi=3$ corresponds to the case when the error of a three-qudit gate is the same as a three-qubit gate [see Eq.~\eqref{Eq:errornDef}]. 
In \cite{Trisciani2025}, it was shown that for qutrits ($\quditd=3$), an $\Oo$-qutrit gate defined in terms of Gell-Mann matrices $\lambda_j$ as 
$e^{i\theta \lambda_{j_1}\lambda_{j_2}\cdots \lambda_{j_{\Oo}} }$ can be decomposed into  
$2^{\Oo-1}+ 2\Oo-3$ two-qutrit {\sc cx} gates.
The error model \eqref{Eq:errornDefQudit} with $\xi=12$ is based on the assumption that the results in \cite{Trisciani2025} can be generalized to qudits and that an $\Oo$-qudit gate can be decomposed into $(\quditd-1)^{\Oo-1}+ 2\Oo-3$ two-qudit gates. When $\quditd=4$ and $\Oo=3$, we have $(\quditd-1)^{\Oo-1}+ 2\Oo-3=12$.

In Fig.~\ref{Fig2_appendix} we show the time-dependent  population in the $|2,0,0\rangle$ state of H$_2$O simulated using the fourth-order  anharmonic potential \eqref{Eq:anharmonicpotfourthorder}. The simulations are conducted using the qubit binary encoding and the qudit encoding at two values of $\xi$, $\xi=3$ [Fig.~\ref{Fig2_appendix}(a)] and $\xi=12$ [Fig.~\ref{Fig2_appendix}(b)]. For each value of $\xi$, we carry out simulations using two values of the two-qubit/qudit gate error, $\errortwoq=10^{-3}$ and $\errortwoq=10^{-5}$.
We can draw two conclusions from the results shown in Fig.~\ref{Fig2_appendix}. The first is that for $\errortwoq=10^{-3}$ (filled symbols), corresponding to the gate error of current qubit-based  ion trap quantum computers, both of the qubit and qudit encodings result in quickly decaying populations without oscillations. This implies that current ion trap quantum computers are not suited for the simulation of  vibrational population dynamics in an anharmonic potential having a contribution from the fourth-order anharmonicity.
The second conclusion is that at the smaller gate error of $\errortwoq=10^{-5}$ (open symbols), oscillating populations are obtained for both of the qubit and qudit encodings. As can be seen in   Fig.~\ref{Fig2_appendix}(a),  there is an advantage of using qudits when the three-qudit gate error is the same as the three-qubit gate error ($\xi=3$).   On the other hand, as shown in Fig.~\ref{Fig2_appendix}(b), when the three-qudit gate error is four times larger than the three-qubit gate error ($\xi=12$), the  qubit and qudit encodings result in equally accurate vibrational populations.

\section{Vibrational energy levels of H$_2$O}\label{Sec:VibEnergiesH2O}
In Table~\ref{Table1}, we show the vibrational energy levels of H$_2$O with energy smaller than $13000$ cm$^{-1}$, obtained by diagonalization of $H_{\text{q}}(\vmax=3)$ (including up to the third-order anharmonic coupling). Because 
the symmetry of the $\nu_1$ and $\nu_2$ modes is $A_1$ and that of the $\nu_3$ mode is $B_2$ in the $C_{2v}$ point group,
the symmetry of an eigenstate having even $v_3$ is $A_1$ and that of an eigenstate having odd $v_3$ is $B_2$.
The vibrational levels having symmetry $A_1$ correspond  to para-H$_2$O (singlet proton spin state), and those having $B_2$ symmetry correspond to ortho-H$_2$O (triplet proton spin state). 

\begingroup
\begin{table}
\caption{\label{Table1} Vibrational energy levels of H$_2$O ($\vmax =3$), labeled by their symmetry species in the $C_{2v}$ point group. In the last column, we list the dominant basis functions in the expansion
\eqref{Eq:Happroxpsi} for expansion coefficients $\psi_{\bm{v}n}^2\ge 0.2$. }
\begin{tabular}{ccdl}
\hline
\multicolumn{1}{c}{$n$} & \multicolumn{1}{c}{Symmetry} & \multicolumn{1}{c}{$E_n/\text{cm}^{-1}$} & \multicolumn{1}{c}{$v_1v_2v_3(\psi_{\bm{v}n}^2)$}  \\
\hline
0 & $a_1$ & -130.87 &  000(0.97)  \\
1 & $a_1$ & 1542.32 &  010(0.98)  \\
2 & $a_1$ & 3171.32 &  020(0.83)  \\
3 & $b_2$ & 3349.39 &  001(0.83)  \\
4 & $a_1$ & 3386.98 &  100(0.69)  \\
5 & $a_1$ & 4815.56 &  030(0.77)  \\
6 & $b_2$ & 5085.57 &  011(0.84)  \\
7 & $a_1$ & 5138.58 &  110(0.66), 030(0.22)  \\
8 & $b_2$ & 6489.93 &  101(0.46), 201(0.21)  \\
9 & $a_1$ & 6704.16 &  120(0.48)  \\
10 & $b_2$ & 6830.69 &  021(0.83)  \\
11 & $a_1$ & 6838.06 &  120(0.37), 002(0.28)  \\
12 & $a_1$ & 7180.56 &  200(0.44), 002(0.35)  \\
13 & $b_2$ & 8206.29 &  111(0.45)  \\
14 & $a_1$ & 8352.48 &  130(0.52), 210(0.23)  \\
15 & $b_2$ & 8576.36 &  031(0.78)  \\
16 & $a_1$ & 8608.96 &  012(0.33), 130(0.31)  \\
17 & $a_1$ & 8933.74 &  210(0.40), 012(0.37)  \\
18 & $b_2$ & 9987.84 &  121(0.52)  \\
19 & $b_2$ & 10207.94 &  003(0.32), 103(0.24), 101(0.21)  \\
20 & $a_1$ & 10224.16 &  220(0.41), 022(0.21)  \\
21 & $a_1$ & 10431.68 &  102(0.30), 202(0.27)  \\
22 & $a_1$ & 10648.78 &  022(0.55), 220(0.25)  \\
23 & $b_2$ & 10721.64 &  201(0.37), 003(0.28)  \\
24 & $b_2$ & 11687.50 &  131(0.52)  \\
25 & $a_1$ & 11915.24 &  230(0.45)  \\
26 & $b_2$ & 11994.01 &  013(0.26)  \\
27 & $a_1$ & 12165.94 &  112(0.27), 212(0.21)  \\
28 & $a_1$ & 12403.94 &  300(0.79)  \\
29 & $a_1$ & 12424.51 &  032(0.56)  \\
30 & $b_2$ & 12449.26 &  013(0.35), 211(0.34)  \\
\hline
\end{tabular}
\end{table}
\endgroup

For reference, we compare in Table~\ref{Table2} the  energy difference $\Delta E_n$ relative to the vibrational ground state, defined as 
\begin{equation}
\Delta E_n = E_n-E_0,
\end{equation}
with the experimentally measured energy differences \cite{Furtenbacher2020a} for the four lowest excited states. 
We include $\Delta E_n$ calculated using the third-order anharmonic potential  defined in Eq.~\eqref{Eq:anharmonicpot} and  $\Delta E_n$ calculated using the fourth-order anharmonic potential  defined in Eqs.~\eqref{Eq:anharmonicpotfourthorder}.
   As can be seen in Table~\ref{Table2}, the energy differences obtained   using the H$_2$O model defined by 
   the third-order anharmonic potential differ by more than 100 cm$^{-1}$ from the experimentally measured energy differences,
   but when the fourth-order anharmonic coupling terms are included, the discrepancies become smaller than 80 cm$^{-1}$.
\begin{table}
\caption{\label{Table2}Excitation energies $\Delta E_n$  in cm$^{-1}$ of the lowest four excited vibrational states of 
H$_2$O ($\vmax=3$), for the Hamiltonians defined by the third-order and fourth-order anharmonic potentials.
Each column is labeled by $n\gamma(v_1v_2v_3)$, where $n$ is the state number from 
Table~\ref{Table1}, $\gamma$ is the symmetry species, and $v_1v_2v_3$ is the dominant basis function. 
We also show the  experimentally measured  energies differences  in the rotational ground state 
whose total angular momentum is $J=0$.}
\begin{tabular}{ldddd}
\hline
\multicolumn{1}{c}{$\;$}      & \multicolumn{1}{c}{$1a_1( 0  1  0)$} & \multicolumn{1}{c}{$2a_1(0  2  0)$} & \multicolumn{1}{c}{$3b_2 (0  0  1)$} & \multicolumn{1}{c}{$4a_1(1  0  0)$}  \\
\hline
 Third order                         &                                         1673.20     &                                          3302.19   &                                          3480.27    &                                          3517.85     \\
 Fourth order                        &                        1574.49    &                                         3135.63   &                                         3790.29    &                                          3732.69     \\
Exp.\footnote{W2020  database \cite{Furtenbacher2020a}.}   
                                      &                                               1594.75     &                                           3151.63    &                                        3755.93    &                                           3657.05             \\
\hline
\end{tabular}
\end{table}

\newpage
\clearpage


\begin{thebibliography}{111}%
\makeatletter
\providecommand \@ifxundefined [1]{%
 \@ifx{#1\undefined}
}%
\providecommand \@ifnum [1]{%
 \ifnum #1\expandafter \@firstoftwo
 \else \expandafter \@secondoftwo
 \fi
}%
\providecommand \@ifx [1]{%
 \ifx #1\expandafter \@firstoftwo
 \else \expandafter \@secondoftwo
 \fi
}%
\providecommand \natexlab [1]{#1}%
\providecommand \enquote  [1]{``#1''}%
\providecommand \bibnamefont  [1]{#1}%
\providecommand \bibfnamefont [1]{#1}%
\providecommand \citenamefont [1]{#1}%
\providecommand \href@noop [0]{\@secondoftwo}%
\providecommand \href [0]{\begingroup \@sanitize@url \@href}%
\providecommand \@href[1]{\@@startlink{#1}\@@href}%
\providecommand \@@href[1]{\endgroup#1\@@endlink}%
\providecommand \@sanitize@url [0]{\catcode `\\12\catcode `\$12\catcode
  `\&12\catcode `\#12\catcode `\^12\catcode `\_12\catcode `\%12\relax}%
\providecommand \@@startlink[1]{}%
\providecommand \@@endlink[0]{}%
\providecommand \url  [0]{\begingroup\@sanitize@url \@url }%
\providecommand \@url [1]{\endgroup\@href {#1}{\urlprefix }}%
\providecommand \urlprefix  [0]{URL }%
\providecommand \Eprint [0]{\href }%
\providecommand \doibase [0]{http://dx.doi.org/}%
\providecommand \selectlanguage [0]{\@gobble}%
\providecommand \bibinfo  [0]{\@secondoftwo}%
\providecommand \bibfield  [0]{\@secondoftwo}%
\providecommand \translation [1]{[#1]}%
\providecommand \BibitemOpen [0]{}%
\providecommand \bibitemStop [0]{}%
\providecommand \bibitemNoStop [0]{.\EOS\space}%
\providecommand \EOS [0]{\spacefactor3000\relax}%
\providecommand \BibitemShut  [1]{\csname bibitem#1\endcsname}%
\let\auto@bib@innerbib\@empty
\bibitem [{\citenamefont {Bravyi}\ \emph {et~al.}(2022)\citenamefont {Bravyi},
  \citenamefont {Dial}, \citenamefont {Gambetta}, \citenamefont {Gil},\ and\
  \citenamefont {Nazario}}]{Bravyi2022}%
  \BibitemOpen
  \bibfield  {author} {\bibinfo {author} {\bibfnamefont {S.}~\bibnamefont
  {Bravyi}}, \bibinfo {author} {\bibfnamefont {O.}~\bibnamefont {Dial}},
  \bibinfo {author} {\bibfnamefont {J.~M.}\ \bibnamefont {Gambetta}}, \bibinfo
  {author} {\bibfnamefont {D.}~\bibnamefont {Gil}}, \ and\ \bibinfo {author}
  {\bibfnamefont {Z.}~\bibnamefont {Nazario}},\ }\bibfield  {title} {\enquote
  {\bibinfo {title} {The future of quantum computing with superconducting
  qubits},}\ }\href {http://dx.doi.org/10.1063/5.0082975} {\bibfield  {journal}
  {\bibinfo  {journal} {\emph {J. Appl. Phys.}}\ }\textbf {\bibinfo {volume}
  {132}},\ \bibinfo {pages} {160902} (\bibinfo {year} {2022})}\BibitemShut
  {NoStop}%
\bibitem [{\citenamefont {Moses}\ \emph {et~al.}(2023)\citenamefont {Moses},
  \citenamefont {Baldwin}, \citenamefont {Allman}, \citenamefont {Ancona},
  \citenamefont {Ascarrunz}, \citenamefont {Barnes}, \citenamefont
  {Bartolotta}, \citenamefont {Bjork}, \citenamefont {Blanchard}, \citenamefont
  {Bohn}, \citenamefont {Bohnet}, \citenamefont {Brown}, \citenamefont
  {Burdick}, \citenamefont {Burton}, \citenamefont {Campbell}, \citenamefont
  {Campora}, \citenamefont {Carron}, \citenamefont {Chambers}, \citenamefont
  {Chan}, \citenamefont {Chen}, \citenamefont {Chernoguzov}, \citenamefont
  {Chertkov}, \citenamefont {Colina}, \citenamefont {Curtis}, \citenamefont
  {Daniel}, \citenamefont {DeCross}, \citenamefont {Deen}, \citenamefont
  {Delaney}, \citenamefont {Dreiling}, \citenamefont {Ertsgaard}, \citenamefont
  {Esposito}, \citenamefont {Estey}, \citenamefont {Fabrikant}, \citenamefont
  {Figgatt}, \citenamefont {Foltz}, \citenamefont {Foss-Feig}, \citenamefont
  {Francois}, \citenamefont {Gaebler}, \citenamefont {Gatterman}, \citenamefont
  {Gilbreth}, \citenamefont {Giles}, \citenamefont {Glynn}, \citenamefont
  {Hall}, \citenamefont {Hankin}, \citenamefont {Hansen}, \citenamefont
  {Hayes}, \citenamefont {Higashi}, \citenamefont {Hoffman}, \citenamefont
  {Horning}, \citenamefont {Hout}, \citenamefont {Jacobs}, \citenamefont
  {Johansen}, \citenamefont {Jones}, \citenamefont {Karcz}, \citenamefont
  {Klein}, \citenamefont {Lauria}, \citenamefont {Lee}, \citenamefont {Liefer},
  \citenamefont {Lu}, \citenamefont {Lucchetti}, \citenamefont {Lytle},
  \citenamefont {Malm}, \citenamefont {Matheny}, \citenamefont {Mathewson},
  \citenamefont {Mayer}, \citenamefont {Miller}, \citenamefont {Mills},
  \citenamefont {Neyenhuis}, \citenamefont {Nugent}, \citenamefont {Olson},
  \citenamefont {Parks}, \citenamefont {Price}, \citenamefont {Price},
  \citenamefont {Pugh}, \citenamefont {Ransford}, \citenamefont {Reed},
  \citenamefont {Roman}, \citenamefont {Rowe}, \citenamefont {Ryan-Anderson},
  \citenamefont {Sanders}, \citenamefont {Sedlacek}, \citenamefont {Shevchuk},
  \citenamefont {Siegfried}, \citenamefont {Skripka}, \citenamefont {Spaun},
  \citenamefont {Sprenkle}, \citenamefont {Stutz}, \citenamefont {Swallows},
  \citenamefont {Tobey}, \citenamefont {Tran}, \citenamefont {Tran},
  \citenamefont {Vogt}, \citenamefont {Volin}, \citenamefont {Walker},
  \citenamefont {Zolot},\ and\ \citenamefont {Pino}}]{Moses2023_PRX}%
  \BibitemOpen
  \bibfield  {author} {\bibinfo {author} {\bibfnamefont {S.~A.}\ \bibnamefont
  {Moses}}, \bibinfo {author} {\bibfnamefont {C.~H.}\ \bibnamefont {Baldwin}},
  \bibinfo {author} {\bibfnamefont {M.~S.}\ \bibnamefont {Allman}}, \bibinfo
  {author} {\bibfnamefont {R.}~\bibnamefont {Ancona}}, \bibinfo {author}
  {\bibfnamefont {L.}~\bibnamefont {Ascarrunz}}, \bibinfo {author}
  {\bibfnamefont {C.}~\bibnamefont {Barnes}}, \bibinfo {author} {\bibfnamefont
  {J.}~\bibnamefont {Bartolotta}}, \bibinfo {author} {\bibfnamefont
  {B.}~\bibnamefont {Bjork}}, \bibinfo {author} {\bibfnamefont
  {P.}~\bibnamefont {Blanchard}}, \bibinfo {author} {\bibfnamefont
  {M.}~\bibnamefont {Bohn}}, \bibinfo {author} {\bibfnamefont {J.~G.}\
  \bibnamefont {Bohnet}}, \bibinfo {author} {\bibfnamefont {N.~C.}\
  \bibnamefont {Brown}}, \bibinfo {author} {\bibfnamefont {N.~Q.}\ \bibnamefont
  {Burdick}}, \bibinfo {author} {\bibfnamefont {W.~C.}\ \bibnamefont {Burton}},
  \bibinfo {author} {\bibfnamefont {S.~L.}\ \bibnamefont {Campbell}}, \bibinfo
  {author} {\bibfnamefont {J.~P.}\ \bibnamefont {Campora}}, \bibinfo {author}
  {\bibfnamefont {C.}~\bibnamefont {Carron}}, \bibinfo {author} {\bibfnamefont
  {J.}~\bibnamefont {Chambers}}, \bibinfo {author} {\bibfnamefont {J.~W.}\
  \bibnamefont {Chan}}, \bibinfo {author} {\bibfnamefont {Y.~H.}\ \bibnamefont
  {Chen}}, \bibinfo {author} {\bibfnamefont {A.}~\bibnamefont {Chernoguzov}},
  \bibinfo {author} {\bibfnamefont {E.}~\bibnamefont {Chertkov}}, \bibinfo
  {author} {\bibfnamefont {J.}~\bibnamefont {Colina}}, \bibinfo {author}
  {\bibfnamefont {J.~P.}\ \bibnamefont {Curtis}}, \bibinfo {author}
  {\bibfnamefont {R.}~\bibnamefont {Daniel}}, \bibinfo {author} {\bibfnamefont
  {M.}~\bibnamefont {DeCross}}, \bibinfo {author} {\bibfnamefont
  {D.}~\bibnamefont {Deen}}, \bibinfo {author} {\bibfnamefont {C.}~\bibnamefont
  {Delaney}}, \bibinfo {author} {\bibfnamefont {J.~M.}\ \bibnamefont
  {Dreiling}}, \bibinfo {author} {\bibfnamefont {C.~T.}\ \bibnamefont
  {Ertsgaard}}, \bibinfo {author} {\bibfnamefont {J.}~\bibnamefont {Esposito}},
  \bibinfo {author} {\bibfnamefont {B.}~\bibnamefont {Estey}}, \bibinfo
  {author} {\bibfnamefont {M.}~\bibnamefont {Fabrikant}}, \bibinfo {author}
  {\bibfnamefont {C.}~\bibnamefont {Figgatt}}, \bibinfo {author} {\bibfnamefont
  {C.}~\bibnamefont {Foltz}}, \bibinfo {author} {\bibfnamefont
  {M.}~\bibnamefont {Foss-Feig}}, \bibinfo {author} {\bibfnamefont
  {D.}~\bibnamefont {Francois}}, \bibinfo {author} {\bibfnamefont {J.~P.}\
  \bibnamefont {Gaebler}}, \bibinfo {author} {\bibfnamefont {T.~M.}\
  \bibnamefont {Gatterman}}, \bibinfo {author} {\bibfnamefont {C.~N.}\
  \bibnamefont {Gilbreth}}, \bibinfo {author} {\bibfnamefont {J.}~\bibnamefont
  {Giles}}, \bibinfo {author} {\bibfnamefont {E.}~\bibnamefont {Glynn}},
  \bibinfo {author} {\bibfnamefont {A.}~\bibnamefont {Hall}}, \bibinfo {author}
  {\bibfnamefont {A.~M.}\ \bibnamefont {Hankin}}, \bibinfo {author}
  {\bibfnamefont {A.}~\bibnamefont {Hansen}}, \bibinfo {author} {\bibfnamefont
  {D.}~\bibnamefont {Hayes}}, \bibinfo {author} {\bibfnamefont
  {B.}~\bibnamefont {Higashi}}, \bibinfo {author} {\bibfnamefont {I.~M.}\
  \bibnamefont {Hoffman}}, \bibinfo {author} {\bibfnamefont {B.}~\bibnamefont
  {Horning}}, \bibinfo {author} {\bibfnamefont {J.~J.}\ \bibnamefont {Hout}},
  \bibinfo {author} {\bibfnamefont {R.}~\bibnamefont {Jacobs}}, \bibinfo
  {author} {\bibfnamefont {J.}~\bibnamefont {Johansen}}, \bibinfo {author}
  {\bibfnamefont {L.}~\bibnamefont {Jones}}, \bibinfo {author} {\bibfnamefont
  {J.}~\bibnamefont {Karcz}}, \bibinfo {author} {\bibfnamefont
  {T.}~\bibnamefont {Klein}}, \bibinfo {author} {\bibfnamefont
  {P.}~\bibnamefont {Lauria}}, \bibinfo {author} {\bibfnamefont
  {P.}~\bibnamefont {Lee}}, \bibinfo {author} {\bibfnamefont {D.}~\bibnamefont
  {Liefer}}, \bibinfo {author} {\bibfnamefont {S.~T.}\ \bibnamefont {Lu}},
  \bibinfo {author} {\bibfnamefont {D.}~\bibnamefont {Lucchetti}}, \bibinfo
  {author} {\bibfnamefont {C.}~\bibnamefont {Lytle}}, \bibinfo {author}
  {\bibfnamefont {A.}~\bibnamefont {Malm}}, \bibinfo {author} {\bibfnamefont
  {M.}~\bibnamefont {Matheny}}, \bibinfo {author} {\bibfnamefont
  {B.}~\bibnamefont {Mathewson}}, \bibinfo {author} {\bibfnamefont
  {K.}~\bibnamefont {Mayer}}, \bibinfo {author} {\bibfnamefont {D.~B.}\
  \bibnamefont {Miller}}, \bibinfo {author} {\bibfnamefont {M.}~\bibnamefont
  {Mills}}, \bibinfo {author} {\bibfnamefont {B.}~\bibnamefont {Neyenhuis}},
  \bibinfo {author} {\bibfnamefont {L.}~\bibnamefont {Nugent}}, \bibinfo
  {author} {\bibfnamefont {S.}~\bibnamefont {Olson}}, \bibinfo {author}
  {\bibfnamefont {J.}~\bibnamefont {Parks}}, \bibinfo {author} {\bibfnamefont
  {G.~N.}\ \bibnamefont {Price}}, \bibinfo {author} {\bibfnamefont
  {Z.}~\bibnamefont {Price}}, \bibinfo {author} {\bibfnamefont
  {M.}~\bibnamefont {Pugh}}, \bibinfo {author} {\bibfnamefont {A.}~\bibnamefont
  {Ransford}}, \bibinfo {author} {\bibfnamefont {A.~P.}\ \bibnamefont {Reed}},
  \bibinfo {author} {\bibfnamefont {C.}~\bibnamefont {Roman}}, \bibinfo
  {author} {\bibfnamefont {M.}~\bibnamefont {Rowe}}, \bibinfo {author}
  {\bibfnamefont {C.}~\bibnamefont {Ryan-Anderson}}, \bibinfo {author}
  {\bibfnamefont {S.}~\bibnamefont {Sanders}}, \bibinfo {author} {\bibfnamefont
  {J.}~\bibnamefont {Sedlacek}}, \bibinfo {author} {\bibfnamefont
  {P.}~\bibnamefont {Shevchuk}}, \bibinfo {author} {\bibfnamefont
  {P.}~\bibnamefont {Siegfried}}, \bibinfo {author} {\bibfnamefont
  {T.}~\bibnamefont {Skripka}}, \bibinfo {author} {\bibfnamefont
  {B.}~\bibnamefont {Spaun}}, \bibinfo {author} {\bibfnamefont {R.~T.}\
  \bibnamefont {Sprenkle}}, \bibinfo {author} {\bibfnamefont {R.~P.}\
  \bibnamefont {Stutz}}, \bibinfo {author} {\bibfnamefont {M.}~\bibnamefont
  {Swallows}}, \bibinfo {author} {\bibfnamefont {R.~I.}\ \bibnamefont {Tobey}},
  \bibinfo {author} {\bibfnamefont {A.}~\bibnamefont {Tran}}, \bibinfo {author}
  {\bibfnamefont {T.}~\bibnamefont {Tran}}, \bibinfo {author} {\bibfnamefont
  {E.}~\bibnamefont {Vogt}}, \bibinfo {author} {\bibfnamefont {C.}~\bibnamefont
  {Volin}}, \bibinfo {author} {\bibfnamefont {J.}~\bibnamefont {Walker}},
  \bibinfo {author} {\bibfnamefont {A.~M.}\ \bibnamefont {Zolot}}, \ and\
  \bibinfo {author} {\bibfnamefont {J.~M.}\ \bibnamefont {Pino}},\ }\bibfield
  {title} {\enquote {\bibinfo {title} {A race-track trapped-ion quantum
  processor},}\ }\href {http://dx.doi.org/10.1103/PhysRevX.13.041052}
  {\bibfield  {journal} {\bibinfo  {journal} {\emph {Phys. Rev. X}}\ }\textbf
  {\bibinfo {volume} {13}},\ \bibinfo {pages} {041052} (\bibinfo {year}
  {2023})}\BibitemShut {NoStop}%
\bibitem [{\citenamefont {Chen}\ \emph {et~al.}(2024)\citenamefont {Chen},
  \citenamefont {Nielsen}, \citenamefont {Ebert}, \citenamefont {Inlek},
  \citenamefont {Wright}, \citenamefont {Chaplin}, \citenamefont {Maksymov},
  \citenamefont {P\'{a}ez}, \citenamefont {Poudel}, \citenamefont {Maunz},\
  and\ \citenamefont {Gamble}}]{Chen2024}%
  \BibitemOpen
  \bibfield  {author} {\bibinfo {author} {\bibfnamefont {J.-S.}\ \bibnamefont
  {Chen}}, \bibinfo {author} {\bibfnamefont {E.}~\bibnamefont {Nielsen}},
  \bibinfo {author} {\bibfnamefont {M.}~\bibnamefont {Ebert}}, \bibinfo
  {author} {\bibfnamefont {V.}~\bibnamefont {Inlek}}, \bibinfo {author}
  {\bibfnamefont {K.}~\bibnamefont {Wright}}, \bibinfo {author} {\bibfnamefont
  {V.}~\bibnamefont {Chaplin}}, \bibinfo {author} {\bibfnamefont
  {A.}~\bibnamefont {Maksymov}}, \bibinfo {author} {\bibfnamefont
  {E.}~\bibnamefont {P\'{a}ez}}, \bibinfo {author} {\bibfnamefont
  {A.}~\bibnamefont {Poudel}}, \bibinfo {author} {\bibfnamefont
  {P.}~\bibnamefont {Maunz}}, \ and\ \bibinfo {author} {\bibfnamefont
  {J.}~\bibnamefont {Gamble}},\ }\bibfield  {title} {\enquote {\bibinfo {title}
  {Benchmarking a trapped-ion quantum computer with 30 qubits},}\ }\href
  {http://dx.doi.org/10.22331/q-2024-11-07-1516} {\bibfield  {journal}
  {\bibinfo  {journal} {\emph {Quantum}}\ }\textbf {\bibinfo {volume} {8}},\
  \bibinfo {pages} {1516} (\bibinfo {year} {2024})}\BibitemShut {NoStop}%
\bibitem [{\citenamefont {Chew}\ \emph {et~al.}(2022)\citenamefont {Chew},
  \citenamefont {Tomita}, \citenamefont {Mahesh}, \citenamefont {Sugawa},
  \citenamefont {de~L{\'{e}}s{\'{e}}leuc},\ and\ \citenamefont
  {Ohmori}}]{Chew2022}%
  \BibitemOpen
  \bibfield  {author} {\bibinfo {author} {\bibfnamefont {Y.}~\bibnamefont
  {Chew}}, \bibinfo {author} {\bibfnamefont {T.}~\bibnamefont {Tomita}},
  \bibinfo {author} {\bibfnamefont {T.~P.}\ \bibnamefont {Mahesh}}, \bibinfo
  {author} {\bibfnamefont {S.}~\bibnamefont {Sugawa}}, \bibinfo {author}
  {\bibfnamefont {S.}~\bibnamefont {de~L{\'{e}}s{\'{e}}leuc}}, \ and\ \bibinfo
  {author} {\bibfnamefont {K.}~\bibnamefont {Ohmori}},\ }\bibfield  {title}
  {\enquote {\bibinfo {title} {Ultrafast energy exchange between two single
  {R}ydberg atoms on a nanosecond timescale},}\ }\href
  {http://dx.doi.org/10.1038/s41566-022-01047-2} {\bibfield  {journal}
  {\bibinfo  {journal} {\emph {Nature Photonics}}\ }\textbf {\bibinfo {volume}
  {16}},\ \bibinfo {pages} {724} (\bibinfo {year} {2022})}\BibitemShut
  {NoStop}%
\bibitem [{\citenamefont {Bluvstein}\ \emph {et~al.}(2024)\citenamefont
  {Bluvstein}, \citenamefont {Evered}, \citenamefont {Geim}, \citenamefont
  {Li}, \citenamefont {Zhou}, \citenamefont {Manovitz}, \citenamefont {Ebadi},
  \citenamefont {Cain}, \citenamefont {Kalinowski}, \citenamefont {Hangleiter},
  \citenamefont {Ataides}, \citenamefont {Maskara}, \citenamefont {Cong},
  \citenamefont {Gao}, \citenamefont {Rodriguez}, \citenamefont {Karolyshyn},
  \citenamefont {Semeghini}, \citenamefont {Gullans}, \citenamefont {Greiner},
  \citenamefont {Vuleti\'{c}},\ and\ \citenamefont {Lukin}}]{Bluvstein2023}%
  \BibitemOpen
  \bibfield  {author} {\bibinfo {author} {\bibfnamefont {D.}~\bibnamefont
  {Bluvstein}}, \bibinfo {author} {\bibfnamefont {S.~J.}\ \bibnamefont
  {Evered}}, \bibinfo {author} {\bibfnamefont {A.~A.}\ \bibnamefont {Geim}},
  \bibinfo {author} {\bibfnamefont {S.~H.}\ \bibnamefont {Li}}, \bibinfo
  {author} {\bibfnamefont {H.}~\bibnamefont {Zhou}}, \bibinfo {author}
  {\bibfnamefont {T.}~\bibnamefont {Manovitz}}, \bibinfo {author}
  {\bibfnamefont {S.}~\bibnamefont {Ebadi}}, \bibinfo {author} {\bibfnamefont
  {M.}~\bibnamefont {Cain}}, \bibinfo {author} {\bibfnamefont {M.}~\bibnamefont
  {Kalinowski}}, \bibinfo {author} {\bibfnamefont {D.}~\bibnamefont
  {Hangleiter}}, \bibinfo {author} {\bibfnamefont {J.~P.~B.}\ \bibnamefont
  {Ataides}}, \bibinfo {author} {\bibfnamefont {N.}~\bibnamefont {Maskara}},
  \bibinfo {author} {\bibfnamefont {I.}~\bibnamefont {Cong}}, \bibinfo {author}
  {\bibfnamefont {X.}~\bibnamefont {Gao}}, \bibinfo {author} {\bibfnamefont
  {P.~S.}\ \bibnamefont {Rodriguez}}, \bibinfo {author} {\bibfnamefont
  {T.}~\bibnamefont {Karolyshyn}}, \bibinfo {author} {\bibfnamefont
  {G.}~\bibnamefont {Semeghini}}, \bibinfo {author} {\bibfnamefont {M.~J.}\
  \bibnamefont {Gullans}}, \bibinfo {author} {\bibfnamefont {M.}~\bibnamefont
  {Greiner}}, \bibinfo {author} {\bibfnamefont {V.}~\bibnamefont
  {Vuleti\'{c}}}, \ and\ \bibinfo {author} {\bibfnamefont {M.~D.}\ \bibnamefont
  {Lukin}},\ }\bibfield  {title} {\enquote {\bibinfo {title} {Logical quantum
  processor based on reconfigurable atom arrays},}\ }\href
  {http://dx.doi.org/10.1038/s41586-023-06927-3} {\bibfield  {journal}
  {\bibinfo  {journal} {\emph {Nature}}\ }\textbf {\bibinfo {volume} {626}},\
  \bibinfo {pages} {58} (\bibinfo {year} {2024})}\BibitemShut {NoStop}%
\bibitem [{\citenamefont {Shao}\ \emph {et~al.}(2024)\citenamefont {Shao},
  \citenamefont {Su}, \citenamefont {Li}, \citenamefont {Nath}, \citenamefont
  {Wu},\ and\ \citenamefont {Li}}]{Shao2024}%
  \BibitemOpen
  \bibfield  {author} {\bibinfo {author} {\bibfnamefont {X.-Q.}\ \bibnamefont
  {Shao}}, \bibinfo {author} {\bibfnamefont {S.-L.}\ \bibnamefont {Su}},
  \bibinfo {author} {\bibfnamefont {L.}~\bibnamefont {Li}}, \bibinfo {author}
  {\bibfnamefont {R.}~\bibnamefont {Nath}}, \bibinfo {author} {\bibfnamefont
  {J.-H.}\ \bibnamefont {Wu}}, \ and\ \bibinfo {author} {\bibfnamefont
  {W.}~\bibnamefont {Li}},\ }\bibfield  {title} {\enquote {\bibinfo {title}
  {Rydberg superatoms: {A}n artificial quantum system for quantum information
  processing and quantum optics},}\ }\href
  {http://dx.doi.org/10.1063/5.0211071} {\bibfield  {journal} {\bibinfo
  {journal} {\emph {Appl. Phys. Rev.}}\ }\textbf {\bibinfo {volume} {11}},\
  \bibinfo {pages} {031320} (\bibinfo {year} {2024})}\BibitemShut {NoStop}%
\bibitem [{\citenamefont {{IBM Quantum Platform}}()}]{ibmqcloud}%
  \BibitemOpen
  \bibfield  {author} {\bibinfo {author} {\bibnamefont {{IBM Quantum
  Platform}}},\ }\href {https://quantum.cloud.ibm.com/} {}\bibinfo
  {howpublished} {https://quantum.cloud.ibm.com/}\BibitemShut {NoStop}%
\bibitem [{\citenamefont {Ransford}\ \emph {et~al.}(2026)\citenamefont
  {Ransford}, \citenamefont {Allman}, \citenamefont {Arkinstall}, \citenamefont
  {Campora}, \citenamefont {Cooper}, \citenamefont {Delaney}, \citenamefont
  {Dreiling}, \citenamefont {Estey}, \citenamefont {Figgatt}, \citenamefont
  {Hall}, \citenamefont {Husain}, \citenamefont {Isanaka}, \citenamefont
  {Kennedy}, \citenamefont {Kotibhaskar}, \citenamefont {Madjarov},
  \citenamefont {Mayer}, \citenamefont {Milne}, \citenamefont {Park},
  \citenamefont {Reed}, \citenamefont {Ancona}, \citenamefont {Andersen},
  \citenamefont {Andres-Martinez}, \citenamefont {Angenent}, \citenamefont
  {Argueta}, \citenamefont {Arkin}, \citenamefont {Ascarrunz}, \citenamefont
  {Baker}, \citenamefont {Barnes}, \citenamefont {Bartolotta}, \citenamefont
  {Berg}, \citenamefont {Besand}, \citenamefont {Bjork}, \citenamefont {Blain},
  \citenamefont {Blanchard}, \citenamefont {Blume-Kohout}, \citenamefont
  {Bohn}, \citenamefont {Borgna}, \citenamefont {Botamanenko}, \citenamefont
  {Boutelle}, \citenamefont {Brown}, \citenamefont {Buckingham}, \citenamefont
  {Burdick}, \citenamefont {Burton}, \citenamefont {Carey}, \citenamefont
  {Carron}, \citenamefont {Chambers}, \citenamefont {Chan}, \citenamefont
  {Children}, \citenamefont {Colussi}, \citenamefont {Crepinsek}, \citenamefont
  {Cureton}, \citenamefont {Davies}, \citenamefont {Davis}, \citenamefont
  {DeCross}, \citenamefont {Deen}, \citenamefont {Delaney}, \citenamefont
  {DelVento}, \citenamefont {DeSalvo}, \citenamefont {Dominy}, \citenamefont
  {Drotar}, \citenamefont {Duncan}, \citenamefont {Eccles}, \citenamefont
  {Edgington}, \citenamefont {Erickson}, \citenamefont {Erickson},
  \citenamefont {Ertsgaard}, \citenamefont {Esposito}, \citenamefont {Evans},
  \citenamefont {Evans}, \citenamefont {Fabrikant}, \citenamefont {Fischer},
  \citenamefont {Foltz}, \citenamefont {Foss-Feig}, \citenamefont {Francois},
  \citenamefont {Freyberg}, \citenamefont {Gao}, \citenamefont {Garay},
  \citenamefont {Garvin}, \citenamefont {Gaudiosi}, \citenamefont {Gilbreth},
  \citenamefont {Giles}, \citenamefont {Glynn}, \citenamefont {Graves},
  \citenamefont {Hansen}, \citenamefont {Hayes}, \citenamefont {Heidemann},
  \citenamefont {Higashi}, \citenamefont {Hilbun}, \citenamefont {Hines},
  \citenamefont {Hlavaty}, \citenamefont {Hoffman}, \citenamefont {Hoffman},
  \citenamefont {Holliman}, \citenamefont {Hooper}, \citenamefont {Horning},
  \citenamefont {Hostetter}, \citenamefont {Hothem}, \citenamefont {Houlton},
  \citenamefont {Hout}, \citenamefont {Hutson}, \citenamefont {Jacobs},
  \citenamefont {Jacobs}, \citenamefont {Johannsen}, \citenamefont {Johansen},
  \citenamefont {Jones}, \citenamefont {Julian}, \citenamefont {Jung},
  \citenamefont {Keay}, \citenamefont {Klein}, \citenamefont {Koch},
  \citenamefont {Kondo}, \citenamefont {Kong}, \citenamefont {Kosto},
  \citenamefont {Lawrence}, \citenamefont {Liefer}, \citenamefont {Lollie},
  \citenamefont {Lucchetti}, \citenamefont {Lysne}, \citenamefont {Lytle},
  \citenamefont {MacPherson}, \citenamefont {Malm}, \citenamefont {Mather},
  \citenamefont {Mathewson}, \citenamefont {Maxwell}, \citenamefont
  {McCaffrey}, \citenamefont {McDougall}, \citenamefont {Mendoza},
  \citenamefont {Miller}, \citenamefont {Mills}, \citenamefont {Morrison},
  \citenamefont {Narmour}, \citenamefont {Nguyen}, \citenamefont {Nugent},
  \citenamefont {Olson}, \citenamefont {Ouellette}, \citenamefont {Parks},
  \citenamefont {Peters}, \citenamefont {Peterson}, \citenamefont {Petricka},
  \citenamefont {Pino}, \citenamefont {Polito}, \citenamefont {Potter},
  \citenamefont {Preidl}, \citenamefont {Price}, \citenamefont {Proctor},
  \citenamefont {Pugh}, \citenamefont {Ratcliff}, \citenamefont {Raymondson},
  \citenamefont {Rhodes}, \citenamefont {Roman}, \citenamefont {Roy},
  \citenamefont {Ryan-Anderson}, \citenamefont {Sanchez}, \citenamefont
  {Sangiolo}, \citenamefont {Sawadski}, \citenamefont {Schaffer}, \citenamefont
  {Schow}, \citenamefont {Sedlacek}, \citenamefont {Semenenko}, \citenamefont
  {Shevchuk}, \citenamefont {Shore}, \citenamefont {Siegfried}, \citenamefont
  {Singhal}, \citenamefont {Sivarajah}, \citenamefont {Skripka}, \citenamefont
  {Sletten}, \citenamefont {Spaun}, \citenamefont {Sprenkle}, \citenamefont
  {Stoufer}, \citenamefont {Tader}, \citenamefont {Taylor}, \citenamefont
  {Thompson}, \citenamefont {Tobey}, \citenamefont {Tran}, \citenamefont
  {Tran}, \citenamefont {Vittorini}, \citenamefont {Volin}, \citenamefont
  {Walker}, \citenamefont {White}, \citenamefont {Williams}, \citenamefont
  {Wilson}, \citenamefont {Wolf}, \citenamefont {Wringe}, \citenamefont
  {Young}, \citenamefont {Zheng}, \citenamefont {Zuraski}, \citenamefont
  {Baldwin}, \citenamefont {Chernoguzov}, \citenamefont {Gaebler},
  \citenamefont {Sanders}, \citenamefont {Neyenhuis}, \citenamefont {Stutz},\
  and\ \citenamefont {Bohnet}}]{Ransford2026}%
  \BibitemOpen
  \bibfield  {author} {\bibinfo {author} {\bibfnamefont {A.}~\bibnamefont
  {Ransford}}, \bibinfo {author} {\bibfnamefont {M.~S.}\ \bibnamefont
  {Allman}}, \bibinfo {author} {\bibfnamefont {J.}~\bibnamefont {Arkinstall}},
  \bibinfo {author} {\bibfnamefont {J.~P.}\ \bibnamefont {Campora}}, \bibinfo
  {author} {\bibfnamefont {S.~F.}\ \bibnamefont {Cooper}}, \bibinfo {author}
  {\bibfnamefont {R.~D.}\ \bibnamefont {Delaney}}, \bibinfo {author}
  {\bibfnamefont {J.~M.}\ \bibnamefont {Dreiling}}, \bibinfo {author}
  {\bibfnamefont {B.}~\bibnamefont {Estey}}, \bibinfo {author} {\bibfnamefont
  {C.}~\bibnamefont {Figgatt}}, \bibinfo {author} {\bibfnamefont
  {A.}~\bibnamefont {Hall}}, \bibinfo {author} {\bibfnamefont {A.~A.}\
  \bibnamefont {Husain}}, \bibinfo {author} {\bibfnamefont {A.}~\bibnamefont
  {Isanaka}}, \bibinfo {author} {\bibfnamefont {C.~J.}\ \bibnamefont
  {Kennedy}}, \bibinfo {author} {\bibfnamefont {N.}~\bibnamefont
  {Kotibhaskar}}, \bibinfo {author} {\bibfnamefont {I.~S.}\ \bibnamefont
  {Madjarov}}, \bibinfo {author} {\bibfnamefont {K.}~\bibnamefont {Mayer}},
  \bibinfo {author} {\bibfnamefont {A.~R.}\ \bibnamefont {Milne}}, \bibinfo
  {author} {\bibfnamefont {A.~J.}\ \bibnamefont {Park}}, \bibinfo {author}
  {\bibfnamefont {A.~P.}\ \bibnamefont {Reed}}, \bibinfo {author}
  {\bibfnamefont {R.}~\bibnamefont {Ancona}}, \bibinfo {author} {\bibfnamefont
  {M.~P.}\ \bibnamefont {Andersen}}, \bibinfo {author} {\bibfnamefont
  {P.}~\bibnamefont {Andres-Martinez}}, \bibinfo {author} {\bibfnamefont
  {W.}~\bibnamefont {Angenent}}, \bibinfo {author} {\bibfnamefont
  {L.}~\bibnamefont {Argueta}}, \bibinfo {author} {\bibfnamefont
  {B.}~\bibnamefont {Arkin}}, \bibinfo {author} {\bibfnamefont
  {L.}~\bibnamefont {Ascarrunz}}, \bibinfo {author} {\bibfnamefont
  {W.}~\bibnamefont {Baker}}, \bibinfo {author} {\bibfnamefont
  {C.}~\bibnamefont {Barnes}}, \bibinfo {author} {\bibfnamefont
  {J.}~\bibnamefont {Bartolotta}}, \bibinfo {author} {\bibfnamefont
  {J.}~\bibnamefont {Berg}}, \bibinfo {author} {\bibfnamefont {R.}~\bibnamefont
  {Besand}}, \bibinfo {author} {\bibfnamefont {B.}~\bibnamefont {Bjork}},
  \bibinfo {author} {\bibfnamefont {M.}~\bibnamefont {Blain}}, \bibinfo
  {author} {\bibfnamefont {P.}~\bibnamefont {Blanchard}}, \bibinfo {author}
  {\bibfnamefont {R.}~\bibnamefont {Blume-Kohout}}, \bibinfo {author}
  {\bibfnamefont {M.}~\bibnamefont {Bohn}}, \bibinfo {author} {\bibfnamefont
  {A.}~\bibnamefont {Borgna}}, \bibinfo {author} {\bibfnamefont {D.~Y.}\
  \bibnamefont {Botamanenko}}, \bibinfo {author} {\bibfnamefont
  {R.}~\bibnamefont {Boutelle}}, \bibinfo {author} {\bibfnamefont
  {N.}~\bibnamefont {Brown}}, \bibinfo {author} {\bibfnamefont {G.~T.}\
  \bibnamefont {Buckingham}}, \bibinfo {author} {\bibfnamefont {N.~Q.}\
  \bibnamefont {Burdick}}, \bibinfo {author} {\bibfnamefont {W.~C.}\
  \bibnamefont {Burton}}, \bibinfo {author} {\bibfnamefont {V.}~\bibnamefont
  {Carey}}, \bibinfo {author} {\bibfnamefont {C.~J.}\ \bibnamefont {Carron}},
  \bibinfo {author} {\bibfnamefont {J.}~\bibnamefont {Chambers}}, \bibinfo
  {author} {\bibfnamefont {J.~W.}\ \bibnamefont {Chan}}, \bibinfo {author}
  {\bibfnamefont {J.}~\bibnamefont {Children}}, \bibinfo {author}
  {\bibfnamefont {V.~E.}\ \bibnamefont {Colussi}}, \bibinfo {author}
  {\bibfnamefont {S.}~\bibnamefont {Crepinsek}}, \bibinfo {author}
  {\bibfnamefont {A.}~\bibnamefont {Cureton}}, \bibinfo {author} {\bibfnamefont
  {J.}~\bibnamefont {Davies}}, \bibinfo {author} {\bibfnamefont
  {D.}~\bibnamefont {Davis}}, \bibinfo {author} {\bibfnamefont
  {M.}~\bibnamefont {DeCross}}, \bibinfo {author} {\bibfnamefont
  {D.}~\bibnamefont {Deen}}, \bibinfo {author} {\bibfnamefont {C.}~\bibnamefont
  {Delaney}}, \bibinfo {author} {\bibfnamefont {D.}~\bibnamefont {DelVento}},
  \bibinfo {author} {\bibfnamefont {B.~J.}\ \bibnamefont {DeSalvo}}, \bibinfo
  {author} {\bibfnamefont {J.}~\bibnamefont {Dominy}}, \bibinfo {author}
  {\bibfnamefont {S.}~\bibnamefont {Drotar}}, \bibinfo {author} {\bibfnamefont
  {R.}~\bibnamefont {Duncan}}, \bibinfo {author} {\bibfnamefont
  {V.}~\bibnamefont {Eccles}}, \bibinfo {author} {\bibfnamefont
  {A.}~\bibnamefont {Edgington}}, \bibinfo {author} {\bibfnamefont
  {N.}~\bibnamefont {Erickson}}, \bibinfo {author} {\bibfnamefont
  {S.}~\bibnamefont {Erickson}}, \bibinfo {author} {\bibfnamefont {C.~T.}\
  \bibnamefont {Ertsgaard}}, \bibinfo {author} {\bibfnamefont {J.}~\bibnamefont
  {Esposito}}, \bibinfo {author} {\bibfnamefont {B.}~\bibnamefont {Evans}},
  \bibinfo {author} {\bibfnamefont {T.}~\bibnamefont {Evans}}, \bibinfo
  {author} {\bibfnamefont {M.~I.}\ \bibnamefont {Fabrikant}}, \bibinfo {author}
  {\bibfnamefont {A.}~\bibnamefont {Fischer}}, \bibinfo {author} {\bibfnamefont
  {C.}~\bibnamefont {Foltz}}, \bibinfo {author} {\bibfnamefont
  {M.}~\bibnamefont {Foss-Feig}}, \bibinfo {author} {\bibfnamefont
  {D.}~\bibnamefont {Francois}}, \bibinfo {author} {\bibfnamefont
  {B.}~\bibnamefont {Freyberg}}, \bibinfo {author} {\bibfnamefont
  {C.}~\bibnamefont {Gao}}, \bibinfo {author} {\bibfnamefont {R.}~\bibnamefont
  {Garay}}, \bibinfo {author} {\bibfnamefont {J.}~\bibnamefont {Garvin}},
  \bibinfo {author} {\bibfnamefont {D.~M.}\ \bibnamefont {Gaudiosi}}, \bibinfo
  {author} {\bibfnamefont {C.~N.}\ \bibnamefont {Gilbreth}}, \bibinfo {author}
  {\bibfnamefont {J.}~\bibnamefont {Giles}}, \bibinfo {author} {\bibfnamefont
  {E.}~\bibnamefont {Glynn}}, \bibinfo {author} {\bibfnamefont
  {J.}~\bibnamefont {Graves}}, \bibinfo {author} {\bibfnamefont
  {A.}~\bibnamefont {Hansen}}, \bibinfo {author} {\bibfnamefont
  {D.}~\bibnamefont {Hayes}}, \bibinfo {author} {\bibfnamefont
  {L.}~\bibnamefont {Heidemann}}, \bibinfo {author} {\bibfnamefont
  {B.}~\bibnamefont {Higashi}}, \bibinfo {author} {\bibfnamefont
  {T.}~\bibnamefont {Hilbun}}, \bibinfo {author} {\bibfnamefont
  {J.}~\bibnamefont {Hines}}, \bibinfo {author} {\bibfnamefont
  {A.}~\bibnamefont {Hlavaty}}, \bibinfo {author} {\bibfnamefont
  {K.}~\bibnamefont {Hoffman}}, \bibinfo {author} {\bibfnamefont {I.~M.}\
  \bibnamefont {Hoffman}}, \bibinfo {author} {\bibfnamefont {C.}~\bibnamefont
  {Holliman}}, \bibinfo {author} {\bibfnamefont {I.}~\bibnamefont {Hooper}},
  \bibinfo {author} {\bibfnamefont {B.}~\bibnamefont {Horning}}, \bibinfo
  {author} {\bibfnamefont {J.}~\bibnamefont {Hostetter}}, \bibinfo {author}
  {\bibfnamefont {D.}~\bibnamefont {Hothem}}, \bibinfo {author} {\bibfnamefont
  {J.}~\bibnamefont {Houlton}}, \bibinfo {author} {\bibfnamefont
  {J.}~\bibnamefont {Hout}}, \bibinfo {author} {\bibfnamefont {R.}~\bibnamefont
  {Hutson}}, \bibinfo {author} {\bibfnamefont {R.~T.}\ \bibnamefont {Jacobs}},
  \bibinfo {author} {\bibfnamefont {T.}~\bibnamefont {Jacobs}}, \bibinfo
  {author} {\bibfnamefont {M.}~\bibnamefont {Johannsen}}, \bibinfo {author}
  {\bibfnamefont {J.}~\bibnamefont {Johansen}}, \bibinfo {author}
  {\bibfnamefont {L.}~\bibnamefont {Jones}}, \bibinfo {author} {\bibfnamefont
  {S.}~\bibnamefont {Julian}}, \bibinfo {author} {\bibfnamefont
  {R.}~\bibnamefont {Jung}}, \bibinfo {author} {\bibfnamefont {A.}~\bibnamefont
  {Keay}}, \bibinfo {author} {\bibfnamefont {T.}~\bibnamefont {Klein}},
  \bibinfo {author} {\bibfnamefont {M.}~\bibnamefont {Koch}}, \bibinfo {author}
  {\bibfnamefont {R.}~\bibnamefont {Kondo}}, \bibinfo {author} {\bibfnamefont
  {C.}~\bibnamefont {Kong}}, \bibinfo {author} {\bibfnamefont {A.}~\bibnamefont
  {Kosto}}, \bibinfo {author} {\bibfnamefont {A.}~\bibnamefont {Lawrence}},
  \bibinfo {author} {\bibfnamefont {D.}~\bibnamefont {Liefer}}, \bibinfo
  {author} {\bibfnamefont {M.}~\bibnamefont {Lollie}}, \bibinfo {author}
  {\bibfnamefont {D.}~\bibnamefont {Lucchetti}}, \bibinfo {author}
  {\bibfnamefont {N.~K.}\ \bibnamefont {Lysne}}, \bibinfo {author}
  {\bibfnamefont {C.}~\bibnamefont {Lytle}}, \bibinfo {author} {\bibfnamefont
  {C.}~\bibnamefont {MacPherson}}, \bibinfo {author} {\bibfnamefont
  {A.}~\bibnamefont {Malm}}, \bibinfo {author} {\bibfnamefont {S.}~\bibnamefont
  {Mather}}, \bibinfo {author} {\bibfnamefont {B.}~\bibnamefont {Mathewson}},
  \bibinfo {author} {\bibfnamefont {D.}~\bibnamefont {Maxwell}}, \bibinfo
  {author} {\bibfnamefont {L.}~\bibnamefont {McCaffrey}}, \bibinfo {author}
  {\bibfnamefont {H.}~\bibnamefont {McDougall}}, \bibinfo {author}
  {\bibfnamefont {R.}~\bibnamefont {Mendoza}}, \bibinfo {author} {\bibfnamefont
  {D.~B.}\ \bibnamefont {Miller}}, \bibinfo {author} {\bibfnamefont
  {M.}~\bibnamefont {Mills}}, \bibinfo {author} {\bibfnamefont
  {R.}~\bibnamefont {Morrison}}, \bibinfo {author} {\bibfnamefont
  {L.}~\bibnamefont {Narmour}}, \bibinfo {author} {\bibfnamefont
  {N.}~\bibnamefont {Nguyen}}, \bibinfo {author} {\bibfnamefont
  {L.}~\bibnamefont {Nugent}}, \bibinfo {author} {\bibfnamefont
  {S.}~\bibnamefont {Olson}}, \bibinfo {author} {\bibfnamefont
  {D.}~\bibnamefont {Ouellette}}, \bibinfo {author} {\bibfnamefont
  {J.}~\bibnamefont {Parks}}, \bibinfo {author} {\bibfnamefont
  {Z.}~\bibnamefont {Peters}}, \bibinfo {author} {\bibfnamefont {T.~A.}\
  \bibnamefont {Peterson}}, \bibinfo {author} {\bibfnamefont {J.}~\bibnamefont
  {Petricka}}, \bibinfo {author} {\bibfnamefont {J.~M.}\ \bibnamefont {Pino}},
  \bibinfo {author} {\bibfnamefont {F.}~\bibnamefont {Polito}}, \bibinfo
  {author} {\bibfnamefont {A.~C.}\ \bibnamefont {Potter}}, \bibinfo {author}
  {\bibfnamefont {M.}~\bibnamefont {Preidl}}, \bibinfo {author} {\bibfnamefont
  {G.}~\bibnamefont {Price}}, \bibinfo {author} {\bibfnamefont
  {T.}~\bibnamefont {Proctor}}, \bibinfo {author} {\bibfnamefont
  {M.}~\bibnamefont {Pugh}}, \bibinfo {author} {\bibfnamefont {N.}~\bibnamefont
  {Ratcliff}}, \bibinfo {author} {\bibfnamefont {D.}~\bibnamefont
  {Raymondson}}, \bibinfo {author} {\bibfnamefont {P.}~\bibnamefont {Rhodes}},
  \bibinfo {author} {\bibfnamefont {C.}~\bibnamefont {Roman}}, \bibinfo
  {author} {\bibfnamefont {C.}~\bibnamefont {Roy}}, \bibinfo {author}
  {\bibfnamefont {C.}~\bibnamefont {Ryan-Anderson}}, \bibinfo {author}
  {\bibfnamefont {F.~B.}\ \bibnamefont {Sanchez}}, \bibinfo {author}
  {\bibfnamefont {G.}~\bibnamefont {Sangiolo}}, \bibinfo {author}
  {\bibfnamefont {T.}~\bibnamefont {Sawadski}}, \bibinfo {author}
  {\bibfnamefont {A.}~\bibnamefont {Schaffer}}, \bibinfo {author}
  {\bibfnamefont {P.}~\bibnamefont {Schow}}, \bibinfo {author} {\bibfnamefont
  {J.}~\bibnamefont {Sedlacek}}, \bibinfo {author} {\bibfnamefont
  {H.}~\bibnamefont {Semenenko}}, \bibinfo {author} {\bibfnamefont
  {P.}~\bibnamefont {Shevchuk}}, \bibinfo {author} {\bibfnamefont
  {S.}~\bibnamefont {Shore}}, \bibinfo {author} {\bibfnamefont
  {P.}~\bibnamefont {Siegfried}}, \bibinfo {author} {\bibfnamefont
  {K.}~\bibnamefont {Singhal}}, \bibinfo {author} {\bibfnamefont
  {S.}~\bibnamefont {Sivarajah}}, \bibinfo {author} {\bibfnamefont
  {T.}~\bibnamefont {Skripka}}, \bibinfo {author} {\bibfnamefont
  {L.}~\bibnamefont {Sletten}}, \bibinfo {author} {\bibfnamefont
  {B.}~\bibnamefont {Spaun}}, \bibinfo {author} {\bibfnamefont {R.~T.}\
  \bibnamefont {Sprenkle}}, \bibinfo {author} {\bibfnamefont {P.}~\bibnamefont
  {Stoufer}}, \bibinfo {author} {\bibfnamefont {M.}~\bibnamefont {Tader}},
  \bibinfo {author} {\bibfnamefont {S.~F.}\ \bibnamefont {Taylor}}, \bibinfo
  {author} {\bibfnamefont {T.~H.}\ \bibnamefont {Thompson}}, \bibinfo {author}
  {\bibfnamefont {R.}~\bibnamefont {Tobey}}, \bibinfo {author} {\bibfnamefont
  {A.}~\bibnamefont {Tran}}, \bibinfo {author} {\bibfnamefont {T.}~\bibnamefont
  {Tran}}, \bibinfo {author} {\bibfnamefont {G.}~\bibnamefont {Vittorini}},
  \bibinfo {author} {\bibfnamefont {C.}~\bibnamefont {Volin}}, \bibinfo
  {author} {\bibfnamefont {J.}~\bibnamefont {Walker}}, \bibinfo {author}
  {\bibfnamefont {S.}~\bibnamefont {White}}, \bibinfo {author} {\bibfnamefont
  {G.~R.}\ \bibnamefont {Williams}}, \bibinfo {author} {\bibfnamefont
  {D.}~\bibnamefont {Wilson}}, \bibinfo {author} {\bibfnamefont
  {Q.}~\bibnamefont {Wolf}}, \bibinfo {author} {\bibfnamefont {C.}~\bibnamefont
  {Wringe}}, \bibinfo {author} {\bibfnamefont {K.}~\bibnamefont {Young}},
  \bibinfo {author} {\bibfnamefont {J.}~\bibnamefont {Zheng}}, \bibinfo
  {author} {\bibfnamefont {K.}~\bibnamefont {Zuraski}}, \bibinfo {author}
  {\bibfnamefont {C.~H.}\ \bibnamefont {Baldwin}}, \bibinfo {author}
  {\bibfnamefont {A.}~\bibnamefont {Chernoguzov}}, \bibinfo {author}
  {\bibfnamefont {J.~P.}\ \bibnamefont {Gaebler}}, \bibinfo {author}
  {\bibfnamefont {S.~J.}\ \bibnamefont {Sanders}}, \bibinfo {author}
  {\bibfnamefont {B.}~\bibnamefont {Neyenhuis}}, \bibinfo {author}
  {\bibfnamefont {R.}~\bibnamefont {Stutz}}, \ and\ \bibinfo {author}
  {\bibfnamefont {J.~G.}\ \bibnamefont {Bohnet}},\ }\bibfield  {title}
  {\enquote {\bibinfo {title} {A 98-qubit trapped-ion quantum computer with
  all-to-all connectivity},}\ }\href
  {http://dx.doi.org/10.1038/s41586-026-10676-4} {\bibfield  {journal}
  {\bibinfo  {journal} {\emph {Nature}}\ }\textbf {\bibinfo {volume} {655}},\
  \bibinfo {pages} {81} (\bibinfo {year} {2026})}\BibitemShut {NoStop}%
\bibitem [{\citenamefont {Cao}\ \emph {et~al.}(2019)\citenamefont {Cao},
  \citenamefont {Romero}, \citenamefont {Olson}, \citenamefont {Degroote},
  \citenamefont {Johnson}, \citenamefont {Kieferov\'a}, \citenamefont
  {Kivlichan}, \citenamefont {Menke}, \citenamefont {Peropadre}, \citenamefont
  {Sawaya}, \citenamefont {Sim}, \citenamefont {Veis},\ and\ \citenamefont
  {Aspuru-Guzik}}]{Caoetal_review2019}%
  \BibitemOpen
  \bibfield  {author} {\bibinfo {author} {\bibfnamefont {Y.}~\bibnamefont
  {Cao}}, \bibinfo {author} {\bibfnamefont {J.}~\bibnamefont {Romero}},
  \bibinfo {author} {\bibfnamefont {J.~P.}\ \bibnamefont {Olson}}, \bibinfo
  {author} {\bibfnamefont {M.}~\bibnamefont {Degroote}}, \bibinfo {author}
  {\bibfnamefont {P.~D.}\ \bibnamefont {Johnson}}, \bibinfo {author}
  {\bibfnamefont {M.}~\bibnamefont {Kieferov\'a}}, \bibinfo {author}
  {\bibfnamefont {I.~D.}\ \bibnamefont {Kivlichan}}, \bibinfo {author}
  {\bibfnamefont {T.}~\bibnamefont {Menke}}, \bibinfo {author} {\bibfnamefont
  {B.}~\bibnamefont {Peropadre}}, \bibinfo {author} {\bibfnamefont {N.~P.~D.}\
  \bibnamefont {Sawaya}}, \bibinfo {author} {\bibfnamefont {S.}~\bibnamefont
  {Sim}}, \bibinfo {author} {\bibfnamefont {L.}~\bibnamefont {Veis}}, \ and\
  \bibinfo {author} {\bibfnamefont {A.}~\bibnamefont {Aspuru-Guzik}},\
  }\bibfield  {title} {\enquote {\bibinfo {title} {Quantum chemistry in the age
  of quantum computing},}\ }\href
  {http://dx.doi.org/10.1021/acs.chemrev.8b00803} {\bibfield  {journal}
  {\bibinfo  {journal} {\emph {Chem. Rev.}}\ }\textbf {\bibinfo {volume}
  {119}},\ \bibinfo {pages} {10856} (\bibinfo {year} {2019})}\BibitemShut
  {NoStop}%
\bibitem [{\citenamefont {Bauer}\ \emph {et~al.}(2020)\citenamefont {Bauer},
  \citenamefont {Bravyi}, \citenamefont {Motta},\ and\ \citenamefont
  {Chan}}]{Baueretal2020}%
  \BibitemOpen
  \bibfield  {author} {\bibinfo {author} {\bibfnamefont {B.}~\bibnamefont
  {Bauer}}, \bibinfo {author} {\bibfnamefont {S.}~\bibnamefont {Bravyi}},
  \bibinfo {author} {\bibfnamefont {M.}~\bibnamefont {Motta}}, \ and\ \bibinfo
  {author} {\bibfnamefont {G.~K.-L.}\ \bibnamefont {Chan}},\ }\bibfield
  {title} {\enquote {\bibinfo {title} {Quantum algorithms for quantum chemistry
  and quantum materials science},}\ }\href
  {http://dx.doi.org/10.1021/acs.chemrev.9b00829} {\bibfield  {journal}
  {\bibinfo  {journal} {\emph {Chem. Rev.}}\ }\textbf {\bibinfo {volume}
  {120}},\ \bibinfo {pages} {12685} (\bibinfo {year} {2020})}\BibitemShut
  {NoStop}%
\bibitem [{\citenamefont {McArdle}\ \emph {et~al.}(2020)\citenamefont
  {McArdle}, \citenamefont {Endo}, \citenamefont {Aspuru-Guzik}, \citenamefont
  {Benjamin},\ and\ \citenamefont {Yuan}}]{McArdleetal_review2020}%
  \BibitemOpen
  \bibfield  {author} {\bibinfo {author} {\bibfnamefont {S.}~\bibnamefont
  {McArdle}}, \bibinfo {author} {\bibfnamefont {S.}~\bibnamefont {Endo}},
  \bibinfo {author} {\bibfnamefont {A.}~\bibnamefont {Aspuru-Guzik}}, \bibinfo
  {author} {\bibfnamefont {S.~C.}\ \bibnamefont {Benjamin}}, \ and\ \bibinfo
  {author} {\bibfnamefont {X.}~\bibnamefont {Yuan}},\ }\bibfield  {title}
  {\enquote {\bibinfo {title} {Quantum computational chemistry},}\ }\href
  {http://dx.doi.org/10.1103/RevModPhys.92.015003} {\bibfield  {journal}
  {\bibinfo  {journal} {\emph {Rev. Mod. Phys.}}\ }\textbf {\bibinfo {volume}
  {92}},\ \bibinfo {pages} {015003} (\bibinfo {year} {2020})}\BibitemShut
  {NoStop}%
\bibitem [{\citenamefont {Weidman}\ \emph {et~al.}(2024)\citenamefont
  {Weidman}, \citenamefont {Sajjan}, \citenamefont {Mikolas}, \citenamefont
  {Stewart}, \citenamefont {Pollanen}, \citenamefont {Kais},\ and\
  \citenamefont {Wilson}}]{Weidman2024}%
  \BibitemOpen
  \bibfield  {author} {\bibinfo {author} {\bibfnamefont {J.~D.}\ \bibnamefont
  {Weidman}}, \bibinfo {author} {\bibfnamefont {M.}~\bibnamefont {Sajjan}},
  \bibinfo {author} {\bibfnamefont {C.}~\bibnamefont {Mikolas}}, \bibinfo
  {author} {\bibfnamefont {Z.~J.}\ \bibnamefont {Stewart}}, \bibinfo {author}
  {\bibfnamefont {J.}~\bibnamefont {Pollanen}}, \bibinfo {author}
  {\bibfnamefont {S.}~\bibnamefont {Kais}}, \ and\ \bibinfo {author}
  {\bibfnamefont {A.~K.}\ \bibnamefont {Wilson}},\ }\bibfield  {title}
  {\enquote {\bibinfo {title} {Quantum computing and chemistry},}\ }\href
  {http://dx.doi.org/10.1016/j.xcrp.2024.102105} {\bibfield  {journal}
  {\bibinfo  {journal} {\emph {Cell Rep. Phys. Sci.}}\ }\textbf {\bibinfo
  {volume} {5}},\ \bibinfo {pages} {102105} (\bibinfo {year}
  {2024})}\BibitemShut {NoStop}%
\bibitem [{\citenamefont {Eddins}\ \emph {et~al.}(2022)\citenamefont {Eddins},
  \citenamefont {Motta}, \citenamefont {Gujarati}, \citenamefont {Bravyi},
  \citenamefont {Mezzacapo}, \citenamefont {Hadfield},\ and\ \citenamefont
  {Sheldon}}]{Eddinsetal2021}%
  \BibitemOpen
  \bibfield  {author} {\bibinfo {author} {\bibfnamefont {A.}~\bibnamefont
  {Eddins}}, \bibinfo {author} {\bibfnamefont {M.}~\bibnamefont {Motta}},
  \bibinfo {author} {\bibfnamefont {T.~P.}\ \bibnamefont {Gujarati}}, \bibinfo
  {author} {\bibfnamefont {S.}~\bibnamefont {Bravyi}}, \bibinfo {author}
  {\bibfnamefont {A.}~\bibnamefont {Mezzacapo}}, \bibinfo {author}
  {\bibfnamefont {C.}~\bibnamefont {Hadfield}}, \ and\ \bibinfo {author}
  {\bibfnamefont {S.}~\bibnamefont {Sheldon}},\ }\bibfield  {title} {\enquote
  {\bibinfo {title} {Doubling the size of quantum simulators by entanglement
  forging},}\ }\href {http://dx.doi.org/10.1103/PRXQuantum.3.010309} {\bibfield
   {journal} {\bibinfo  {journal} {\emph {PRX Quantum}}\ }\textbf {\bibinfo
  {volume} {3}},\ \bibinfo {pages} {010309} (\bibinfo {year}
  {2022})}\BibitemShut {NoStop}%
\bibitem [{\citenamefont {Guo}\ \emph {et~al.}(2024)\citenamefont {Guo},
  \citenamefont {Sun}, \citenamefont {Qian}, \citenamefont {Gong},
  \citenamefont {Zhang}, \citenamefont {Chen}, \citenamefont {Ye},
  \citenamefont {Wu}, \citenamefont {Cao}, \citenamefont {Liu}, \citenamefont
  {Zha}, \citenamefont {Ying}, \citenamefont {Zhu}, \citenamefont {Huang},
  \citenamefont {Zhao}, \citenamefont {Li}, \citenamefont {Wang}, \citenamefont
  {Yu}, \citenamefont {Fan}, \citenamefont {Wu}, \citenamefont {Su},
  \citenamefont {Deng}, \citenamefont {Rong}, \citenamefont {Li}, \citenamefont
  {Zhang}, \citenamefont {Chung}, \citenamefont {Liang}, \citenamefont {Lin},
  \citenamefont {Xu}, \citenamefont {Sun}, \citenamefont {Guo}, \citenamefont
  {Li}, \citenamefont {Huo}, \citenamefont {Peng}, \citenamefont {Lu},
  \citenamefont {Yuan}, \citenamefont {Zhu},\ and\ \citenamefont
  {Pan}}]{Guo2024}%
  \BibitemOpen
  \bibfield  {author} {\bibinfo {author} {\bibfnamefont {S.}~\bibnamefont
  {Guo}}, \bibinfo {author} {\bibfnamefont {J.}~\bibnamefont {Sun}}, \bibinfo
  {author} {\bibfnamefont {H.}~\bibnamefont {Qian}}, \bibinfo {author}
  {\bibfnamefont {M.}~\bibnamefont {Gong}}, \bibinfo {author} {\bibfnamefont
  {Y.}~\bibnamefont {Zhang}}, \bibinfo {author} {\bibfnamefont
  {F.}~\bibnamefont {Chen}}, \bibinfo {author} {\bibfnamefont {Y.}~\bibnamefont
  {Ye}}, \bibinfo {author} {\bibfnamefont {Y.}~\bibnamefont {Wu}}, \bibinfo
  {author} {\bibfnamefont {S.}~\bibnamefont {Cao}}, \bibinfo {author}
  {\bibfnamefont {K.}~\bibnamefont {Liu}}, \bibinfo {author} {\bibfnamefont
  {C.}~\bibnamefont {Zha}}, \bibinfo {author} {\bibfnamefont {C.}~\bibnamefont
  {Ying}}, \bibinfo {author} {\bibfnamefont {Q.}~\bibnamefont {Zhu}}, \bibinfo
  {author} {\bibfnamefont {H.-L.}\ \bibnamefont {Huang}}, \bibinfo {author}
  {\bibfnamefont {Y.}~\bibnamefont {Zhao}}, \bibinfo {author} {\bibfnamefont
  {S.}~\bibnamefont {Li}}, \bibinfo {author} {\bibfnamefont {S.}~\bibnamefont
  {Wang}}, \bibinfo {author} {\bibfnamefont {J.}~\bibnamefont {Yu}}, \bibinfo
  {author} {\bibfnamefont {D.}~\bibnamefont {Fan}}, \bibinfo {author}
  {\bibfnamefont {D.}~\bibnamefont {Wu}}, \bibinfo {author} {\bibfnamefont
  {H.}~\bibnamefont {Su}}, \bibinfo {author} {\bibfnamefont {H.}~\bibnamefont
  {Deng}}, \bibinfo {author} {\bibfnamefont {H.}~\bibnamefont {Rong}}, \bibinfo
  {author} {\bibfnamefont {Y.}~\bibnamefont {Li}}, \bibinfo {author}
  {\bibfnamefont {K.}~\bibnamefont {Zhang}}, \bibinfo {author} {\bibfnamefont
  {T.-H.}\ \bibnamefont {Chung}}, \bibinfo {author} {\bibfnamefont
  {F.}~\bibnamefont {Liang}}, \bibinfo {author} {\bibfnamefont
  {J.}~\bibnamefont {Lin}}, \bibinfo {author} {\bibfnamefont {Y.}~\bibnamefont
  {Xu}}, \bibinfo {author} {\bibfnamefont {L.}~\bibnamefont {Sun}}, \bibinfo
  {author} {\bibfnamefont {C.}~\bibnamefont {Guo}}, \bibinfo {author}
  {\bibfnamefont {N.}~\bibnamefont {Li}}, \bibinfo {author} {\bibfnamefont
  {Y.-H.}\ \bibnamefont {Huo}}, \bibinfo {author} {\bibfnamefont {C.-Z.}\
  \bibnamefont {Peng}}, \bibinfo {author} {\bibfnamefont {C.-Y.}\ \bibnamefont
  {Lu}}, \bibinfo {author} {\bibfnamefont {X.}~\bibnamefont {Yuan}}, \bibinfo
  {author} {\bibfnamefont {X.}~\bibnamefont {Zhu}}, \ and\ \bibinfo {author}
  {\bibfnamefont {J.-W.}\ \bibnamefont {Pan}},\ }\bibfield  {title} {\enquote
  {\bibinfo {title} {Experimental quantum computational chemistry with
  optimized unitary coupled cluster ansatz},}\ }\href
  {http://dx.doi.org/10.1038/s41567-024-02530-z} {\bibfield  {journal}
  {\bibinfo  {journal} {\emph {Nat. Phys.}}\ }\textbf {\bibinfo {volume}
  {20}},\ \bibinfo {pages} {1240} (\bibinfo {year} {2024})}\BibitemShut
  {NoStop}%
\bibitem [{\citenamefont {Tranter}\ \emph {et~al.}(2018)\citenamefont
  {Tranter}, \citenamefont {Love}, \citenamefont {Mintert},\ and\ \citenamefont
  {Coveney}}]{Tranteretal_2018}%
  \BibitemOpen
  \bibfield  {author} {\bibinfo {author} {\bibfnamefont {A.}~\bibnamefont
  {Tranter}}, \bibinfo {author} {\bibfnamefont {P.~J.}\ \bibnamefont {Love}},
  \bibinfo {author} {\bibfnamefont {F.}~\bibnamefont {Mintert}}, \ and\
  \bibinfo {author} {\bibfnamefont {P.~V.}\ \bibnamefont {Coveney}},\
  }\bibfield  {title} {\enquote {\bibinfo {title} {A comparison of the
  {Bravyi-Kitaev} and {Jordan-Wigner} transformations for the quantum
  simulation of quantum chemistry},}\ }\href
  {http://dx.doi.org/10.1021/acs.jctc.8b00450} {\bibfield  {journal} {\bibinfo
  {journal} {\emph {J. Chem. Theory Comput.}}\ }\textbf {\bibinfo {volume}
  {14}},\ \bibinfo {pages} {5617} (\bibinfo {year} {2018})}\BibitemShut
  {NoStop}%
\bibitem [{\citenamefont {Alexeev}\ \emph {et~al.}(2024)\citenamefont
  {Alexeev}, \citenamefont {Amsler}, \citenamefont {Barroca}, \citenamefont
  {Bassini}, \citenamefont {Battelle}, \citenamefont {Camps}, \citenamefont
  {Casanova}, \citenamefont {Choi}, \citenamefont {Chong}, \citenamefont
  {Chung}, \citenamefont {Codella}, \citenamefont {C\'orcoles}, \citenamefont
  {Cruise}, \citenamefont {Di~Meglio}, \citenamefont {Duran}, \citenamefont
  {Eckl}, \citenamefont {Economou}, \citenamefont {Eidenbenz}, \citenamefont
  {Elmegreen}, \citenamefont {Fare}, \citenamefont {Faro}, \citenamefont
  {Fern\'andez}, \citenamefont {Ferreira}, \citenamefont {Fuji}, \citenamefont
  {Fuller}, \citenamefont {Gagliardi}, \citenamefont {Galli}, \citenamefont
  {Glick}, \citenamefont {Gobbi}, \citenamefont {Gokhale}, \citenamefont {de~la
  Puente~Gonzalez}, \citenamefont {Greiner}, \citenamefont {Gropp},
  \citenamefont {Grossi}, \citenamefont {Gull}, \citenamefont {Healy},
  \citenamefont {Hermes}, \citenamefont {Huang}, \citenamefont {Humble},
  \citenamefont {Ito}, \citenamefont {Izmaylov}, \citenamefont {Javadi-Abhari},
  \citenamefont {Jennewein}, \citenamefont {Jha}, \citenamefont {Jiang},
  \citenamefont {Jones}, \citenamefont {de~Jong}, \citenamefont {Jurcevic},
  \citenamefont {Kirby}, \citenamefont {Kister}, \citenamefont {Kitagawa},
  \citenamefont {Klassen}, \citenamefont {Klymko}, \citenamefont {Koh},
  \citenamefont {Kondo}, \citenamefont {K\"urk\c{c}\"uo\~glu}, \citenamefont
  {Kurowski}, \citenamefont {Laino}, \citenamefont {Landfield}, \citenamefont
  {Leininger}, \citenamefont {Leyton-Ortega}, \citenamefont {Li}, \citenamefont
  {Lin}, \citenamefont {Liu}, \citenamefont {Lorente}, \citenamefont {Luckow},
  \citenamefont {Martiel}, \citenamefont {Martin-Fernandez}, \citenamefont
  {Martonosi}, \citenamefont {Marvinney}, \citenamefont {Medina}, \citenamefont
  {Merten}, \citenamefont {Mezzacapo}, \citenamefont {Michielsen},
  \citenamefont {Mitra}, \citenamefont {Mittal}, \citenamefont {Moon},
  \citenamefont {Moore}, \citenamefont {Mostame}, \citenamefont {Motta},
  \citenamefont {Na}, \citenamefont {Nam}, \citenamefont {Narang},
  \citenamefont {Ohnishi}, \citenamefont {Ottaviani}, \citenamefont {Otten},
  \citenamefont {Pakin}, \citenamefont {Pascuzzi}, \citenamefont {Pednault},
  \citenamefont {Piontek}, \citenamefont {Pitera}, \citenamefont {Rall},
  \citenamefont {Ravi}, \citenamefont {Robertson}, \citenamefont {Rossi},
  \citenamefont {Rydlichowski}, \citenamefont {Ryu}, \citenamefont
  {Samsonidze}, \citenamefont {Sato}, \citenamefont {Saurabh}, \citenamefont
  {Sharma}, \citenamefont {Sharma}, \citenamefont {Shin}, \citenamefont
  {Slessman}, \citenamefont {Steiner}, \citenamefont {Sitdikov}, \citenamefont
  {Suh}, \citenamefont {Switzer}, \citenamefont {Tang}, \citenamefont
  {Thompson}, \citenamefont {Todo}, \citenamefont {Tran}, \citenamefont
  {Trenev}, \citenamefont {Trott}, \citenamefont {Tseng}, \citenamefont
  {Tubman}, \citenamefont {Tureci}, \citenamefont {Vali\~nas}, \citenamefont
  {Vallecorsa}, \citenamefont {Wever}, \citenamefont {Wojciechowski},
  \citenamefont {Wu}, \citenamefont {Yoo}, \citenamefont {Yoshioka},
  \citenamefont {Yu}, \citenamefont {Yunoki}, \citenamefont {Zhuk},\ and\
  \citenamefont {Zubarev}}]{Alexeev2024}%
  \BibitemOpen
  \bibfield  {author} {\bibinfo {author} {\bibfnamefont {Y.}~\bibnamefont
  {Alexeev}}, \bibinfo {author} {\bibfnamefont {M.}~\bibnamefont {Amsler}},
  \bibinfo {author} {\bibfnamefont {M.~A.}\ \bibnamefont {Barroca}}, \bibinfo
  {author} {\bibfnamefont {S.}~\bibnamefont {Bassini}}, \bibinfo {author}
  {\bibfnamefont {T.}~\bibnamefont {Battelle}}, \bibinfo {author}
  {\bibfnamefont {D.}~\bibnamefont {Camps}}, \bibinfo {author} {\bibfnamefont
  {D.}~\bibnamefont {Casanova}}, \bibinfo {author} {\bibfnamefont {Y.~J.}\
  \bibnamefont {Choi}}, \bibinfo {author} {\bibfnamefont {F.~T.}\ \bibnamefont
  {Chong}}, \bibinfo {author} {\bibfnamefont {C.}~\bibnamefont {Chung}},
  \bibinfo {author} {\bibfnamefont {C.}~\bibnamefont {Codella}}, \bibinfo
  {author} {\bibfnamefont {A.~D.}\ \bibnamefont {C\'orcoles}}, \bibinfo
  {author} {\bibfnamefont {J.}~\bibnamefont {Cruise}}, \bibinfo {author}
  {\bibfnamefont {A.}~\bibnamefont {Di~Meglio}}, \bibinfo {author}
  {\bibfnamefont {I.}~\bibnamefont {Duran}}, \bibinfo {author} {\bibfnamefont
  {T.}~\bibnamefont {Eckl}}, \bibinfo {author} {\bibfnamefont {S.}~\bibnamefont
  {Economou}}, \bibinfo {author} {\bibfnamefont {S.}~\bibnamefont {Eidenbenz}},
  \bibinfo {author} {\bibfnamefont {B.}~\bibnamefont {Elmegreen}}, \bibinfo
  {author} {\bibfnamefont {C.}~\bibnamefont {Fare}}, \bibinfo {author}
  {\bibfnamefont {I.}~\bibnamefont {Faro}}, \bibinfo {author} {\bibfnamefont
  {C.~S.}\ \bibnamefont {Fern\'andez}}, \bibinfo {author} {\bibfnamefont
  {R.~N.~B.}\ \bibnamefont {Ferreira}}, \bibinfo {author} {\bibfnamefont
  {K.}~\bibnamefont {Fuji}}, \bibinfo {author} {\bibfnamefont {B.}~\bibnamefont
  {Fuller}}, \bibinfo {author} {\bibfnamefont {L.}~\bibnamefont {Gagliardi}},
  \bibinfo {author} {\bibfnamefont {G.}~\bibnamefont {Galli}}, \bibinfo
  {author} {\bibfnamefont {J.~R.}\ \bibnamefont {Glick}}, \bibinfo {author}
  {\bibfnamefont {I.}~\bibnamefont {Gobbi}}, \bibinfo {author} {\bibfnamefont
  {P.}~\bibnamefont {Gokhale}}, \bibinfo {author} {\bibfnamefont
  {S.}~\bibnamefont {de~la Puente~Gonzalez}}, \bibinfo {author} {\bibfnamefont
  {J.}~\bibnamefont {Greiner}}, \bibinfo {author} {\bibfnamefont
  {B.}~\bibnamefont {Gropp}}, \bibinfo {author} {\bibfnamefont
  {M.}~\bibnamefont {Grossi}}, \bibinfo {author} {\bibfnamefont
  {E.}~\bibnamefont {Gull}}, \bibinfo {author} {\bibfnamefont {B.}~\bibnamefont
  {Healy}}, \bibinfo {author} {\bibfnamefont {M.~R.}\ \bibnamefont {Hermes}},
  \bibinfo {author} {\bibfnamefont {B.}~\bibnamefont {Huang}}, \bibinfo
  {author} {\bibfnamefont {T.~S.}\ \bibnamefont {Humble}}, \bibinfo {author}
  {\bibfnamefont {N.}~\bibnamefont {Ito}}, \bibinfo {author} {\bibfnamefont
  {A.~F.}\ \bibnamefont {Izmaylov}}, \bibinfo {author} {\bibfnamefont
  {A.}~\bibnamefont {Javadi-Abhari}}, \bibinfo {author} {\bibfnamefont
  {D.}~\bibnamefont {Jennewein}}, \bibinfo {author} {\bibfnamefont
  {S.}~\bibnamefont {Jha}}, \bibinfo {author} {\bibfnamefont {L.}~\bibnamefont
  {Jiang}}, \bibinfo {author} {\bibfnamefont {B.}~\bibnamefont {Jones}},
  \bibinfo {author} {\bibfnamefont {W.~A.}\ \bibnamefont {de~Jong}}, \bibinfo
  {author} {\bibfnamefont {P.}~\bibnamefont {Jurcevic}}, \bibinfo {author}
  {\bibfnamefont {W.}~\bibnamefont {Kirby}}, \bibinfo {author} {\bibfnamefont
  {S.}~\bibnamefont {Kister}}, \bibinfo {author} {\bibfnamefont
  {M.}~\bibnamefont {Kitagawa}}, \bibinfo {author} {\bibfnamefont
  {J.}~\bibnamefont {Klassen}}, \bibinfo {author} {\bibfnamefont
  {K.}~\bibnamefont {Klymko}}, \bibinfo {author} {\bibfnamefont
  {K.}~\bibnamefont {Koh}}, \bibinfo {author} {\bibfnamefont {M.}~\bibnamefont
  {Kondo}}, \bibinfo {author} {\bibfnamefont {D.~M.}\ \bibnamefont
  {K\"urk\c{c}\"uo\~glu}}, \bibinfo {author} {\bibfnamefont {K.}~\bibnamefont
  {Kurowski}}, \bibinfo {author} {\bibfnamefont {T.}~\bibnamefont {Laino}},
  \bibinfo {author} {\bibfnamefont {R.}~\bibnamefont {Landfield}}, \bibinfo
  {author} {\bibfnamefont {M.}~\bibnamefont {Leininger}}, \bibinfo {author}
  {\bibfnamefont {V.}~\bibnamefont {Leyton-Ortega}}, \bibinfo {author}
  {\bibfnamefont {A.}~\bibnamefont {Li}}, \bibinfo {author} {\bibfnamefont
  {M.}~\bibnamefont {Lin}}, \bibinfo {author} {\bibfnamefont {J.}~\bibnamefont
  {Liu}}, \bibinfo {author} {\bibfnamefont {N.}~\bibnamefont {Lorente}},
  \bibinfo {author} {\bibfnamefont {A.}~\bibnamefont {Luckow}}, \bibinfo
  {author} {\bibfnamefont {S.}~\bibnamefont {Martiel}}, \bibinfo {author}
  {\bibfnamefont {F.}~\bibnamefont {Martin-Fernandez}}, \bibinfo {author}
  {\bibfnamefont {M.}~\bibnamefont {Martonosi}}, \bibinfo {author}
  {\bibfnamefont {C.}~\bibnamefont {Marvinney}}, \bibinfo {author}
  {\bibfnamefont {A.~C.}\ \bibnamefont {Medina}}, \bibinfo {author}
  {\bibfnamefont {D.}~\bibnamefont {Merten}}, \bibinfo {author} {\bibfnamefont
  {A.}~\bibnamefont {Mezzacapo}}, \bibinfo {author} {\bibfnamefont
  {K.}~\bibnamefont {Michielsen}}, \bibinfo {author} {\bibfnamefont
  {A.}~\bibnamefont {Mitra}}, \bibinfo {author} {\bibfnamefont
  {T.}~\bibnamefont {Mittal}}, \bibinfo {author} {\bibfnamefont
  {K.}~\bibnamefont {Moon}}, \bibinfo {author} {\bibfnamefont {J.}~\bibnamefont
  {Moore}}, \bibinfo {author} {\bibfnamefont {S.}~\bibnamefont {Mostame}},
  \bibinfo {author} {\bibfnamefont {M.}~\bibnamefont {Motta}}, \bibinfo
  {author} {\bibfnamefont {Y.-H.}\ \bibnamefont {Na}}, \bibinfo {author}
  {\bibfnamefont {Y.}~\bibnamefont {Nam}}, \bibinfo {author} {\bibfnamefont
  {P.}~\bibnamefont {Narang}}, \bibinfo {author} {\bibfnamefont {Y.-y.}\
  \bibnamefont {Ohnishi}}, \bibinfo {author} {\bibfnamefont {D.}~\bibnamefont
  {Ottaviani}}, \bibinfo {author} {\bibfnamefont {M.}~\bibnamefont {Otten}},
  \bibinfo {author} {\bibfnamefont {S.}~\bibnamefont {Pakin}}, \bibinfo
  {author} {\bibfnamefont {V.~R.}\ \bibnamefont {Pascuzzi}}, \bibinfo {author}
  {\bibfnamefont {E.}~\bibnamefont {Pednault}}, \bibinfo {author}
  {\bibfnamefont {T.}~\bibnamefont {Piontek}}, \bibinfo {author} {\bibfnamefont
  {J.}~\bibnamefont {Pitera}}, \bibinfo {author} {\bibfnamefont
  {P.}~\bibnamefont {Rall}}, \bibinfo {author} {\bibfnamefont {G.~S.}\
  \bibnamefont {Ravi}}, \bibinfo {author} {\bibfnamefont {N.}~\bibnamefont
  {Robertson}}, \bibinfo {author} {\bibfnamefont {M.~A.}\ \bibnamefont
  {Rossi}}, \bibinfo {author} {\bibfnamefont {P.}~\bibnamefont {Rydlichowski}},
  \bibinfo {author} {\bibfnamefont {H.}~\bibnamefont {Ryu}}, \bibinfo {author}
  {\bibfnamefont {G.}~\bibnamefont {Samsonidze}}, \bibinfo {author}
  {\bibfnamefont {M.}~\bibnamefont {Sato}}, \bibinfo {author} {\bibfnamefont
  {N.}~\bibnamefont {Saurabh}}, \bibinfo {author} {\bibfnamefont
  {V.}~\bibnamefont {Sharma}}, \bibinfo {author} {\bibfnamefont
  {K.}~\bibnamefont {Sharma}}, \bibinfo {author} {\bibfnamefont
  {S.}~\bibnamefont {Shin}}, \bibinfo {author} {\bibfnamefont {G.}~\bibnamefont
  {Slessman}}, \bibinfo {author} {\bibfnamefont {M.}~\bibnamefont {Steiner}},
  \bibinfo {author} {\bibfnamefont {I.}~\bibnamefont {Sitdikov}}, \bibinfo
  {author} {\bibfnamefont {I.-S.}\ \bibnamefont {Suh}}, \bibinfo {author}
  {\bibfnamefont {E.~D.}\ \bibnamefont {Switzer}}, \bibinfo {author}
  {\bibfnamefont {W.}~\bibnamefont {Tang}}, \bibinfo {author} {\bibfnamefont
  {J.}~\bibnamefont {Thompson}}, \bibinfo {author} {\bibfnamefont
  {S.}~\bibnamefont {Todo}}, \bibinfo {author} {\bibfnamefont {M.~C.}\
  \bibnamefont {Tran}}, \bibinfo {author} {\bibfnamefont {D.}~\bibnamefont
  {Trenev}}, \bibinfo {author} {\bibfnamefont {C.}~\bibnamefont {Trott}},
  \bibinfo {author} {\bibfnamefont {H.-H.}\ \bibnamefont {Tseng}}, \bibinfo
  {author} {\bibfnamefont {N.~M.}\ \bibnamefont {Tubman}}, \bibinfo {author}
  {\bibfnamefont {E.}~\bibnamefont {Tureci}}, \bibinfo {author} {\bibfnamefont
  {D.~G.}\ \bibnamefont {Vali\~nas}}, \bibinfo {author} {\bibfnamefont
  {S.}~\bibnamefont {Vallecorsa}}, \bibinfo {author} {\bibfnamefont
  {C.}~\bibnamefont {Wever}}, \bibinfo {author} {\bibfnamefont
  {K.}~\bibnamefont {Wojciechowski}}, \bibinfo {author} {\bibfnamefont
  {X.}~\bibnamefont {Wu}}, \bibinfo {author} {\bibfnamefont {S.}~\bibnamefont
  {Yoo}}, \bibinfo {author} {\bibfnamefont {N.}~\bibnamefont {Yoshioka}},
  \bibinfo {author} {\bibfnamefont {V.~W.-z.}\ \bibnamefont {Yu}}, \bibinfo
  {author} {\bibfnamefont {S.}~\bibnamefont {Yunoki}}, \bibinfo {author}
  {\bibfnamefont {S.}~\bibnamefont {Zhuk}}, \ and\ \bibinfo {author}
  {\bibfnamefont {D.}~\bibnamefont {Zubarev}},\ }\bibfield  {title} {\enquote
  {\bibinfo {title} {Quantum-centric supercomputing for materials science: {A}
  perspective on challenges and future directions},}\ }\href
  {http://dx.doi.org/10.1016/j.future.2024.04.060} {\bibfield  {journal}
  {\bibinfo  {journal} {\emph {Future Gener. Comput. Syst.}}\ }\textbf
  {\bibinfo {volume} {160}},\ \bibinfo {pages} {666} (\bibinfo {year}
  {2024})}\BibitemShut {NoStop}%
\bibitem [{\citenamefont {Kanno}\ \emph {et~al.}(2026)\citenamefont {Kanno},
  \citenamefont {Kohda}, \citenamefont {Imai}, \citenamefont {Koh},
  \citenamefont {Mitarai}, \citenamefont {Mizukami},\ and\ \citenamefont
  {Nakagawa}}]{Kanno2026}%
  \BibitemOpen
  \bibfield  {author} {\bibinfo {author} {\bibfnamefont {K.}~\bibnamefont
  {Kanno}}, \bibinfo {author} {\bibfnamefont {M.}~\bibnamefont {Kohda}},
  \bibinfo {author} {\bibfnamefont {R.}~\bibnamefont {Imai}}, \bibinfo {author}
  {\bibfnamefont {S.}~\bibnamefont {Koh}}, \bibinfo {author} {\bibfnamefont
  {K.}~\bibnamefont {Mitarai}}, \bibinfo {author} {\bibfnamefont
  {W.}~\bibnamefont {Mizukami}}, \ and\ \bibinfo {author} {\bibfnamefont
  {Y.~O.}\ \bibnamefont {Nakagawa}},\ }\bibfield  {title} {\enquote {\bibinfo
  {title} {Quantum-selected configuration interaction: Classical
  diagonalization of {H}amiltonians in subspaces selected by quantum
  computers},}\ }\href {http://dx.doi.org/10.1103/dmn4-snfx} {\bibfield
  {journal} {\bibinfo  {journal} {\emph {Phys. Rev. Res.}}\ }\textbf {\bibinfo
  {volume} {8}},\ \bibinfo {pages} {023268} (\bibinfo {year}
  {2026})}\BibitemShut {NoStop}%
\bibitem [{\citenamefont {Nakagawa}\ \emph {et~al.}(2024)\citenamefont
  {Nakagawa}, \citenamefont {Kamoshita}, \citenamefont {Mizukami},
  \citenamefont {Sudo},\ and\ \citenamefont {Ohnishi}}]{Nakagawa2024}%
  \BibitemOpen
  \bibfield  {author} {\bibinfo {author} {\bibfnamefont {Y.~O.}\ \bibnamefont
  {Nakagawa}}, \bibinfo {author} {\bibfnamefont {M.}~\bibnamefont {Kamoshita}},
  \bibinfo {author} {\bibfnamefont {W.}~\bibnamefont {Mizukami}}, \bibinfo
  {author} {\bibfnamefont {S.}~\bibnamefont {Sudo}}, \ and\ \bibinfo {author}
  {\bibfnamefont {Y.-y.}\ \bibnamefont {Ohnishi}},\ }\bibfield  {title}
  {\enquote {\bibinfo {title} {{ADAPT-QSCI:} {A}daptive construction of an
  input state for quantum-selected configuration interaction},}\ }\href
  {http://dx.doi.org/10.1021/acs.jctc.4c00846} {\bibfield  {journal} {\bibinfo
  {journal} {\emph {J. Chem. Theory Comput.}}\ }\textbf {\bibinfo {volume}
  {20}},\ \bibinfo {pages} {10817} (\bibinfo {year} {2024})}\BibitemShut
  {NoStop}%
\bibitem [{\citenamefont {Robledo-Moreno}\ \emph {et~al.}(2025)\citenamefont
  {Robledo-Moreno}, \citenamefont {Motta}, \citenamefont {Haas}, \citenamefont
  {Javadi-Abhari}, \citenamefont {Jurcevic}, \citenamefont {Kirby},
  \citenamefont {Martiel}, \citenamefont {Sharma}, \citenamefont {Sharma},
  \citenamefont {Shirakawa}, \citenamefont {Sitdikov}, \citenamefont {Sun},
  \citenamefont {Sung}, \citenamefont {Takita}, \citenamefont {Tran},
  \citenamefont {Yunoki},\ and\ \citenamefont {Mezzacapo}}]{RobledoMoreno2025}%
  \BibitemOpen
  \bibfield  {author} {\bibinfo {author} {\bibfnamefont {J.}~\bibnamefont
  {Robledo-Moreno}}, \bibinfo {author} {\bibfnamefont {M.}~\bibnamefont
  {Motta}}, \bibinfo {author} {\bibfnamefont {H.}~\bibnamefont {Haas}},
  \bibinfo {author} {\bibfnamefont {A.}~\bibnamefont {Javadi-Abhari}}, \bibinfo
  {author} {\bibfnamefont {P.}~\bibnamefont {Jurcevic}}, \bibinfo {author}
  {\bibfnamefont {W.}~\bibnamefont {Kirby}}, \bibinfo {author} {\bibfnamefont
  {S.}~\bibnamefont {Martiel}}, \bibinfo {author} {\bibfnamefont
  {K.}~\bibnamefont {Sharma}}, \bibinfo {author} {\bibfnamefont
  {S.}~\bibnamefont {Sharma}}, \bibinfo {author} {\bibfnamefont
  {T.}~\bibnamefont {Shirakawa}}, \bibinfo {author} {\bibfnamefont
  {I.}~\bibnamefont {Sitdikov}}, \bibinfo {author} {\bibfnamefont {R.-Y.}\
  \bibnamefont {Sun}}, \bibinfo {author} {\bibfnamefont {K.~J.}\ \bibnamefont
  {Sung}}, \bibinfo {author} {\bibfnamefont {M.}~\bibnamefont {Takita}},
  \bibinfo {author} {\bibfnamefont {M.~C.}\ \bibnamefont {Tran}}, \bibinfo
  {author} {\bibfnamefont {S.}~\bibnamefont {Yunoki}}, \ and\ \bibinfo {author}
  {\bibfnamefont {A.}~\bibnamefont {Mezzacapo}},\ }\bibfield  {title} {\enquote
  {\bibinfo {title} {Chemistry beyond the scale of exact diagonalization on a
  quantum-centric supercomputer},}\ }\href
  {http://dx.doi.org/10.1126/sciadv.adu9991} {\bibfield  {journal} {\bibinfo
  {journal} {\emph {Science Advances}}\ }\textbf {\bibinfo {volume} {11}},\
  \bibinfo {pages} {eadu9991} (\bibinfo {year} {2025})}\BibitemShut {NoStop}%
\bibitem [{\citenamefont {Muthukrishnan}\ and\ \citenamefont
  {Stroud}(2000)}]{Muthukrishnan2000}%
  \BibitemOpen
  \bibfield  {author} {\bibinfo {author} {\bibfnamefont {A.}~\bibnamefont
  {Muthukrishnan}}\ and\ \bibinfo {author} {\bibfnamefont {C.~R.}\ \bibnamefont
  {Stroud}},\ }\bibfield  {title} {\enquote {\bibinfo {title} {Multivalued
  logic gates for quantum computation},}\ }\href
  {http://dx.doi.org/10.1103/physreva.62.052309} {\bibfield  {journal}
  {\bibinfo  {journal} {\emph {Phys. Rev. A}}\ }\textbf {\bibinfo {volume}
  {62}},\ \bibinfo {pages} {052309} (\bibinfo {year} {2000})}\BibitemShut
  {NoStop}%
\bibitem [{\citenamefont {Vlasov}(2002)}]{Vlasov2002}%
  \BibitemOpen
  \bibfield  {author} {\bibinfo {author} {\bibfnamefont {A.~Y.}\ \bibnamefont
  {Vlasov}},\ }\bibfield  {title} {\enquote {\bibinfo {title} {Noncommutative
  tori and universal sets of nonbinary quantum gates},}\ }\href
  {http://dx.doi.org/10.1063/1.1476391} {\bibfield  {journal} {\bibinfo
  {journal} {\emph {J. Math. Phys.}}\ }\textbf {\bibinfo {volume} {43}},\
  \bibinfo {pages} {2959} (\bibinfo {year} {2002})}\BibitemShut {NoStop}%
\bibitem [{\citenamefont {Wang}\ \emph {et~al.}(2020)\citenamefont {Wang},
  \citenamefont {Hu}, \citenamefont {Sanders},\ and\ \citenamefont
  {Kais}}]{Wang2020}%
  \BibitemOpen
  \bibfield  {author} {\bibinfo {author} {\bibfnamefont {Y.}~\bibnamefont
  {Wang}}, \bibinfo {author} {\bibfnamefont {Z.}~\bibnamefont {Hu}}, \bibinfo
  {author} {\bibfnamefont {B.~C.}\ \bibnamefont {Sanders}}, \ and\ \bibinfo
  {author} {\bibfnamefont {S.}~\bibnamefont {Kais}},\ }\bibfield  {title}
  {\enquote {\bibinfo {title} {Qudits and high-dimensional quantum
  computing},}\ }\href {http://dx.doi.org/10.3389/fphy.2020.589504} {\bibfield
  {journal} {\bibinfo  {journal} {\emph {Front. Phys.}}\ }\textbf {\bibinfo
  {volume} {8}},\ \bibinfo {pages} {589504} (\bibinfo {year}
  {2020})}\BibitemShut {NoStop}%
\bibitem [{\citenamefont {Bianchetti}\ \emph {et~al.}(2010)\citenamefont
  {Bianchetti}, \citenamefont {Filipp}, \citenamefont {Baur}, \citenamefont
  {Fink}, \citenamefont {Lang}, \citenamefont {Steffen}, \citenamefont
  {Boissonneault}, \citenamefont {Blais},\ and\ \citenamefont
  {Wallraff}}]{Bianchetti2010}%
  \BibitemOpen
  \bibfield  {author} {\bibinfo {author} {\bibfnamefont {R.}~\bibnamefont
  {Bianchetti}}, \bibinfo {author} {\bibfnamefont {S.}~\bibnamefont {Filipp}},
  \bibinfo {author} {\bibfnamefont {M.}~\bibnamefont {Baur}}, \bibinfo {author}
  {\bibfnamefont {J.~M.}\ \bibnamefont {Fink}}, \bibinfo {author}
  {\bibfnamefont {C.}~\bibnamefont {Lang}}, \bibinfo {author} {\bibfnamefont
  {L.}~\bibnamefont {Steffen}}, \bibinfo {author} {\bibfnamefont
  {M.}~\bibnamefont {Boissonneault}}, \bibinfo {author} {\bibfnamefont
  {A.}~\bibnamefont {Blais}}, \ and\ \bibinfo {author} {\bibfnamefont
  {A.}~\bibnamefont {Wallraff}},\ }\bibfield  {title} {\enquote {\bibinfo
  {title} {Control and tomography of a three level superconducting artificial
  atom},}\ }\href {http://dx.doi.org/10.1103/physrevlett.105.223601} {\bibfield
   {journal} {\bibinfo  {journal} {\emph {Phys. Rev. Lett.}}\ }\textbf
  {\bibinfo {volume} {105}},\ \bibinfo {pages} {223601} (\bibinfo {year}
  {2010})}\BibitemShut {NoStop}%
\bibitem [{\citenamefont {Blok}\ \emph {et~al.}(2021)\citenamefont {Blok},
  \citenamefont {Ramasesh}, \citenamefont {Schuster}, \citenamefont {O'Brien},
  \citenamefont {Kreikebaum}, \citenamefont {Dahlen}, \citenamefont {Morvan},
  \citenamefont {Yoshida}, \citenamefont {Yao},\ and\ \citenamefont
  {Siddiqi}}]{Blok2021}%
  \BibitemOpen
  \bibfield  {author} {\bibinfo {author} {\bibfnamefont {M.~S.}\ \bibnamefont
  {Blok}}, \bibinfo {author} {\bibfnamefont {V.~V.}\ \bibnamefont {Ramasesh}},
  \bibinfo {author} {\bibfnamefont {T.}~\bibnamefont {Schuster}}, \bibinfo
  {author} {\bibfnamefont {K.}~\bibnamefont {O'Brien}}, \bibinfo {author}
  {\bibfnamefont {J.~M.}\ \bibnamefont {Kreikebaum}}, \bibinfo {author}
  {\bibfnamefont {D.}~\bibnamefont {Dahlen}}, \bibinfo {author} {\bibfnamefont
  {A.}~\bibnamefont {Morvan}}, \bibinfo {author} {\bibfnamefont
  {B.}~\bibnamefont {Yoshida}}, \bibinfo {author} {\bibfnamefont {N.~Y.}\
  \bibnamefont {Yao}}, \ and\ \bibinfo {author} {\bibfnamefont
  {I.}~\bibnamefont {Siddiqi}},\ }\bibfield  {title} {\enquote {\bibinfo
  {title} {Quantum information scrambling on a superconducting qutrit
  processor},}\ }\href {http://dx.doi.org/10.1103/PhysRevX.11.021010}
  {\bibfield  {journal} {\bibinfo  {journal} {\emph {Phys. Rev. X}}\ }\textbf
  {\bibinfo {volume} {11}},\ \bibinfo {pages} {021010} (\bibinfo {year}
  {2021})}\BibitemShut {NoStop}%
\bibitem [{\citenamefont {Luo}\ \emph {et~al.}(2023)\citenamefont {Luo},
  \citenamefont {Huang}, \citenamefont {Tao}, \citenamefont {Zhang},
  \citenamefont {Zhou}, \citenamefont {Chu}, \citenamefont {Liu}, \citenamefont
  {Wang}, \citenamefont {Cui}, \citenamefont {Liu}, \citenamefont {Yan},
  \citenamefont {Yung}, \citenamefont {Chen}, \citenamefont {Yan},\ and\
  \citenamefont {Yu}}]{Luo2023}%
  \BibitemOpen
  \bibfield  {author} {\bibinfo {author} {\bibfnamefont {K.}~\bibnamefont
  {Luo}}, \bibinfo {author} {\bibfnamefont {W.}~\bibnamefont {Huang}}, \bibinfo
  {author} {\bibfnamefont {Z.}~\bibnamefont {Tao}}, \bibinfo {author}
  {\bibfnamefont {L.}~\bibnamefont {Zhang}}, \bibinfo {author} {\bibfnamefont
  {Y.}~\bibnamefont {Zhou}}, \bibinfo {author} {\bibfnamefont {J.}~\bibnamefont
  {Chu}}, \bibinfo {author} {\bibfnamefont {W.}~\bibnamefont {Liu}}, \bibinfo
  {author} {\bibfnamefont {B.}~\bibnamefont {Wang}}, \bibinfo {author}
  {\bibfnamefont {J.}~\bibnamefont {Cui}}, \bibinfo {author} {\bibfnamefont
  {S.}~\bibnamefont {Liu}}, \bibinfo {author} {\bibfnamefont {F.}~\bibnamefont
  {Yan}}, \bibinfo {author} {\bibfnamefont {M.-H.}\ \bibnamefont {Yung}},
  \bibinfo {author} {\bibfnamefont {Y.}~\bibnamefont {Chen}}, \bibinfo {author}
  {\bibfnamefont {T.}~\bibnamefont {Yan}}, \ and\ \bibinfo {author}
  {\bibfnamefont {D.}~\bibnamefont {Yu}},\ }\bibfield  {title} {\enquote
  {\bibinfo {title} {Experimental realization of two qutrits gate with tunable
  coupling in superconducting circuits},}\ }\href
  {http://dx.doi.org/10.1103/physrevlett.130.030603} {\bibfield  {journal}
  {\bibinfo  {journal} {\emph {Phys. Rev. Lett.}}\ }\textbf {\bibinfo {volume}
  {130}},\ \bibinfo {pages} {030603} (\bibinfo {year} {2023})}\BibitemShut
  {NoStop}%
\bibitem [{\citenamefont {Roy}\ \emph {et~al.}(2023)\citenamefont {Roy},
  \citenamefont {Li}, \citenamefont {Kapit},\ and\ \citenamefont
  {Schuster}}]{Roy2023}%
  \BibitemOpen
  \bibfield  {author} {\bibinfo {author} {\bibfnamefont {T.}~\bibnamefont
  {Roy}}, \bibinfo {author} {\bibfnamefont {Z.}~\bibnamefont {Li}}, \bibinfo
  {author} {\bibfnamefont {E.}~\bibnamefont {Kapit}}, \ and\ \bibinfo {author}
  {\bibfnamefont {D.}~\bibnamefont {Schuster}},\ }\bibfield  {title} {\enquote
  {\bibinfo {title} {Two-qutrit quantum algorithms on a programmable
  superconducting processor},}\ }\href
  {http://dx.doi.org/10.1103/physrevapplied.19.064024} {\bibfield  {journal}
  {\bibinfo  {journal} {\emph {Phys. Rev. Appl.}}\ }\textbf {\bibinfo {volume}
  {19}},\ \bibinfo {pages} {064024} (\bibinfo {year} {2023})}\BibitemShut
  {NoStop}%
\bibitem [{\citenamefont {Goss}\ \emph {et~al.}(2024)\citenamefont {Goss},
  \citenamefont {Ferracin}, \citenamefont {Hashim}, \citenamefont
  {Carignan-Dugas}, \citenamefont {Kreikebaum}, \citenamefont {Naik},
  \citenamefont {Santiago},\ and\ \citenamefont {Siddiqi}}]{Goss2024}%
  \BibitemOpen
  \bibfield  {author} {\bibinfo {author} {\bibfnamefont {N.}~\bibnamefont
  {Goss}}, \bibinfo {author} {\bibfnamefont {S.}~\bibnamefont {Ferracin}},
  \bibinfo {author} {\bibfnamefont {A.}~\bibnamefont {Hashim}}, \bibinfo
  {author} {\bibfnamefont {A.}~\bibnamefont {Carignan-Dugas}}, \bibinfo
  {author} {\bibfnamefont {J.~M.}\ \bibnamefont {Kreikebaum}}, \bibinfo
  {author} {\bibfnamefont {R.~K.}\ \bibnamefont {Naik}}, \bibinfo {author}
  {\bibfnamefont {D.~I.}\ \bibnamefont {Santiago}}, \ and\ \bibinfo {author}
  {\bibfnamefont {I.}~\bibnamefont {Siddiqi}},\ }\bibfield  {title} {\enquote
  {\bibinfo {title} {Extending the computational reach of a superconducting
  qutrit processor},}\ }\href {http://dx.doi.org/10.1038/s41534-024-00892-z}
  {\bibfield  {journal} {\bibinfo  {journal} {\emph {npj Quantum Inf.}}\
  }\textbf {\bibinfo {volume} {10}},\ \bibinfo {pages} {101} (\bibinfo {year}
  {2024})}\BibitemShut {NoStop}%
\bibitem [{\citenamefont {Wang}\ \emph {et~al.}(2025)\citenamefont {Wang},
  \citenamefont {Parker}, \citenamefont {Champion},\ and\ \citenamefont
  {Blok}}]{WangParker2025}%
  \BibitemOpen
  \bibfield  {author} {\bibinfo {author} {\bibfnamefont {Z.}~\bibnamefont
  {Wang}}, \bibinfo {author} {\bibfnamefont {R.~W.}\ \bibnamefont {Parker}},
  \bibinfo {author} {\bibfnamefont {E.}~\bibnamefont {Champion}}, \ and\
  \bibinfo {author} {\bibfnamefont {M.~S.}\ \bibnamefont {Blok}},\ }\bibfield
  {title} {\enquote {\bibinfo {title} {High-{$E_J/E_C$} transmon qudits with up
  to 12 levels},}\ }\href {http://dx.doi.org/10.1103/physrevapplied.23.034046}
  {\bibfield  {journal} {\bibinfo  {journal} {\emph {Phys. Rev. Appl.}}\
  }\textbf {\bibinfo {volume} {23}},\ \bibinfo {pages} {034046} (\bibinfo
  {year} {2025})}\BibitemShut {NoStop}%
\bibitem [{\citenamefont {Randall}\ \emph {et~al.}(2015)\citenamefont
  {Randall}, \citenamefont {Weidt}, \citenamefont {Standing}, \citenamefont
  {Lake}, \citenamefont {Webster}, \citenamefont {Murgia}, \citenamefont
  {Navickas}, \citenamefont {Roth},\ and\ \citenamefont
  {Hensinger}}]{Randall2015}%
  \BibitemOpen
  \bibfield  {author} {\bibinfo {author} {\bibfnamefont {J.}~\bibnamefont
  {Randall}}, \bibinfo {author} {\bibfnamefont {S.}~\bibnamefont {Weidt}},
  \bibinfo {author} {\bibfnamefont {E.~D.}\ \bibnamefont {Standing}}, \bibinfo
  {author} {\bibfnamefont {K.}~\bibnamefont {Lake}}, \bibinfo {author}
  {\bibfnamefont {S.~C.}\ \bibnamefont {Webster}}, \bibinfo {author}
  {\bibfnamefont {D.~F.}\ \bibnamefont {Murgia}}, \bibinfo {author}
  {\bibfnamefont {T.}~\bibnamefont {Navickas}}, \bibinfo {author}
  {\bibfnamefont {K.}~\bibnamefont {Roth}}, \ and\ \bibinfo {author}
  {\bibfnamefont {W.~K.}\ \bibnamefont {Hensinger}},\ }\bibfield  {title}
  {\enquote {\bibinfo {title} {Efficient preparation and detection of microwave
  dressed-state qubits and qutrits with trapped ions},}\ }\href
  {http://dx.doi.org/10.1103/physreva.91.012322} {\bibfield  {journal}
  {\bibinfo  {journal} {\emph {Phys. Rev. A}}\ }\textbf {\bibinfo {volume}
  {91}},\ \bibinfo {pages} {012322} (\bibinfo {year} {2015})}\BibitemShut
  {NoStop}%
\bibitem [{\citenamefont {Senko}\ \emph {et~al.}(2015)\citenamefont {Senko},
  \citenamefont {Richerme}, \citenamefont {Smith}, \citenamefont {Lee},
  \citenamefont {Cohen}, \citenamefont {Retzker},\ and\ \citenamefont
  {Monroe}}]{Senko2015}%
  \BibitemOpen
  \bibfield  {author} {\bibinfo {author} {\bibfnamefont {C.}~\bibnamefont
  {Senko}}, \bibinfo {author} {\bibfnamefont {P.}~\bibnamefont {Richerme}},
  \bibinfo {author} {\bibfnamefont {J.}~\bibnamefont {Smith}}, \bibinfo
  {author} {\bibfnamefont {A.}~\bibnamefont {Lee}}, \bibinfo {author}
  {\bibfnamefont {I.}~\bibnamefont {Cohen}}, \bibinfo {author} {\bibfnamefont
  {A.}~\bibnamefont {Retzker}}, \ and\ \bibinfo {author} {\bibfnamefont
  {C.}~\bibnamefont {Monroe}},\ }\bibfield  {title} {\enquote {\bibinfo {title}
  {Realization of a quantum integer-spin chain with controllable
  interactions},}\ }\href {http://dx.doi.org/10.1103/physrevx.5.021026}
  {\bibfield  {journal} {\bibinfo  {journal} {\emph {Phys. Rev. X}}\ }\textbf
  {\bibinfo {volume} {5}},\ \bibinfo {pages} {021026} (\bibinfo {year}
  {2015})}\BibitemShut {NoStop}%
\bibitem [{\citenamefont {Ringbauer}\ \emph {et~al.}(2022)\citenamefont
  {Ringbauer}, \citenamefont {Meth}, \citenamefont {Postler}, \citenamefont
  {Stricker}, \citenamefont {Blatt}, \citenamefont {Schindler},\ and\
  \citenamefont {Monz}}]{Ringbauer2022}%
  \BibitemOpen
  \bibfield  {author} {\bibinfo {author} {\bibfnamefont {M.}~\bibnamefont
  {Ringbauer}}, \bibinfo {author} {\bibfnamefont {M.}~\bibnamefont {Meth}},
  \bibinfo {author} {\bibfnamefont {L.}~\bibnamefont {Postler}}, \bibinfo
  {author} {\bibfnamefont {R.}~\bibnamefont {Stricker}}, \bibinfo {author}
  {\bibfnamefont {R.}~\bibnamefont {Blatt}}, \bibinfo {author} {\bibfnamefont
  {P.}~\bibnamefont {Schindler}}, \ and\ \bibinfo {author} {\bibfnamefont
  {T.}~\bibnamefont {Monz}},\ }\bibfield  {title} {\enquote {\bibinfo {title}
  {A universal qudit quantum processor with trapped ions},}\ }\href
  {http://dx.doi.org/10.1038/s41567-022-01658-0} {\bibfield  {journal}
  {\bibinfo  {journal} {\emph {Nat. Phys.}}\ }\textbf {\bibinfo {volume}
  {18}},\ \bibinfo {pages} {1053} (\bibinfo {year} {2022})}\BibitemShut
  {NoStop}%
\bibitem [{\citenamefont {Aksenov}\ \emph {et~al.}(2023)\citenamefont
  {Aksenov}, \citenamefont {Zalivako}, \citenamefont {Semerikov}, \citenamefont
  {Borisenko}, \citenamefont {Semenin}, \citenamefont {Sidorov}, \citenamefont
  {Fedorov}, \citenamefont {Khabarova},\ and\ \citenamefont
  {Kolachevsky}}]{Aksenov2023}%
  \BibitemOpen
  \bibfield  {author} {\bibinfo {author} {\bibfnamefont {M.~A.}\ \bibnamefont
  {Aksenov}}, \bibinfo {author} {\bibfnamefont {I.~V.}\ \bibnamefont
  {Zalivako}}, \bibinfo {author} {\bibfnamefont {I.~A.}\ \bibnamefont
  {Semerikov}}, \bibinfo {author} {\bibfnamefont {A.~S.}\ \bibnamefont
  {Borisenko}}, \bibinfo {author} {\bibfnamefont {N.~V.}\ \bibnamefont
  {Semenin}}, \bibinfo {author} {\bibfnamefont {P.~L.}\ \bibnamefont
  {Sidorov}}, \bibinfo {author} {\bibfnamefont {A.~K.}\ \bibnamefont
  {Fedorov}}, \bibinfo {author} {\bibfnamefont {K.~Y.}\ \bibnamefont
  {Khabarova}}, \ and\ \bibinfo {author} {\bibfnamefont {N.~N.}\ \bibnamefont
  {Kolachevsky}},\ }\bibfield  {title} {\enquote {\bibinfo {title} {Realizing
  quantum gates with optically addressable {${}^{171}$Yb$^+$} ion qudits},}\
  }\href {http://dx.doi.org/10.1103/physreva.107.052612} {\bibfield  {journal}
  {\bibinfo  {journal} {\emph {Phys. Rev. A}}\ }\textbf {\bibinfo {volume}
  {107}},\ \bibinfo {pages} {052612} (\bibinfo {year} {2023})}\BibitemShut
  {NoStop}%
\bibitem [{\citenamefont {Hrmo}\ \emph {et~al.}(2023)\citenamefont {Hrmo},
  \citenamefont {Wilhelm}, \citenamefont {Gerster}, \citenamefont {van Mourik},
  \citenamefont {Huber}, \citenamefont {Blatt}, \citenamefont {Schindler},
  \citenamefont {Monz},\ and\ \citenamefont {Ringbauer}}]{Hrmo2023}%
  \BibitemOpen
  \bibfield  {author} {\bibinfo {author} {\bibfnamefont {P.}~\bibnamefont
  {Hrmo}}, \bibinfo {author} {\bibfnamefont {B.}~\bibnamefont {Wilhelm}},
  \bibinfo {author} {\bibfnamefont {L.}~\bibnamefont {Gerster}}, \bibinfo
  {author} {\bibfnamefont {M.~W.}\ \bibnamefont {van Mourik}}, \bibinfo
  {author} {\bibfnamefont {M.}~\bibnamefont {Huber}}, \bibinfo {author}
  {\bibfnamefont {R.}~\bibnamefont {Blatt}}, \bibinfo {author} {\bibfnamefont
  {P.}~\bibnamefont {Schindler}}, \bibinfo {author} {\bibfnamefont
  {T.}~\bibnamefont {Monz}}, \ and\ \bibinfo {author} {\bibfnamefont
  {M.}~\bibnamefont {Ringbauer}},\ }\bibfield  {title} {\enquote {\bibinfo
  {title} {Native qudit entanglement in a trapped ion quantum processor},}\
  }\href {http://dx.doi.org/10.1038/s41467-023-37375-2} {\bibfield  {journal}
  {\bibinfo  {journal} {\emph {Nat. Commun.}}\ }\textbf {\bibinfo {volume}
  {14}},\ \bibinfo {pages} {2242} (\bibinfo {year} {2023})}\BibitemShut
  {NoStop}%
\bibitem [{\citenamefont {Edmunds}\ \emph {et~al.}(2025)\citenamefont
  {Edmunds}, \citenamefont {Rico}, \citenamefont {Arrazola}, \citenamefont
  {Brennen}, \citenamefont {Meth}, \citenamefont {Blatt},\ and\ \citenamefont
  {Ringbauer}}]{Edmunds2025}%
  \BibitemOpen
  \bibfield  {author} {\bibinfo {author} {\bibfnamefont {C.}~\bibnamefont
  {Edmunds}}, \bibinfo {author} {\bibfnamefont {E.}~\bibnamefont {Rico}},
  \bibinfo {author} {\bibfnamefont {I.}~\bibnamefont {Arrazola}}, \bibinfo
  {author} {\bibfnamefont {G.}~\bibnamefont {Brennen}}, \bibinfo {author}
  {\bibfnamefont {M.}~\bibnamefont {Meth}}, \bibinfo {author} {\bibfnamefont
  {R.}~\bibnamefont {Blatt}}, \ and\ \bibinfo {author} {\bibfnamefont
  {M.}~\bibnamefont {Ringbauer}},\ }\bibfield  {title} {\enquote {\bibinfo
  {title} {Symmetry-protected topological {H}aldane phase on a qudit quantum
  processor},}\ }\href {http://dx.doi.org/10.1103/prxquantum.6.020349}
  {\bibfield  {journal} {\bibinfo  {journal} {\emph {PRX Quantum}}\ }\textbf
  {\bibinfo {volume} {6}},\ \bibinfo {pages} {020349} (\bibinfo {year}
  {2025})}\BibitemShut {NoStop}%
\bibitem [{\citenamefont {Nikolaeva}\ \emph
  {et~al.}(2024{\natexlab{a}})\citenamefont {Nikolaeva}, \citenamefont
  {Kiktenko},\ and\ \citenamefont {Fedorov}}]{Nikolaeva2024}%
  \BibitemOpen
  \bibfield  {author} {\bibinfo {author} {\bibfnamefont {A.~S.}\ \bibnamefont
  {Nikolaeva}}, \bibinfo {author} {\bibfnamefont {E.~O.}\ \bibnamefont
  {Kiktenko}}, \ and\ \bibinfo {author} {\bibfnamefont {A.~K.}\ \bibnamefont
  {Fedorov}},\ }\bibfield  {title} {\enquote {\bibinfo {title} {Universal
  quantum computing with qubits embedded in trapped-ion qudits},}\ }\href
  {http://dx.doi.org/10.1103/physreva.109.022615} {\bibfield  {journal}
  {\bibinfo  {journal} {\emph {Phys. Rev. A}}\ }\textbf {\bibinfo {volume}
  {109}},\ \bibinfo {pages} {022615} (\bibinfo {year}
  {2024}{\natexlab{a}})}\BibitemShut {NoStop}%
\bibitem [{\citenamefont {Meth}\ \emph {et~al.}(2025)\citenamefont {Meth},
  \citenamefont {Zhang}, \citenamefont {Haase}, \citenamefont {Edmunds},
  \citenamefont {Postler}, \citenamefont {Jena}, \citenamefont {Steiner},
  \citenamefont {Dellantonio}, \citenamefont {Blatt}, \citenamefont {Zoller},
  \citenamefont {Monz}, \citenamefont {Schindler}, \citenamefont {Muschik},\
  and\ \citenamefont {Ringbauer}}]{Meth2025}%
  \BibitemOpen
  \bibfield  {author} {\bibinfo {author} {\bibfnamefont {M.}~\bibnamefont
  {Meth}}, \bibinfo {author} {\bibfnamefont {J.}~\bibnamefont {Zhang}},
  \bibinfo {author} {\bibfnamefont {J.~F.}\ \bibnamefont {Haase}}, \bibinfo
  {author} {\bibfnamefont {C.}~\bibnamefont {Edmunds}}, \bibinfo {author}
  {\bibfnamefont {L.}~\bibnamefont {Postler}}, \bibinfo {author} {\bibfnamefont
  {A.~J.}\ \bibnamefont {Jena}}, \bibinfo {author} {\bibfnamefont
  {A.}~\bibnamefont {Steiner}}, \bibinfo {author} {\bibfnamefont
  {L.}~\bibnamefont {Dellantonio}}, \bibinfo {author} {\bibfnamefont
  {R.}~\bibnamefont {Blatt}}, \bibinfo {author} {\bibfnamefont
  {P.}~\bibnamefont {Zoller}}, \bibinfo {author} {\bibfnamefont
  {T.}~\bibnamefont {Monz}}, \bibinfo {author} {\bibfnamefont {P.}~\bibnamefont
  {Schindler}}, \bibinfo {author} {\bibfnamefont {C.}~\bibnamefont {Muschik}},
  \ and\ \bibinfo {author} {\bibfnamefont {M.}~\bibnamefont {Ringbauer}},\
  }\bibfield  {title} {\enquote {\bibinfo {title} {Simulating two-dimensional
  lattice gauge theories on a qudit quantum computer},}\ }\href
  {http://dx.doi.org/10.1038/s41567-025-02797-w} {\bibfield  {journal}
  {\bibinfo  {journal} {\emph {Nat. Phys.}}\ }\textbf {\bibinfo {volume}
  {21}},\ \bibinfo {pages} {570} (\bibinfo {year} {2025})}\BibitemShut
  {NoStop}%
\bibitem [{\citenamefont {Low}\ \emph {et~al.}(2025)\citenamefont {Low},
  \citenamefont {White},\ and\ \citenamefont {Senko}}]{Low2025}%
  \BibitemOpen
  \bibfield  {author} {\bibinfo {author} {\bibfnamefont {P.~J.}\ \bibnamefont
  {Low}}, \bibinfo {author} {\bibfnamefont {B.}~\bibnamefont {White}}, \ and\
  \bibinfo {author} {\bibfnamefont {C.}~\bibnamefont {Senko}},\ }\bibfield
  {title} {\enquote {\bibinfo {title} {Control and readout of a 13-level
  trapped ion qudit},}\ }\href {http://dx.doi.org/10.1038/s41534-025-01031-y}
  {\bibfield  {journal} {\bibinfo  {journal} {\emph {npj Quantum Information}}\
  }\textbf {\bibinfo {volume} {11}},\ \bibinfo {pages} {85} (\bibinfo {year}
  {2025})}\BibitemShut {NoStop}%
\bibitem [{\citenamefont {Zalivako}\ \emph {et~al.}(2025)\citenamefont
  {Zalivako}, \citenamefont {Nikolaeva}, \citenamefont {Borisenko},
  \citenamefont {Korolkov}, \citenamefont {Sidorov}, \citenamefont {Galstyan},
  \citenamefont {Semenin}, \citenamefont {Smirnov}, \citenamefont {Aksenov},
  \citenamefont {Makushin}, \citenamefont {Kiktenko}, \citenamefont {Fedorov},
  \citenamefont {Semerikov}, \citenamefont {Khabarova},\ and\ \citenamefont
  {Kolachevsky}}]{Zalivako2025}%
  \BibitemOpen
  \bibfield  {author} {\bibinfo {author} {\bibfnamefont {I.~V.}\ \bibnamefont
  {Zalivako}}, \bibinfo {author} {\bibfnamefont {A.~S.}\ \bibnamefont
  {Nikolaeva}}, \bibinfo {author} {\bibfnamefont {A.~S.}\ \bibnamefont
  {Borisenko}}, \bibinfo {author} {\bibfnamefont {A.~E.}\ \bibnamefont
  {Korolkov}}, \bibinfo {author} {\bibfnamefont {P.~L.}\ \bibnamefont
  {Sidorov}}, \bibinfo {author} {\bibfnamefont {K.~P.}\ \bibnamefont
  {Galstyan}}, \bibinfo {author} {\bibfnamefont {N.~V.}\ \bibnamefont
  {Semenin}}, \bibinfo {author} {\bibfnamefont {V.~N.}\ \bibnamefont
  {Smirnov}}, \bibinfo {author} {\bibfnamefont {M.~A.}\ \bibnamefont
  {Aksenov}}, \bibinfo {author} {\bibfnamefont {K.~M.}\ \bibnamefont
  {Makushin}}, \bibinfo {author} {\bibfnamefont {E.~O.}\ \bibnamefont
  {Kiktenko}}, \bibinfo {author} {\bibfnamefont {A.~K.}\ \bibnamefont
  {Fedorov}}, \bibinfo {author} {\bibfnamefont {I.~A.}\ \bibnamefont
  {Semerikov}}, \bibinfo {author} {\bibfnamefont {K.~Y.}\ \bibnamefont
  {Khabarova}}, \ and\ \bibinfo {author} {\bibfnamefont {N.~N.}\ \bibnamefont
  {Kolachevsky}},\ }\bibfield  {title} {\enquote {\bibinfo {title} {Towards a
  multiqudit quantum processor based on a {${}^{171}$Yb$^+$} ion string:
  Realizing basic quantum algorithms},}\ }\href
  {http://dx.doi.org/10.3390/quantum7020019} {\bibfield  {journal} {\bibinfo
  {journal} {\emph {Quantum Rep.}}\ }\textbf {\bibinfo {volume} {7}},\ \bibinfo
  {pages} {19} (\bibinfo {year} {2025})}\BibitemShut {NoStop}%
\bibitem [{\citenamefont {John}\ \emph {et~al.}(2026)\citenamefont {John},
  \citenamefont {Pareek}, \citenamefont {Tirler}, \citenamefont {Gollerthan},
  \citenamefont {Meth}, \citenamefont {Gerster}, \citenamefont {Zoller},
  \citenamefont {Gonz\'alez-Cuadra}, \citenamefont {Zache},\ and\ \citenamefont
  {Ringbauer}}]{John2026}%
  \BibitemOpen
  \bibfield  {author} {\bibinfo {author} {\bibfnamefont {M.}~\bibnamefont
  {John}}, \bibinfo {author} {\bibfnamefont {K.}~\bibnamefont {Pareek}},
  \bibinfo {author} {\bibfnamefont {P.}~\bibnamefont {Tirler}}, \bibinfo
  {author} {\bibfnamefont {T.}~\bibnamefont {Gollerthan}}, \bibinfo {author}
  {\bibfnamefont {M.}~\bibnamefont {Meth}}, \bibinfo {author} {\bibfnamefont
  {L.}~\bibnamefont {Gerster}}, \bibinfo {author} {\bibfnamefont
  {P.}~\bibnamefont {Zoller}}, \bibinfo {author} {\bibfnamefont
  {D.}~\bibnamefont {Gonz\'alez-Cuadra}}, \bibinfo {author} {\bibfnamefont
  {T.~V.}\ \bibnamefont {Zache}}, \ and\ \bibinfo {author} {\bibfnamefont
  {M.}~\bibnamefont {Ringbauer}},\ }\href
  {http://dx.doi.org/10.48550/ARXIV.2605.05841} {\enquote {\bibinfo {title}
  {Non-abelian string-breaking dynamics on a qudit quantum computer},}\ }
  (\bibinfo {year} {2026}),\ \bibinfo {note} {arXiv:2605.05841
  [quant-ph]}\BibitemShut {NoStop}%
\bibitem [{\citenamefont {Lanyon}\ \emph {et~al.}(2008)\citenamefont {Lanyon},
  \citenamefont {Weinhold}, \citenamefont {Langford}, \citenamefont {O'Brien},
  \citenamefont {Resch}, \citenamefont {Gilchrist},\ and\ \citenamefont
  {White}}]{Lanyon2008}%
  \BibitemOpen
  \bibfield  {author} {\bibinfo {author} {\bibfnamefont {B.~P.}\ \bibnamefont
  {Lanyon}}, \bibinfo {author} {\bibfnamefont {T.~J.}\ \bibnamefont
  {Weinhold}}, \bibinfo {author} {\bibfnamefont {N.~K.}\ \bibnamefont
  {Langford}}, \bibinfo {author} {\bibfnamefont {J.~L.}\ \bibnamefont
  {O'Brien}}, \bibinfo {author} {\bibfnamefont {K.~J.}\ \bibnamefont {Resch}},
  \bibinfo {author} {\bibfnamefont {A.}~\bibnamefont {Gilchrist}}, \ and\
  \bibinfo {author} {\bibfnamefont {A.~G.}\ \bibnamefont {White}},\ }\bibfield
  {title} {\enquote {\bibinfo {title} {Manipulating biphotonic qutrits},}\
  }\href {http://dx.doi.org/10.1103/physrevlett.100.060504} {\bibfield
  {journal} {\bibinfo  {journal} {\emph {Phys. Rev. Lett.}}\ }\textbf {\bibinfo
  {volume} {100}},\ \bibinfo {pages} {060504} (\bibinfo {year}
  {2008})}\BibitemShut {NoStop}%
\bibitem [{\citenamefont {Lanyon}\ \emph {et~al.}(2009)\citenamefont {Lanyon},
  \citenamefont {Barbieri}, \citenamefont {Almeida}, \citenamefont {Jennewein},
  \citenamefont {Ralph}, \citenamefont {Resch}, \citenamefont {Pryde},
  \citenamefont {O'Brien}, \citenamefont {Gilchrist},\ and\ \citenamefont
  {White}}]{Lanyon2009}%
  \BibitemOpen
  \bibfield  {author} {\bibinfo {author} {\bibfnamefont {B.~P.}\ \bibnamefont
  {Lanyon}}, \bibinfo {author} {\bibfnamefont {M.}~\bibnamefont {Barbieri}},
  \bibinfo {author} {\bibfnamefont {M.~P.}\ \bibnamefont {Almeida}}, \bibinfo
  {author} {\bibfnamefont {T.}~\bibnamefont {Jennewein}}, \bibinfo {author}
  {\bibfnamefont {T.~C.}\ \bibnamefont {Ralph}}, \bibinfo {author}
  {\bibfnamefont {K.~J.}\ \bibnamefont {Resch}}, \bibinfo {author}
  {\bibfnamefont {G.~J.}\ \bibnamefont {Pryde}}, \bibinfo {author}
  {\bibfnamefont {J.~L.}\ \bibnamefont {O'Brien}}, \bibinfo {author}
  {\bibfnamefont {A.}~\bibnamefont {Gilchrist}}, \ and\ \bibinfo {author}
  {\bibfnamefont {A.~G.}\ \bibnamefont {White}},\ }\bibfield  {title} {\enquote
  {\bibinfo {title} {Simplifying quantum logic using higher-dimensional
  {H}ilbert spaces},}\ }\href {http://dx.doi.org/10.1038/nphys1150} {\bibfield
  {journal} {\bibinfo  {journal} {\emph {Nat. Phys.}}\ }\textbf {\bibinfo
  {volume} {5}},\ \bibinfo {pages} {134} (\bibinfo {year} {2009})}\BibitemShut
  {NoStop}%
\bibitem [{\citenamefont {Chi}\ \emph {et~al.}(2022)\citenamefont {Chi},
  \citenamefont {Huang}, \citenamefont {Zhang}, \citenamefont {Mao},
  \citenamefont {Zhou}, \citenamefont {Chen}, \citenamefont {Zhai},
  \citenamefont {Bao}, \citenamefont {Dai}, \citenamefont {Yuan}, \citenamefont
  {Zhang}, \citenamefont {Dai}, \citenamefont {Tang}, \citenamefont {Yang},
  \citenamefont {Li}, \citenamefont {Ding}, \citenamefont {Oxenl{\o}we},
  \citenamefont {Thompson}, \citenamefont {O'Brien}, \citenamefont {Li},
  \citenamefont {Gong},\ and\ \citenamefont {Wang}}]{Chi2022}%
  \BibitemOpen
  \bibfield  {author} {\bibinfo {author} {\bibfnamefont {Y.}~\bibnamefont
  {Chi}}, \bibinfo {author} {\bibfnamefont {J.}~\bibnamefont {Huang}}, \bibinfo
  {author} {\bibfnamefont {Z.}~\bibnamefont {Zhang}}, \bibinfo {author}
  {\bibfnamefont {J.}~\bibnamefont {Mao}}, \bibinfo {author} {\bibfnamefont
  {Z.}~\bibnamefont {Zhou}}, \bibinfo {author} {\bibfnamefont {X.}~\bibnamefont
  {Chen}}, \bibinfo {author} {\bibfnamefont {C.}~\bibnamefont {Zhai}}, \bibinfo
  {author} {\bibfnamefont {J.}~\bibnamefont {Bao}}, \bibinfo {author}
  {\bibfnamefont {T.}~\bibnamefont {Dai}}, \bibinfo {author} {\bibfnamefont
  {H.}~\bibnamefont {Yuan}}, \bibinfo {author} {\bibfnamefont {M.}~\bibnamefont
  {Zhang}}, \bibinfo {author} {\bibfnamefont {D.}~\bibnamefont {Dai}}, \bibinfo
  {author} {\bibfnamefont {B.}~\bibnamefont {Tang}}, \bibinfo {author}
  {\bibfnamefont {Y.}~\bibnamefont {Yang}}, \bibinfo {author} {\bibfnamefont
  {Z.}~\bibnamefont {Li}}, \bibinfo {author} {\bibfnamefont {Y.}~\bibnamefont
  {Ding}}, \bibinfo {author} {\bibfnamefont {L.~K.}\ \bibnamefont
  {Oxenl{\o}we}}, \bibinfo {author} {\bibfnamefont {M.~G.}\ \bibnamefont
  {Thompson}}, \bibinfo {author} {\bibfnamefont {J.~L.}\ \bibnamefont
  {O'Brien}}, \bibinfo {author} {\bibfnamefont {Y.}~\bibnamefont {Li}},
  \bibinfo {author} {\bibfnamefont {Q.}~\bibnamefont {Gong}}, \ and\ \bibinfo
  {author} {\bibfnamefont {J.}~\bibnamefont {Wang}},\ }\bibfield  {title}
  {\enquote {\bibinfo {title} {A programmable qudit-based quantum processor},}\
  }\href {http://dx.doi.org/10.1038/s41467-022-28767-x} {\bibfield  {journal}
  {\bibinfo  {journal} {\emph {Nat. Commun.}}\ }\textbf {\bibinfo {volume}
  {13}},\ \bibinfo {pages} {1166} (\bibinfo {year} {2022})}\BibitemShut
  {NoStop}%
\bibitem [{\citenamefont {Fern\'{a}ndez~de Fuentes}\ \emph
  {et~al.}(2024)\citenamefont {Fern\'{a}ndez~de Fuentes}, \citenamefont
  {Botzem}, \citenamefont {Johnson}, \citenamefont {Vaartjes}, \citenamefont
  {Asaad}, \citenamefont {Mourik}, \citenamefont {Hudson}, \citenamefont
  {Itoh}, \citenamefont {Johnson}, \citenamefont {Jakob}, \citenamefont
  {McCallum}, \citenamefont {Jamieson}, \citenamefont {Dzurak},\ and\
  \citenamefont {Morello}}]{FernandezdeFuentes2024}%
  \BibitemOpen
  \bibfield  {author} {\bibinfo {author} {\bibfnamefont {I.}~\bibnamefont
  {Fern\'{a}ndez~de Fuentes}}, \bibinfo {author} {\bibfnamefont
  {T.}~\bibnamefont {Botzem}}, \bibinfo {author} {\bibfnamefont {M.~A.~I.}\
  \bibnamefont {Johnson}}, \bibinfo {author} {\bibfnamefont {A.}~\bibnamefont
  {Vaartjes}}, \bibinfo {author} {\bibfnamefont {S.}~\bibnamefont {Asaad}},
  \bibinfo {author} {\bibfnamefont {V.}~\bibnamefont {Mourik}}, \bibinfo
  {author} {\bibfnamefont {F.~E.}\ \bibnamefont {Hudson}}, \bibinfo {author}
  {\bibfnamefont {K.~M.}\ \bibnamefont {Itoh}}, \bibinfo {author}
  {\bibfnamefont {B.~C.}\ \bibnamefont {Johnson}}, \bibinfo {author}
  {\bibfnamefont {A.~M.}\ \bibnamefont {Jakob}}, \bibinfo {author}
  {\bibfnamefont {J.~C.}\ \bibnamefont {McCallum}}, \bibinfo {author}
  {\bibfnamefont {D.~N.}\ \bibnamefont {Jamieson}}, \bibinfo {author}
  {\bibfnamefont {A.~S.}\ \bibnamefont {Dzurak}}, \ and\ \bibinfo {author}
  {\bibfnamefont {A.}~\bibnamefont {Morello}},\ }\bibfield  {title} {\enquote
  {\bibinfo {title} {Navigating the 16-dimensional {H}ilbert space of a
  high-spin donor qudit with electric and magnetic fields},}\ }\href
  {http://dx.doi.org/10.1038/s41467-024-45368-y} {\bibfield  {journal}
  {\bibinfo  {journal} {\emph {Nat. Commun.}}\ }\textbf {\bibinfo {volume}
  {15}},\ \bibinfo {pages} {1380} (\bibinfo {year} {2024})}\BibitemShut
  {NoStop}%
\bibitem [{\citenamefont {Chaudhury}\ \emph {et~al.}(2007)\citenamefont
  {Chaudhury}, \citenamefont {Merkel}, \citenamefont {Herr}, \citenamefont
  {Silberfarb}, \citenamefont {Deutsch},\ and\ \citenamefont
  {Jessen}}]{Chaudhury2007}%
  \BibitemOpen
  \bibfield  {author} {\bibinfo {author} {\bibfnamefont {S.}~\bibnamefont
  {Chaudhury}}, \bibinfo {author} {\bibfnamefont {S.}~\bibnamefont {Merkel}},
  \bibinfo {author} {\bibfnamefont {T.}~\bibnamefont {Herr}}, \bibinfo {author}
  {\bibfnamefont {A.}~\bibnamefont {Silberfarb}}, \bibinfo {author}
  {\bibfnamefont {I.~H.}\ \bibnamefont {Deutsch}}, \ and\ \bibinfo {author}
  {\bibfnamefont {P.~S.}\ \bibnamefont {Jessen}},\ }\bibfield  {title}
  {\enquote {\bibinfo {title} {Quantum control of the hyperfine spin of a {Cs}
  atom ensemble},}\ }\href {http://dx.doi.org/10.1103/physrevlett.99.163002}
  {\bibfield  {journal} {\bibinfo  {journal} {\emph {Phys. Rev. Lett.}}\
  }\textbf {\bibinfo {volume} {99}},\ \bibinfo {pages} {163002} (\bibinfo
  {year} {2007})}\BibitemShut {NoStop}%
\bibitem [{\citenamefont {Omanakuttan}\ \emph {et~al.}(2021)\citenamefont
  {Omanakuttan}, \citenamefont {Mitra}, \citenamefont {Martin},\ and\
  \citenamefont {Deutsch}}]{Omanakuttan2021}%
  \BibitemOpen
  \bibfield  {author} {\bibinfo {author} {\bibfnamefont {S.}~\bibnamefont
  {Omanakuttan}}, \bibinfo {author} {\bibfnamefont {A.}~\bibnamefont {Mitra}},
  \bibinfo {author} {\bibfnamefont {M.~J.}\ \bibnamefont {Martin}}, \ and\
  \bibinfo {author} {\bibfnamefont {I.~H.}\ \bibnamefont {Deutsch}},\
  }\bibfield  {title} {\enquote {\bibinfo {title} {Quantum optimal control of
  ten-level nuclear spin qudits in {${}^{87}$Sr}},}\ }\href
  {http://dx.doi.org/10.1103/physreva.104.l060401} {\bibfield  {journal}
  {\bibinfo  {journal} {\emph {Phys. Rev. A}}\ }\textbf {\bibinfo {volume}
  {104}},\ \bibinfo {pages} {l060401} (\bibinfo {year} {2021})}\BibitemShut
  {NoStop}%
\bibitem [{\citenamefont {Lindon}\ \emph {et~al.}(2023)\citenamefont {Lindon},
  \citenamefont {Tashchilina}, \citenamefont {Cooke},\ and\ \citenamefont
  {LeBlanc}}]{Lindon2023}%
  \BibitemOpen
  \bibfield  {author} {\bibinfo {author} {\bibfnamefont {J.}~\bibnamefont
  {Lindon}}, \bibinfo {author} {\bibfnamefont {A.}~\bibnamefont {Tashchilina}},
  \bibinfo {author} {\bibfnamefont {L.~W.}\ \bibnamefont {Cooke}}, \ and\
  \bibinfo {author} {\bibfnamefont {L.~J.}\ \bibnamefont {LeBlanc}},\
  }\bibfield  {title} {\enquote {\bibinfo {title} {Complete unitary qutrit
  control in ultracold atoms},}\ }\href
  {http://dx.doi.org/10.1103/physrevapplied.19.034089} {\bibfield  {journal}
  {\bibinfo  {journal} {\emph {Phys. Rev. Appl.}}\ }\textbf {\bibinfo {volume}
  {19}},\ \bibinfo {pages} {034089} (\bibinfo {year} {2023})}\BibitemShut
  {NoStop}%
\bibitem [{\citenamefont {Soltamov}\ \emph {et~al.}(2019)\citenamefont
  {Soltamov}, \citenamefont {Kasper}, \citenamefont {Poshakinskiy},
  \citenamefont {Anisimov}, \citenamefont {Mokhov}, \citenamefont {Sperlich},
  \citenamefont {Tarasenko}, \citenamefont {Baranov}, \citenamefont
  {Astakhov},\ and\ \citenamefont {Dyakonov}}]{Soltamov2019}%
  \BibitemOpen
  \bibfield  {author} {\bibinfo {author} {\bibfnamefont {V.~A.}\ \bibnamefont
  {Soltamov}}, \bibinfo {author} {\bibfnamefont {C.}~\bibnamefont {Kasper}},
  \bibinfo {author} {\bibfnamefont {A.~V.}\ \bibnamefont {Poshakinskiy}},
  \bibinfo {author} {\bibfnamefont {A.~N.}\ \bibnamefont {Anisimov}}, \bibinfo
  {author} {\bibfnamefont {E.~N.}\ \bibnamefont {Mokhov}}, \bibinfo {author}
  {\bibfnamefont {A.}~\bibnamefont {Sperlich}}, \bibinfo {author}
  {\bibfnamefont {S.~A.}\ \bibnamefont {Tarasenko}}, \bibinfo {author}
  {\bibfnamefont {P.~G.}\ \bibnamefont {Baranov}}, \bibinfo {author}
  {\bibfnamefont {G.~V.}\ \bibnamefont {Astakhov}}, \ and\ \bibinfo {author}
  {\bibfnamefont {V.}~\bibnamefont {Dyakonov}},\ }\bibfield  {title} {\enquote
  {\bibinfo {title} {Excitation and coherent control of spin qudit modes in
  silicon carbide at room temperature},}\ }\href
  {http://dx.doi.org/10.1038/s41467-019-09429-x} {\bibfield  {journal}
  {\bibinfo  {journal} {\emph {Nat. Commun.}}\ }\textbf {\bibinfo {volume}
  {10}},\ \bibinfo {pages} {1678} (\bibinfo {year} {2019})}\BibitemShut
  {NoStop}%
\bibitem [{\citenamefont {Adambukulam}\ \emph {et~al.}(2024)\citenamefont
  {Adambukulam}, \citenamefont {Johnson}, \citenamefont {Morello},\ and\
  \citenamefont {Laucht}}]{Adambukulam2024}%
  \BibitemOpen
  \bibfield  {author} {\bibinfo {author} {\bibfnamefont {C.}~\bibnamefont
  {Adambukulam}}, \bibinfo {author} {\bibfnamefont {B.~C.}\ \bibnamefont
  {Johnson}}, \bibinfo {author} {\bibfnamefont {A.}~\bibnamefont {Morello}}, \
  and\ \bibinfo {author} {\bibfnamefont {A.}~\bibnamefont {Laucht}},\
  }\bibfield  {title} {\enquote {\bibinfo {title} {Hyperfine spectroscopy and
  fast, all-optical arbitrary state initialization and readout of a single,
  ten-level {${}^{73}$Ge} vacancy nuclear spin qudit in diamond},}\ }\href
  {http://dx.doi.org/10.1103/physrevlett.132.060603} {\bibfield  {journal}
  {\bibinfo  {journal} {\emph {Phys. Rev. Lett.}}\ }\textbf {\bibinfo {volume}
  {132}},\ \bibinfo {pages} {060603} (\bibinfo {year} {2024})}\BibitemShut
  {NoStop}%
\bibitem [{\citenamefont {Chizzini}\ \emph {et~al.}(2024)\citenamefont
  {Chizzini}, \citenamefont {Tacchino}, \citenamefont {Chiesa}, \citenamefont
  {Tavernelli}, \citenamefont {Carretta},\ and\ \citenamefont
  {Santini}}]{Chizzini2024}%
  \BibitemOpen
  \bibfield  {author} {\bibinfo {author} {\bibfnamefont {M.}~\bibnamefont
  {Chizzini}}, \bibinfo {author} {\bibfnamefont {F.}~\bibnamefont {Tacchino}},
  \bibinfo {author} {\bibfnamefont {A.}~\bibnamefont {Chiesa}}, \bibinfo
  {author} {\bibfnamefont {I.}~\bibnamefont {Tavernelli}}, \bibinfo {author}
  {\bibfnamefont {S.}~\bibnamefont {Carretta}}, \ and\ \bibinfo {author}
  {\bibfnamefont {P.}~\bibnamefont {Santini}},\ }\bibfield  {title} {\enquote
  {\bibinfo {title} {Qudit-based quantum simulation of fermionic systems},}\
  }\href {http://dx.doi.org/10.1103/physreva.110.062602} {\bibfield  {journal}
  {\bibinfo  {journal} {\emph {Phys. Rev. A}}\ }\textbf {\bibinfo {volume}
  {110}},\ \bibinfo {pages} {062602} (\bibinfo {year} {2024})}\BibitemShut
  {NoStop}%
\bibitem [{\citenamefont {MacDonell}\ \emph {et~al.}(2021)\citenamefont
  {MacDonell}, \citenamefont {Dickerson}, \citenamefont {Birch}, \citenamefont
  {Kumar}, \citenamefont {Edmunds}, \citenamefont {Biercuk}, \citenamefont
  {Hempel},\ and\ \citenamefont {Kassal}}]{MacDonell2021}%
  \BibitemOpen
  \bibfield  {author} {\bibinfo {author} {\bibfnamefont {R.~J.}\ \bibnamefont
  {MacDonell}}, \bibinfo {author} {\bibfnamefont {C.~E.}\ \bibnamefont
  {Dickerson}}, \bibinfo {author} {\bibfnamefont {C.~J.~T.}\ \bibnamefont
  {Birch}}, \bibinfo {author} {\bibfnamefont {A.}~\bibnamefont {Kumar}},
  \bibinfo {author} {\bibfnamefont {C.~L.}\ \bibnamefont {Edmunds}}, \bibinfo
  {author} {\bibfnamefont {M.~J.}\ \bibnamefont {Biercuk}}, \bibinfo {author}
  {\bibfnamefont {C.}~\bibnamefont {Hempel}}, \ and\ \bibinfo {author}
  {\bibfnamefont {I.}~\bibnamefont {Kassal}},\ }\bibfield  {title} {\enquote
  {\bibinfo {title} {Analog quantum simulation of chemical dynamics},}\ }\href
  {http://dx.doi.org/10.1039/d1sc02142g} {\bibfield  {journal} {\bibinfo
  {journal} {\emph {Chem. Sci.}}\ }\textbf {\bibinfo {volume} {12}},\ \bibinfo
  {pages} {9794} (\bibinfo {year} {2021})}\BibitemShut {NoStop}%
\bibitem [{\citenamefont {Kiktenko}\ \emph {et~al.}(2025)\citenamefont
  {Kiktenko}, \citenamefont {Nikolaeva},\ and\ \citenamefont
  {Fedorov}}]{Kiktenko2025}%
  \BibitemOpen
  \bibfield  {author} {\bibinfo {author} {\bibfnamefont {E.~O.}\ \bibnamefont
  {Kiktenko}}, \bibinfo {author} {\bibfnamefont {A.~S.}\ \bibnamefont
  {Nikolaeva}}, \ and\ \bibinfo {author} {\bibfnamefont {A.~K.}\ \bibnamefont
  {Fedorov}},\ }\bibfield  {title} {\enquote {\bibinfo {title} {Colloquium:
  {Q}udits for decomposing multiqubit gates and realizing quantum
  algorithms},}\ }\href {http://dx.doi.org/10.1103/revmodphys.97.021003}
  {\bibfield  {journal} {\bibinfo  {journal} {\emph {Rev. Mod. Phys.}}\
  }\textbf {\bibinfo {volume} {97}},\ \bibinfo {pages} {021003} (\bibinfo
  {year} {2025})}\BibitemShut {NoStop}%
\bibitem [{\citenamefont {Nikolaeva}\ \emph {et~al.}(2023)\citenamefont
  {Nikolaeva}, \citenamefont {Kiktenko},\ and\ \citenamefont
  {Fedorov}}]{Nikolaeva2023}%
  \BibitemOpen
  \bibfield  {author} {\bibinfo {author} {\bibfnamefont {A.~S.}\ \bibnamefont
  {Nikolaeva}}, \bibinfo {author} {\bibfnamefont {E.~O.}\ \bibnamefont
  {Kiktenko}}, \ and\ \bibinfo {author} {\bibfnamefont {A.~K.}\ \bibnamefont
  {Fedorov}},\ }\bibfield  {title} {\enquote {\bibinfo {title} {Generalized
  {T}offoli gate decomposition using ququints: Towards realizing {G}rover's
  algorithm with qudits},}\ }\href {http://dx.doi.org/10.3390/e25020387}
  {\bibfield  {journal} {\bibinfo  {journal} {\emph {Entropy}}\ }\textbf
  {\bibinfo {volume} {25}},\ \bibinfo {pages} {387} (\bibinfo {year}
  {2023})}\BibitemShut {NoStop}%
\bibitem [{\citenamefont {Nikolaeva}\ \emph
  {et~al.}(2024{\natexlab{b}})\citenamefont {Nikolaeva}, \citenamefont
  {Kiktenko},\ and\ \citenamefont {Fedorov}}]{NikolaevaKiktenkoFedorov2024}%
  \BibitemOpen
  \bibfield  {author} {\bibinfo {author} {\bibfnamefont {A.~S.}\ \bibnamefont
  {Nikolaeva}}, \bibinfo {author} {\bibfnamefont {E.~O.}\ \bibnamefont
  {Kiktenko}}, \ and\ \bibinfo {author} {\bibfnamefont {A.~K.}\ \bibnamefont
  {Fedorov}},\ }\bibfield  {title} {\enquote {\bibinfo {title} {Efficient
  realization of quantum algorithms with qudits},}\ }\href
  {http://dx.doi.org/10.1140/epjqt/s40507-024-00250-0} {\bibfield  {journal}
  {\bibinfo  {journal} {\emph {EPJ Quantum Technol.}}\ }\textbf {\bibinfo
  {volume} {11}},\ \bibinfo {pages} {43} (\bibinfo {year}
  {2024}{\natexlab{b}})}\BibitemShut {NoStop}%
\bibitem [{\citenamefont {Gottesman}(1999)}]{Gottesman1999_qudit}%
  \BibitemOpen
  \bibfield  {author} {\bibinfo {author} {\bibfnamefont {D.}~\bibnamefont
  {Gottesman}},\ }\bibfield  {title} {\enquote {\bibinfo {title}
  {Fault-tolerant quantum computation with higher-dimensional systems},}\
  }\href {http://dx.doi.org/10.1016/s0960-0779(98)00218-5} {\bibfield
  {journal} {\bibinfo  {journal} {\emph {Chaos Solit. Fractals}}\ }\textbf
  {\bibinfo {volume} {10}},\ \bibinfo {pages} {1749} (\bibinfo {year}
  {1999})}\BibitemShut {NoStop}%
\bibitem [{\citenamefont {Campbell}(2014)}]{Campbell2014}%
  \BibitemOpen
  \bibfield  {author} {\bibinfo {author} {\bibfnamefont {E.~T.}\ \bibnamefont
  {Campbell}},\ }\bibfield  {title} {\enquote {\bibinfo {title} {Enhanced
  fault-tolerant quantum computing in {$d$}-level systems},}\ }\href
  {http://dx.doi.org/10.1103/physrevlett.113.230501} {\bibfield  {journal}
  {\bibinfo  {journal} {\emph {Phys. Rev. Lett.}}\ }\textbf {\bibinfo {volume}
  {113}},\ \bibinfo {pages} {230501} (\bibinfo {year} {2014})}\BibitemShut
  {NoStop}%
\bibitem [{\citenamefont {Keppens}\ \emph {et~al.}(2025)\citenamefont
  {Keppens}, \citenamefont {Eggerickx}, \citenamefont {Levajac}, \citenamefont
  {Simion},\ and\ \citenamefont {Sor\'{e}e}}]{Keppens2025}%
  \BibitemOpen
  \bibfield  {author} {\bibinfo {author} {\bibfnamefont {J.}~\bibnamefont
  {Keppens}}, \bibinfo {author} {\bibfnamefont {Q.}~\bibnamefont {Eggerickx}},
  \bibinfo {author} {\bibfnamefont {V.}~\bibnamefont {Levajac}}, \bibinfo
  {author} {\bibfnamefont {G.}~\bibnamefont {Simion}}, \ and\ \bibinfo {author}
  {\bibfnamefont {B.}~\bibnamefont {Sor\'{e}e}},\ }\bibfield  {title} {\enquote
  {\bibinfo {title} {Qudit vs. qubit: Simulated performance of error-correction
  codes in higher dimensions},}\ }\href {http://dx.doi.org/10.1103/2w52-qd2j}
  {\bibfield  {journal} {\bibinfo  {journal} {\emph {Phys. Rev. A}}\ }\textbf
  {\bibinfo {volume} {112}},\ \bibinfo {pages} {032435} (\bibinfo {year}
  {2025})}\BibitemShut {NoStop}%
\bibitem [{\citenamefont {Gottesman}\ \emph {et~al.}(2001)\citenamefont
  {Gottesman}, \citenamefont {Kitaev},\ and\ \citenamefont
  {Preskill}}]{Gottesman2001}%
  \BibitemOpen
  \bibfield  {author} {\bibinfo {author} {\bibfnamefont {D.}~\bibnamefont
  {Gottesman}}, \bibinfo {author} {\bibfnamefont {A.}~\bibnamefont {Kitaev}}, \
  and\ \bibinfo {author} {\bibfnamefont {J.}~\bibnamefont {Preskill}},\
  }\bibfield  {title} {\enquote {\bibinfo {title} {Encoding a qubit in an
  oscillator},}\ }\href {http://dx.doi.org/10.1103/physreva.64.012310}
  {\bibfield  {journal} {\bibinfo  {journal} {\emph {Phys. Rev. A}}\ }\textbf
  {\bibinfo {volume} {64}},\ \bibinfo {pages} {012310} (\bibinfo {year}
  {2001})}\BibitemShut {NoStop}%
\bibitem [{\citenamefont {Brock}\ \emph {et~al.}(2025)\citenamefont {Brock},
  \citenamefont {Singh}, \citenamefont {Eickbusch}, \citenamefont {Sivak},
  \citenamefont {Ding}, \citenamefont {Frunzio}, \citenamefont {Girvin},\ and\
  \citenamefont {Devoret}}]{Brock2025}%
  \BibitemOpen
  \bibfield  {author} {\bibinfo {author} {\bibfnamefont {B.~L.}\ \bibnamefont
  {Brock}}, \bibinfo {author} {\bibfnamefont {S.}~\bibnamefont {Singh}},
  \bibinfo {author} {\bibfnamefont {A.}~\bibnamefont {Eickbusch}}, \bibinfo
  {author} {\bibfnamefont {V.~V.}\ \bibnamefont {Sivak}}, \bibinfo {author}
  {\bibfnamefont {A.~Z.}\ \bibnamefont {Ding}}, \bibinfo {author}
  {\bibfnamefont {L.}~\bibnamefont {Frunzio}}, \bibinfo {author} {\bibfnamefont
  {S.~M.}\ \bibnamefont {Girvin}}, \ and\ \bibinfo {author} {\bibfnamefont
  {M.~H.}\ \bibnamefont {Devoret}},\ }\bibfield  {title} {\enquote {\bibinfo
  {title} {Quantum error correction of qudits beyond break-even},}\ }\href
  {http://dx.doi.org/10.1038/s41586-025-08899-y} {\bibfield  {journal}
  {\bibinfo  {journal} {\emph {Nature}}\ }\textbf {\bibinfo {volume} {641}},\
  \bibinfo {pages} {612} (\bibinfo {year} {2025})}\BibitemShut {NoStop}%
\bibitem [{\citenamefont {Teplukhin}\ \emph {et~al.}(2019)\citenamefont
  {Teplukhin}, \citenamefont {Kendrick},\ and\ \citenamefont
  {Babikov}}]{Teplukhin2019}%
  \BibitemOpen
  \bibfield  {author} {\bibinfo {author} {\bibfnamefont {A.}~\bibnamefont
  {Teplukhin}}, \bibinfo {author} {\bibfnamefont {B.~K.}\ \bibnamefont
  {Kendrick}}, \ and\ \bibinfo {author} {\bibfnamefont {D.}~\bibnamefont
  {Babikov}},\ }\bibfield  {title} {\enquote {\bibinfo {title} {Calculation of
  molecular vibrational spectra on a quantum annealer},}\ }\href
  {http://dx.doi.org/10.1021/acs.jctc.9b00402} {\bibfield  {journal} {\bibinfo
  {journal} {\emph {J. Chem. Theory Comput.}}\ }\textbf {\bibinfo {volume}
  {15}},\ \bibinfo {pages} {4555} (\bibinfo {year} {2019})}\BibitemShut
  {NoStop}%
\bibitem [{\citenamefont {McArdle}\ \emph {et~al.}(2019)\citenamefont
  {McArdle}, \citenamefont {Mayorov}, \citenamefont {Shan}, \citenamefont
  {Benjamin},\ and\ \citenamefont {Yuan}}]{McArdleetal2019}%
  \BibitemOpen
  \bibfield  {author} {\bibinfo {author} {\bibfnamefont {S.}~\bibnamefont
  {McArdle}}, \bibinfo {author} {\bibfnamefont {A.}~\bibnamefont {Mayorov}},
  \bibinfo {author} {\bibfnamefont {X.}~\bibnamefont {Shan}}, \bibinfo {author}
  {\bibfnamefont {S.}~\bibnamefont {Benjamin}}, \ and\ \bibinfo {author}
  {\bibfnamefont {X.}~\bibnamefont {Yuan}},\ }\bibfield  {title} {\enquote
  {\bibinfo {title} {Digital quantum simulation of molecular vibrations},}\
  }\href {http://dx.doi.org/10.1039/c9sc01313j} {\bibfield  {journal} {\bibinfo
   {journal} {\emph {Chem. Sci.}}\ }\textbf {\bibinfo {volume} {10}},\ \bibinfo
  {pages} {5725} (\bibinfo {year} {2019})}\BibitemShut {NoStop}%
\bibitem [{\citenamefont {Sawaya}\ and\ \citenamefont
  {Huh}(2019)}]{SawayaHuh_2019}%
  \BibitemOpen
  \bibfield  {author} {\bibinfo {author} {\bibfnamefont {N.~P.~D.}\
  \bibnamefont {Sawaya}}\ and\ \bibinfo {author} {\bibfnamefont
  {J.}~\bibnamefont {Huh}},\ }\bibfield  {title} {\enquote {\bibinfo {title}
  {Quantum algorithm for calculating molecular vibronic spectra},}\ }\href
  {http://dx.doi.org/10.1021/acs.jpclett.9b01117} {\bibfield  {journal}
  {\bibinfo  {journal} {\emph {J. Phys. Chem. Lett.}}\ }\textbf {\bibinfo
  {volume} {10}},\ \bibinfo {pages} {3586} (\bibinfo {year}
  {2019})}\BibitemShut {NoStop}%
\bibitem [{\citenamefont {L\"otstedt}\ \emph {et~al.}(2021)\citenamefont
  {L\"otstedt}, \citenamefont {Yamanouchi}, \citenamefont {Tsuchiya},\ and\
  \citenamefont {Tachikawa}}]{Loetstedt2021}%
  \BibitemOpen
  \bibfield  {author} {\bibinfo {author} {\bibfnamefont {E.}~\bibnamefont
  {L\"otstedt}}, \bibinfo {author} {\bibfnamefont {K.}~\bibnamefont
  {Yamanouchi}}, \bibinfo {author} {\bibfnamefont {T.}~\bibnamefont
  {Tsuchiya}}, \ and\ \bibinfo {author} {\bibfnamefont {Y.}~\bibnamefont
  {Tachikawa}},\ }\bibfield  {title} {\enquote {\bibinfo {title} {Calculation
  of vibrational eigenenergies on a quantum computer: Application to the
  {F}ermi resonance in {CO$_2$}},}\ }\href
  {http://dx.doi.org/10.1103/PhysRevA.103.062609} {\bibfield  {journal}
  {\bibinfo  {journal} {\emph {Phys. Rev. A}}\ }\textbf {\bibinfo {volume}
  {103}},\ \bibinfo {pages} {062609} (\bibinfo {year} {2021})}\BibitemShut
  {NoStop}%
\bibitem [{\citenamefont {L\"otstedt}\ \emph {et~al.}(2022)\citenamefont
  {L\"otstedt}, \citenamefont {Yamanouchi},\ and\ \citenamefont
  {Tachikawa}}]{Loetstedt2022_copy}%
  \BibitemOpen
  \bibfield  {author} {\bibinfo {author} {\bibfnamefont {E.}~\bibnamefont
  {L\"otstedt}}, \bibinfo {author} {\bibfnamefont {K.}~\bibnamefont
  {Yamanouchi}}, \ and\ \bibinfo {author} {\bibfnamefont {Y.}~\bibnamefont
  {Tachikawa}},\ }\bibfield  {title} {\enquote {\bibinfo {title} {Evaluation of
  vibrational energies and wave functions of {CO}$_2$ on a quantum computer},}\
  }\href {http://dx.doi.org/10.1116/5.0091144} {\bibfield  {journal} {\bibinfo
  {journal} {\emph {{AVS} Quantum Science}}\ }\textbf {\bibinfo {volume} {4}},\
  \bibinfo {pages} {036801} (\bibinfo {year} {2022})}\BibitemShut {NoStop}%
\bibitem [{\citenamefont {Sawaya}\ \emph {et~al.}(2021)\citenamefont {Sawaya},
  \citenamefont {Paesani},\ and\ \citenamefont {Tabor}}]{Sawaya2021}%
  \BibitemOpen
  \bibfield  {author} {\bibinfo {author} {\bibfnamefont {N.~P.~D.}\
  \bibnamefont {Sawaya}}, \bibinfo {author} {\bibfnamefont {F.}~\bibnamefont
  {Paesani}}, \ and\ \bibinfo {author} {\bibfnamefont {D.~P.}\ \bibnamefont
  {Tabor}},\ }\bibfield  {title} {\enquote {\bibinfo {title} {Near- and
  long-term quantum algorithmic approaches for vibrational spectroscopy},}\
  }\href {http://dx.doi.org/10.1103/physreva.104.062419} {\bibfield  {journal}
  {\bibinfo  {journal} {\emph {Phys. Rev. A}}\ }\textbf {\bibinfo {volume}
  {104}},\ \bibinfo {pages} {062419} (\bibinfo {year} {2021})}\BibitemShut
  {NoStop}%
\bibitem [{\citenamefont {Majland}\ \emph {et~al.}(2023)\citenamefont
  {Majland}, \citenamefont {Berg~Jensen}, \citenamefont {H{\o}jlund},
  \citenamefont {Thomas~Zinner},\ and\ \citenamefont
  {Christiansen}}]{Majland2023}%
  \BibitemOpen
  \bibfield  {author} {\bibinfo {author} {\bibfnamefont {M.}~\bibnamefont
  {Majland}}, \bibinfo {author} {\bibfnamefont {R.}~\bibnamefont
  {Berg~Jensen}}, \bibinfo {author} {\bibfnamefont {M.~G.}\ \bibnamefont
  {H{\o}jlund}}, \bibinfo {author} {\bibfnamefont {N.}~\bibnamefont
  {Thomas~Zinner}}, \ and\ \bibinfo {author} {\bibfnamefont {O.}~\bibnamefont
  {Christiansen}},\ }\bibfield  {title} {\enquote {\bibinfo {title} {Optimizing
  the number of measurements for vibrational structure on quantum computers:
  coordinates and measurement schemes},}\ }\href
  {http://dx.doi.org/10.1039/d3sc01984e} {\bibfield  {journal} {\bibinfo
  {journal} {\emph {Chem. Sci.}}\ }\textbf {\bibinfo {volume} {14}},\ \bibinfo
  {pages} {7733} (\bibinfo {year} {2023})}\BibitemShut {NoStop}%
\bibitem [{\citenamefont {Somasundaram}\ \emph {et~al.}(2025)\citenamefont
  {Somasundaram}, \citenamefont {Jayaharish}, \citenamefont {Ramanan},\ and\
  \citenamefont {Chowdhury}}]{Somasundarametal2025}%
  \BibitemOpen
  \bibfield  {author} {\bibinfo {author} {\bibfnamefont {R.}~\bibnamefont
  {Somasundaram}}, \bibinfo {author} {\bibfnamefont {R.}~\bibnamefont
  {Jayaharish}}, \bibinfo {author} {\bibfnamefont {R.}~\bibnamefont {Ramanan}},
  \ and\ \bibinfo {author} {\bibfnamefont {C.}~\bibnamefont {Chowdhury}},\
  }\bibfield  {title} {\enquote {\bibinfo {title} {Quantum computing for
  molecular vibrational energies: {A} comprehensive study},}\ }\href
  {http://dx.doi.org/10.1016/j.mtquan.2025.100031} {\bibfield  {journal}
  {\bibinfo  {journal} {\emph {Mater. Today Quantum}}\ }\textbf {\bibinfo
  {volume} {6}},\ \bibinfo {pages} {100031} (\bibinfo {year}
  {2025})}\BibitemShut {NoStop}%
\bibitem [{\citenamefont {L\"otstedt}\ and\ \citenamefont
  {Szidarovszky}(2026)}]{LotstedtSzidarovszky2026}%
  \BibitemOpen
  \bibfield  {author} {\bibinfo {author} {\bibfnamefont {E.}~\bibnamefont
  {L\"otstedt}}\ and\ \bibinfo {author} {\bibfnamefont {T.}~\bibnamefont
  {Szidarovszky}},\ }\bibfield  {title} {\enquote {\bibinfo {title}
  {Rovibrational energy levels of {H$_2$O} by quantum computing},}\ }\href
  {http://dx.doi.org/10.1063/5.0333947} {\bibfield  {journal} {\bibinfo
  {journal} {\emph {J. Chem. Phys.}}\ }\textbf {\bibinfo {volume} {165}},\
  \bibinfo {pages} {024118} (\bibinfo {year} {2026})}\BibitemShut {NoStop}%
\bibitem [{\citenamefont {Asnaashari}\ \emph {et~al.}(2026)\citenamefont
  {Asnaashari}, \citenamefont {Bondarenko},\ and\ \citenamefont
  {Krems}}]{Asnaashari2026}%
  \BibitemOpen
  \bibfield  {author} {\bibinfo {author} {\bibfnamefont {K.}~\bibnamefont
  {Asnaashari}}, \bibinfo {author} {\bibfnamefont {D.}~\bibnamefont
  {Bondarenko}}, \ and\ \bibinfo {author} {\bibfnamefont {R.~V.}\ \bibnamefont
  {Krems}},\ }\bibfield  {title} {\enquote {\bibinfo {title} {Advantages of
  discrete variable representation in variational quantum eigensolvers for
  vibrational energy calculations},}\ }\href
  {http://dx.doi.org/10.1039/d5cp02860d} {\bibfield  {journal} {\bibinfo
  {journal} {\emph {Phys. Chem. Chem. Phys.}}\ }\textbf {\bibinfo {volume}
  {28}},\ \bibinfo {pages} {7900} (\bibinfo {year} {2026})}\BibitemShut
  {NoStop}%
\bibitem [{\citenamefont {Ollitrault}\ \emph {et~al.}(2020)\citenamefont
  {Ollitrault}, \citenamefont {Baiardi}, \citenamefont {Reiher},\ and\
  \citenamefont {Tavernelli}}]{Ollitraultetal2020}%
  \BibitemOpen
  \bibfield  {author} {\bibinfo {author} {\bibfnamefont {P.~J.}\ \bibnamefont
  {Ollitrault}}, \bibinfo {author} {\bibfnamefont {A.}~\bibnamefont {Baiardi}},
  \bibinfo {author} {\bibfnamefont {M.}~\bibnamefont {Reiher}}, \ and\ \bibinfo
  {author} {\bibfnamefont {I.}~\bibnamefont {Tavernelli}},\ }\bibfield  {title}
  {\enquote {\bibinfo {title} {Hardware efficient quantum algorithms for
  vibrational structure calculations},}\ }\href
  {http://dx.doi.org/10.1039/d0sc01908a} {\bibfield  {journal} {\bibinfo
  {journal} {\emph {Chem. Sci.}}\ }\textbf {\bibinfo {volume} {11}},\ \bibinfo
  {pages} {6842} (\bibinfo {year} {2020})}\BibitemShut {NoStop}%
\bibitem [{\citenamefont {Cs\'{a}sz\'{a}r}(2011)}]{Csaszar2011}%
  \BibitemOpen
  \bibfield  {author} {\bibinfo {author} {\bibfnamefont {A.~G.}\ \bibnamefont
  {Cs\'{a}sz\'{a}r}},\ }\bibfield  {title} {\enquote {\bibinfo {title}
  {Anharmonic molecular force fields},}\ }\href
  {http://dx.doi.org/10.1002/wcms.75} {\bibfield  {journal} {\bibinfo
  {journal} {\emph {WIREs Comput. Mol. Sci.}}\ }\textbf {\bibinfo {volume}
  {2}},\ \bibinfo {pages} {273} (\bibinfo {year} {2011})}\BibitemShut {NoStop}%
\bibitem [{\citenamefont {Sawaya}\ \emph {et~al.}(2020)\citenamefont {Sawaya},
  \citenamefont {Menke}, \citenamefont {Kyaw}, \citenamefont {Johri},
  \citenamefont {Aspuru-Guzik},\ and\ \citenamefont
  {Guerreschi}}]{Sawayaetal2020}%
  \BibitemOpen
  \bibfield  {author} {\bibinfo {author} {\bibfnamefont {N.~P.~D.}\
  \bibnamefont {Sawaya}}, \bibinfo {author} {\bibfnamefont {T.}~\bibnamefont
  {Menke}}, \bibinfo {author} {\bibfnamefont {T.~H.}\ \bibnamefont {Kyaw}},
  \bibinfo {author} {\bibfnamefont {S.}~\bibnamefont {Johri}}, \bibinfo
  {author} {\bibfnamefont {A.}~\bibnamefont {Aspuru-Guzik}}, \ and\ \bibinfo
  {author} {\bibfnamefont {G.~G.}\ \bibnamefont {Guerreschi}},\ }\bibfield
  {title} {\enquote {\bibinfo {title} {Resource-efficient digital quantum
  simulation of $d$-level systems for photonic, vibrational, and spin-$s$
  {H}amiltonians},}\ }\href {http://dx.doi.org/10.1038/s41534-020-0278-0}
  {\bibfield  {journal} {\bibinfo  {journal} {\emph {npj Quantum Inf.}}\
  }\textbf {\bibinfo {volume} {6}},\ \bibinfo {pages} {49} (\bibinfo {year}
  {2020})}\BibitemShut {NoStop}%
\bibitem [{\citenamefont {Hadfield}\ \emph {et~al.}(2019)\citenamefont
  {Hadfield}, \citenamefont {Wang}, \citenamefont {O'Gorman}, \citenamefont
  {Rieffel}, \citenamefont {Venturelli},\ and\ \citenamefont
  {Biswas}}]{Hadfield2019}%
  \BibitemOpen
  \bibfield  {author} {\bibinfo {author} {\bibfnamefont {S.}~\bibnamefont
  {Hadfield}}, \bibinfo {author} {\bibfnamefont {Z.}~\bibnamefont {Wang}},
  \bibinfo {author} {\bibfnamefont {B.}~\bibnamefont {O'Gorman}}, \bibinfo
  {author} {\bibfnamefont {E.~G.}\ \bibnamefont {Rieffel}}, \bibinfo {author}
  {\bibfnamefont {D.}~\bibnamefont {Venturelli}}, \ and\ \bibinfo {author}
  {\bibfnamefont {R.}~\bibnamefont {Biswas}},\ }\bibfield  {title} {\enquote
  {\bibinfo {title} {From the quantum approximate optimization algorithm to a
  quantum alternating operator ansatz},}\ }\href
  {http://dx.doi.org/10.3390/a12020034} {\bibfield  {journal} {\bibinfo
  {journal} {\emph {Algorithms}}\ }\textbf {\bibinfo {volume} {12}},\ \bibinfo
  {pages} {34} (\bibinfo {year} {2019})}\BibitemShut {NoStop}%
\bibitem [{\citenamefont {Luo}\ \emph {et~al.}(2014)\citenamefont {Luo},
  \citenamefont {Chen}, \citenamefont {Yang},\ and\ \citenamefont
  {Wang}}]{Luo2014}%
  \BibitemOpen
  \bibfield  {author} {\bibinfo {author} {\bibfnamefont {M.-X.}\ \bibnamefont
  {Luo}}, \bibinfo {author} {\bibfnamefont {X.-B.}\ \bibnamefont {Chen}},
  \bibinfo {author} {\bibfnamefont {Y.-X.}\ \bibnamefont {Yang}}, \ and\
  \bibinfo {author} {\bibfnamefont {X.}~\bibnamefont {Wang}},\ }\bibfield
  {title} {\enquote {\bibinfo {title} {Geometry of quantum computation with
  qudits},}\ }\href {http://dx.doi.org/10.1038/srep04044} {\bibfield  {journal}
  {\bibinfo  {journal} {\emph {Sci. Rep.}}\ }\textbf {\bibinfo {volume} {4}},\
  \bibinfo {pages} {4044} (\bibinfo {year} {2014})}\BibitemShut {NoStop}%
\bibitem [{\citenamefont {L\"{o}tstedt}\ and\ \citenamefont
  {Yamanouchi}(2025)}]{Loetstedt2025}%
  \BibitemOpen
  \bibfield  {author} {\bibinfo {author} {\bibfnamefont {E.}~\bibnamefont
  {L\"{o}tstedt}}\ and\ \bibinfo {author} {\bibfnamefont {K.}~\bibnamefont
  {Yamanouchi}},\ }\bibfield  {title} {\enquote {\bibinfo {title} {Comparison
  of encoding schemes for quantum computing of {$S>1/2$} spin chains},}\ }\href
  {http://dx.doi.org/10.1103/7mvy-zq7j} {\bibfield  {journal} {\bibinfo
  {journal} {\emph {Phys. Rev. A}}\ }\textbf {\bibinfo {volume} {111}},\
  \bibinfo {pages} {062416} (\bibinfo {year} {2025})}\BibitemShut {NoStop}%
\bibitem [{\citenamefont {Trotter}(1959)}]{Trotter1959}%
  \BibitemOpen
  \bibfield  {author} {\bibinfo {author} {\bibfnamefont {H.~F.}\ \bibnamefont
  {Trotter}},\ }\bibfield  {title} {\enquote {\bibinfo {title} {On the product
  of semi-groups of operators},}\ }\href
  {http://dx.doi.org/10.1090/S0002-9939-1959-0108732-6} {\bibfield  {journal}
  {\bibinfo  {journal} {\emph {Proc. Amer. Math. Soc.}}\ }\textbf {\bibinfo
  {volume} {10}},\ \bibinfo {pages} {545} (\bibinfo {year} {1959})}\BibitemShut
  {NoStop}%
\bibitem [{\citenamefont {Suzuki}(1976)}]{Suzuki1976}%
  \BibitemOpen
  \bibfield  {author} {\bibinfo {author} {\bibfnamefont {M.}~\bibnamefont
  {Suzuki}},\ }\bibfield  {title} {\enquote {\bibinfo {title} {Generalized
  {T}rotter's formula and systematic approximants of exponential operators and
  inner derivations with applications to many-body problems},}\ }\href
  {http://dx.doi.org/10.1007/bf01609348} {\bibfield  {journal} {\bibinfo
  {journal} {\emph {Commun. Math. Phys.}}\ }\textbf {\bibinfo {volume} {51}},\
  \bibinfo {pages} {183} (\bibinfo {year} {1976})}\BibitemShut {NoStop}%
\bibitem [{\citenamefont {Nielsen}\ and\ \citenamefont
  {Chuang}(2010)}]{NielsenChuang}%
  \BibitemOpen
  \bibfield  {author} {\bibinfo {author} {\bibfnamefont {M.~A.}\ \bibnamefont
  {Nielsen}}\ and\ \bibinfo {author} {\bibfnamefont {I.~L.}\ \bibnamefont
  {Chuang}},\ }\href@noop {} {\emph {\bibinfo {title} {Quantum Computation and
  Quantum Information}}}\ (\bibinfo  {publisher} {Cambridge University Press},\
  \bibinfo {address} {Cambridge, UK},\ \bibinfo {year} {2010})\BibitemShut
  {NoStop}%
\bibitem [{\citenamefont {Sriluckshmy}\ \emph {et~al.}(2023)\citenamefont
  {Sriluckshmy}, \citenamefont {Pina-Canelles}, \citenamefont {Ponce},
  \citenamefont {Algaba}, \citenamefont {{{\v S}imkovic IV}},\ and\
  \citenamefont {Leib}}]{Sriluckshmy2023}%
  \BibitemOpen
  \bibfield  {author} {\bibinfo {author} {\bibfnamefont {P.~V.}\ \bibnamefont
  {Sriluckshmy}}, \bibinfo {author} {\bibfnamefont {V.}~\bibnamefont
  {Pina-Canelles}}, \bibinfo {author} {\bibfnamefont {M.}~\bibnamefont
  {Ponce}}, \bibinfo {author} {\bibfnamefont {M.~G.}\ \bibnamefont {Algaba}},
  \bibinfo {author} {\bibfnamefont {F.}~\bibnamefont {{{\v S}imkovic IV}}}, \
  and\ \bibinfo {author} {\bibfnamefont {M.}~\bibnamefont {Leib}},\ }\bibfield
  {title} {\enquote {\bibinfo {title} {Optimal, hardware native decomposition
  of parameterized multi-qubit {P}auli gates},}\ }\href
  {http://dx.doi.org/10.1088/2058-9565/acfa20} {\bibfield  {journal} {\bibinfo
  {journal} {\emph {Quantum Sci. Technol.}}\ }\textbf {\bibinfo {volume} {8}},\
  \bibinfo {pages} {045029} (\bibinfo {year} {2023})}\BibitemShut {NoStop}%
\bibitem [{\citenamefont {Tranter}\ \emph {et~al.}(2019)\citenamefont
  {Tranter}, \citenamefont {Love}, \citenamefont {Mintert}, \citenamefont
  {Wiebe},\ and\ \citenamefont {Coveney}}]{Tranter2019}%
  \BibitemOpen
  \bibfield  {author} {\bibinfo {author} {\bibfnamefont {A.}~\bibnamefont
  {Tranter}}, \bibinfo {author} {\bibfnamefont {P.~J.}\ \bibnamefont {Love}},
  \bibinfo {author} {\bibfnamefont {F.}~\bibnamefont {Mintert}}, \bibinfo
  {author} {\bibfnamefont {N.}~\bibnamefont {Wiebe}}, \ and\ \bibinfo {author}
  {\bibfnamefont {P.~V.}\ \bibnamefont {Coveney}},\ }\bibfield  {title}
  {\enquote {\bibinfo {title} {Ordering of {T}rotterization: {I}mpact on errors
  in quantum simulation of electronic structure},}\ }\href
  {http://dx.doi.org/10.3390/e21121218} {\bibfield  {journal} {\bibinfo
  {journal} {\emph {Entropy}}\ }\textbf {\bibinfo {volume} {21}},\ \bibinfo
  {pages} {1218} (\bibinfo {year} {2019})}\BibitemShut {NoStop}%
\bibitem [{\citenamefont {Hastings}\ \emph {et~al.}(2015)\citenamefont
  {Hastings}, \citenamefont {Wecker}, \citenamefont {Bauer},\ and\
  \citenamefont {Troyer}}]{Hastingsetal2015}%
  \BibitemOpen
  \bibfield  {author} {\bibinfo {author} {\bibfnamefont {M.~B.}\ \bibnamefont
  {Hastings}}, \bibinfo {author} {\bibfnamefont {D.}~\bibnamefont {Wecker}},
  \bibinfo {author} {\bibfnamefont {B.}~\bibnamefont {Bauer}}, \ and\ \bibinfo
  {author} {\bibfnamefont {M.}~\bibnamefont {Troyer}},\ }\bibfield  {title}
  {\enquote {\bibinfo {title} {Improving quantum algorithms for quantum
  chemistry},}\ }\href {https://dl.acm.org/doi/10.5555/2685188.2685189}
  {\bibfield  {journal} {\bibinfo  {journal} {\emph {Quantum Info. Comput.}}\
  }\textbf {\bibinfo {volume} {15}},\ \bibinfo {pages} {1} (\bibinfo {year}
  {2015})}\BibitemShut {NoStop}%
\bibitem [{\citenamefont {Poulin}\ \emph {et~al.}(2015)\citenamefont {Poulin},
  \citenamefont {Hastings}, \citenamefont {Wecker}, \citenamefont {Wiebe},
  \citenamefont {Doberty},\ and\ \citenamefont {Troyer}}]{Poulinetal2015}%
  \BibitemOpen
  \bibfield  {author} {\bibinfo {author} {\bibfnamefont {D.}~\bibnamefont
  {Poulin}}, \bibinfo {author} {\bibfnamefont {M.~B.}\ \bibnamefont
  {Hastings}}, \bibinfo {author} {\bibfnamefont {D.}~\bibnamefont {Wecker}},
  \bibinfo {author} {\bibfnamefont {N.}~\bibnamefont {Wiebe}}, \bibinfo
  {author} {\bibfnamefont {A.~C.}\ \bibnamefont {Doberty}}, \ and\ \bibinfo
  {author} {\bibfnamefont {M.}~\bibnamefont {Troyer}},\ }\bibfield  {title}
  {\enquote {\bibinfo {title} {The {T}rotter step size required for accurate
  quantum simulation of quantum chemistry},}\ }\href
  {https://dl.acm.org/doi/abs/10.5555/2871401.2871402} {\bibfield  {journal}
  {\bibinfo  {journal} {\emph {Quantum Info. Comput.}}\ }\textbf {\bibinfo
  {volume} {15}},\ \bibinfo {pages} {361} (\bibinfo {year} {2015})}\BibitemShut
  {NoStop}%
\bibitem [{\citenamefont {Suzuki}(1968)}]{Suzuki1968}%
  \BibitemOpen
  \bibfield  {author} {\bibinfo {author} {\bibfnamefont {I.}~\bibnamefont
  {Suzuki}},\ }\bibfield  {title} {\enquote {\bibinfo {title} {General
  anharmonic force constants of carbon dioxide},}\ }\href
  {http://dx.doi.org/10.1016/S0022-2852(68)80018-9} {\bibfield  {journal}
  {\bibinfo  {journal} {\emph {J. Mol. Spectroscopy}}\ }\textbf {\bibinfo
  {volume} {25}},\ \bibinfo {pages} {479} (\bibinfo {year} {1968})}\BibitemShut
  {NoStop}%
\bibitem [{\citenamefont {{The MathWorks Inc.}}(2024)}]{MATLAB_2024b}%
  \BibitemOpen
  \bibfield  {author} {\bibinfo {author} {\bibnamefont {{The MathWorks
  Inc.}}},\ }\href {https://www.mathworks.com} {\enquote {\bibinfo {title}
  {{MATLAB} version 24.2.0 {(R2024b)}},}\ } (\bibinfo {year}
  {2024})\BibitemShut {NoStop}%
\bibitem [{\citenamefont {Fermi}(1931)}]{Fermi1931}%
  \BibitemOpen
  \bibfield  {author} {\bibinfo {author} {\bibfnamefont {E.}~\bibnamefont
  {Fermi}},\ }\bibfield  {title} {\enquote {\bibinfo {title} {{\"U}ber den
  {R}amaneffekt des {K}ohlendioxyds},}\ }\href
  {http://dx.doi.org/10.1007/BF01341712} {\bibfield  {journal} {\bibinfo
  {journal} {\emph {Z. Physik}}\ }\textbf {\bibinfo {volume} {71}},\ \bibinfo
  {pages} {250} (\bibinfo {year} {1931})}\BibitemShut {NoStop}%
\bibitem [{\citenamefont {Rodriguez-Garcia}\ \emph {et~al.}(2007)\citenamefont
  {Rodriguez-Garcia}, \citenamefont {Hirata}, \citenamefont {Yagi},
  \citenamefont {Hirao}, \citenamefont {Taketsugu}, \citenamefont
  {Schweigert},\ and\ \citenamefont {Tasumi}}]{RodriguezGarciaetal2007}%
  \BibitemOpen
  \bibfield  {author} {\bibinfo {author} {\bibfnamefont {V.}~\bibnamefont
  {Rodriguez-Garcia}}, \bibinfo {author} {\bibfnamefont {S.}~\bibnamefont
  {Hirata}}, \bibinfo {author} {\bibfnamefont {K.}~\bibnamefont {Yagi}},
  \bibinfo {author} {\bibfnamefont {K.}~\bibnamefont {Hirao}}, \bibinfo
  {author} {\bibfnamefont {T.}~\bibnamefont {Taketsugu}}, \bibinfo {author}
  {\bibfnamefont {I.}~\bibnamefont {Schweigert}}, \ and\ \bibinfo {author}
  {\bibfnamefont {M.}~\bibnamefont {Tasumi}},\ }\bibfield  {title} {\enquote
  {\bibinfo {title} {Fermi resonance in {CO$_2$: A} combined electronic
  coupled-cluster and vibrational configuration-interaction prediction},}\
  }\href {http://dx.doi.org/10.1063/1.2710256} {\bibfield  {journal} {\bibinfo
  {journal} {\emph {J. Chem. Phys.}}\ }\textbf {\bibinfo {volume} {126}},\
  \bibinfo {pages} {124303} (\bibinfo {year} {2007})}\BibitemShut {NoStop}%
\bibitem [{\citenamefont {Sivarajah}\ \emph {et~al.}(2020)\citenamefont
  {Sivarajah}, \citenamefont {Dilkes}, \citenamefont {Cowtan}, \citenamefont
  {Simmons}, \citenamefont {Edgington},\ and\ \citenamefont
  {Duncan}}]{Sivarajah2020}%
  \BibitemOpen
  \bibfield  {author} {\bibinfo {author} {\bibfnamefont {S.}~\bibnamefont
  {Sivarajah}}, \bibinfo {author} {\bibfnamefont {S.}~\bibnamefont {Dilkes}},
  \bibinfo {author} {\bibfnamefont {A.}~\bibnamefont {Cowtan}}, \bibinfo
  {author} {\bibfnamefont {W.}~\bibnamefont {Simmons}}, \bibinfo {author}
  {\bibfnamefont {A.}~\bibnamefont {Edgington}}, \ and\ \bibinfo {author}
  {\bibfnamefont {R.}~\bibnamefont {Duncan}},\ }\bibfield  {title} {\enquote
  {\bibinfo {title} {t$\vert$ket$\rangle$: a retargetable compiler for {NISQ}
  devices},}\ }\href {http://dx.doi.org/10.1088/2058-9565/ab8e92} {\bibfield
  {journal} {\bibinfo  {journal} {\emph {Quantum Sci. Technol.}}\ }\textbf
  {\bibinfo {volume} {6}},\ \bibinfo {pages} {014003} (\bibinfo {year}
  {2020})}\BibitemShut {NoStop}%
\bibitem [{\citenamefont {Kremer}\ \emph {et~al.}(2024)\citenamefont {Kremer},
  \citenamefont {Villar}, \citenamefont {Paik}, \citenamefont {Duran},
  \citenamefont {Faro},\ and\ \citenamefont {Cruz-Benito}}]{Kremer2024}%
  \BibitemOpen
  \bibfield  {author} {\bibinfo {author} {\bibfnamefont {D.}~\bibnamefont
  {Kremer}}, \bibinfo {author} {\bibfnamefont {V.}~\bibnamefont {Villar}},
  \bibinfo {author} {\bibfnamefont {H.}~\bibnamefont {Paik}}, \bibinfo {author}
  {\bibfnamefont {I.}~\bibnamefont {Duran}}, \bibinfo {author} {\bibfnamefont
  {I.}~\bibnamefont {Faro}}, \ and\ \bibinfo {author} {\bibfnamefont
  {J.}~\bibnamefont {Cruz-Benito}},\ }\href
  {http://dx.doi.org/10.48550/ARXIV.2405.13196} {\enquote {\bibinfo {title}
  {Practical and efficient quantum circuit synthesis and transpiling with
  reinforcement learning},}\ } (\bibinfo {year} {2024}),\ \bibinfo {note}
  {arXiv:2405.13196 [quant-ph]}\BibitemShut {NoStop}%
\bibitem [{\citenamefont {L\"otstedt}\ and\ \citenamefont
  {Yamanouchi}(2026)}]{Lotstedt_qubitquditDataset}%
  \BibitemOpen
  \bibfield  {author} {\bibinfo {author} {\bibfnamefont {E.}~\bibnamefont
  {L\"otstedt}}\ and\ \bibinfo {author} {\bibfnamefont {K.}~\bibnamefont
  {Yamanouchi}},\ }\href {http://dx.doi.org/10.5281/zenodo.20115329} {\enquote
  {\bibinfo {title} {Simulation of vibrational dynamics using qubits and
  qudits},}\ }\bibinfo {howpublished} {Zenodo dataset} (\bibinfo {year}
  {2026}),\ \bibinfo {note} {doi:10.5281/zenodo.20115329}\BibitemShut {NoStop}%
\bibitem [{\citenamefont {Ando}\ \emph {et~al.}(2018)\citenamefont {Ando},
  \citenamefont {Iwasaki},\ and\ \citenamefont {Yamanouchi}}]{Ando2018}%
  \BibitemOpen
  \bibfield  {author} {\bibinfo {author} {\bibfnamefont {T.}~\bibnamefont
  {Ando}}, \bibinfo {author} {\bibfnamefont {A.}~\bibnamefont {Iwasaki}}, \
  and\ \bibinfo {author} {\bibfnamefont {K.}~\bibnamefont {Yamanouchi}},\
  }\bibfield  {title} {\enquote {\bibinfo {title} {Strong-field {F}ourier
  transform vibrational spectroscopy of {D$_2^+$} using few-cycle near-infrared
  laser pulses},}\ }\href {http://dx.doi.org/10.1103/physrevlett.120.263002}
  {\bibfield  {journal} {\bibinfo  {journal} {\emph {Phys. Rev. Lett.}}\
  }\textbf {\bibinfo {volume} {120}},\ \bibinfo {pages} {263002} (\bibinfo
  {year} {2018})}\BibitemShut {NoStop}%
\bibitem [{\citenamefont {Ando}\ \emph {et~al.}(2025)\citenamefont {Ando},
  \citenamefont {Yamada}, \citenamefont {Iwasaki},\ and\ \citenamefont
  {Yamanouchi}}]{Ando2025}%
  \BibitemOpen
  \bibfield  {author} {\bibinfo {author} {\bibfnamefont {T.}~\bibnamefont
  {Ando}}, \bibinfo {author} {\bibfnamefont {K.}~\bibnamefont {Yamada}},
  \bibinfo {author} {\bibfnamefont {A.}~\bibnamefont {Iwasaki}}, \ and\
  \bibinfo {author} {\bibfnamefont {K.}~\bibnamefont {Yamanouchi}},\ }\bibfield
   {title} {\enquote {\bibinfo {title} {Isotope shift of fine structure of
  {Kr$^+$} and hyperfine structure of {${}^{83}$Kr$^+$} by strong-field
  ultrahigh-resolution {F}ourier-transform spectroscopy},}\ }\href
  {http://dx.doi.org/10.1103/physrevresearch.7.l022025} {\bibfield  {journal}
  {\bibinfo  {journal} {\emph {Phys. Rev. Res.}}\ }\textbf {\bibinfo {volume}
  {7}},\ \bibinfo {pages} {l022025} (\bibinfo {year} {2025})}\BibitemShut
  {NoStop}%
\bibitem [{\citenamefont {Low}\ \emph {et~al.}(2020)\citenamefont {Low},
  \citenamefont {White}, \citenamefont {Cox}, \citenamefont {Day},\ and\
  \citenamefont {Senko}}]{Low2020}%
  \BibitemOpen
  \bibfield  {author} {\bibinfo {author} {\bibfnamefont {P.~J.}\ \bibnamefont
  {Low}}, \bibinfo {author} {\bibfnamefont {B.~M.}\ \bibnamefont {White}},
  \bibinfo {author} {\bibfnamefont {A.~A.}\ \bibnamefont {Cox}}, \bibinfo
  {author} {\bibfnamefont {M.~L.}\ \bibnamefont {Day}}, \ and\ \bibinfo
  {author} {\bibfnamefont {C.}~\bibnamefont {Senko}},\ }\bibfield  {title}
  {\enquote {\bibinfo {title} {Practical trapped-ion protocols for universal
  qudit-based quantum computing},}\ }\href
  {http://dx.doi.org/10.1103/physrevresearch.2.033128} {\bibfield  {journal}
  {\bibinfo  {journal} {\emph {Phys. Rev. Res.}}\ }\textbf {\bibinfo {volume}
  {2}},\ \bibinfo {pages} {033128} (\bibinfo {year} {2020})}\BibitemShut
  {NoStop}%
\bibitem [{\citenamefont {Cs\'asz\'ar}\ and\ \citenamefont
  {Mills}(1997)}]{Csaszar1997}%
  \BibitemOpen
  \bibfield  {author} {\bibinfo {author} {\bibfnamefont {A.~G.}\ \bibnamefont
  {Cs\'asz\'ar}}\ and\ \bibinfo {author} {\bibfnamefont {I.~M.}\ \bibnamefont
  {Mills}},\ }\bibfield  {title} {\enquote {\bibinfo {title} {Vibrational
  energy levels of water},}\ }\href
  {http://dx.doi.org/10.1016/s1386-1425(97)00020-6} {\bibfield  {journal}
  {\bibinfo  {journal} {\emph {Spectrochim. Acta. A}}\ }\textbf {\bibinfo
  {volume} {53}},\ \bibinfo {pages} {1101} (\bibinfo {year}
  {1997})}\BibitemShut {NoStop}%
\bibitem [{\citenamefont {Endo}\ \emph {et~al.}(2021)\citenamefont {Endo},
  \citenamefont {Cai}, \citenamefont {Benjamin},\ and\ \citenamefont
  {Yuan}}]{EndoCaiBenjaminYuan2021}%
  \BibitemOpen
  \bibfield  {author} {\bibinfo {author} {\bibfnamefont {S.}~\bibnamefont
  {Endo}}, \bibinfo {author} {\bibfnamefont {Z.}~\bibnamefont {Cai}}, \bibinfo
  {author} {\bibfnamefont {S.~C.}\ \bibnamefont {Benjamin}}, \ and\ \bibinfo
  {author} {\bibfnamefont {X.}~\bibnamefont {Yuan}},\ }\bibfield  {title}
  {\enquote {\bibinfo {title} {Hybrid quantum-classical algorithms and quantum
  error mitigation},}\ }\href {http://dx.doi.org/10.7566/JPSJ.90.032001}
  {\bibfield  {journal} {\bibinfo  {journal} {\emph {J. Phys. Soc. Jpn.}}\
  }\textbf {\bibinfo {volume} {90}},\ \bibinfo {pages} {032001} (\bibinfo
  {year} {2021})}\BibitemShut {NoStop}%
\bibitem [{\citenamefont {Cai}\ \emph {et~al.}(2023)\citenamefont {Cai},
  \citenamefont {Babbush}, \citenamefont {Benjamin}, \citenamefont {Endo},
  \citenamefont {Huggins}, \citenamefont {Li}, \citenamefont {McClean},\ and\
  \citenamefont {O'Brien}}]{Cai2023}%
  \BibitemOpen
  \bibfield  {author} {\bibinfo {author} {\bibfnamefont {Z.}~\bibnamefont
  {Cai}}, \bibinfo {author} {\bibfnamefont {R.}~\bibnamefont {Babbush}},
  \bibinfo {author} {\bibfnamefont {S.~C.}\ \bibnamefont {Benjamin}}, \bibinfo
  {author} {\bibfnamefont {S.}~\bibnamefont {Endo}}, \bibinfo {author}
  {\bibfnamefont {W.~J.}\ \bibnamefont {Huggins}}, \bibinfo {author}
  {\bibfnamefont {Y.}~\bibnamefont {Li}}, \bibinfo {author} {\bibfnamefont
  {J.~R.}\ \bibnamefont {McClean}}, \ and\ \bibinfo {author} {\bibfnamefont
  {T.~E.}\ \bibnamefont {O'Brien}},\ }\bibfield  {title} {\enquote {\bibinfo
  {title} {Quantum error mitigation},}\ }\href
  {http://dx.doi.org/10.1103/revmodphys.95.045005} {\bibfield  {journal}
  {\bibinfo  {journal} {\emph {Rev. Mod. Phys.}}\ }\textbf {\bibinfo {volume}
  {95}},\ \bibinfo {pages} {045005} (\bibinfo {year} {2023})}\BibitemShut
  {NoStop}%
\bibitem [{\citenamefont {Temme}\ \emph {et~al.}(2017)\citenamefont {Temme},
  \citenamefont {Bravyi},\ and\ \citenamefont
  {Gambetta}}]{TemmeBravyiGambetta2017}%
  \BibitemOpen
  \bibfield  {author} {\bibinfo {author} {\bibfnamefont {K.}~\bibnamefont
  {Temme}}, \bibinfo {author} {\bibfnamefont {S.}~\bibnamefont {Bravyi}}, \
  and\ \bibinfo {author} {\bibfnamefont {J.~M.}\ \bibnamefont {Gambetta}},\
  }\bibfield  {title} {\enquote {\bibinfo {title} {Error mitigation for
  short-depth quantum circuits},}\ }\href
  {http://dx.doi.org/10.1103/PhysRevLett.119.180509} {\bibfield  {journal}
  {\bibinfo  {journal} {\emph {Phys. Rev. Lett.}}\ }\textbf {\bibinfo {volume}
  {119}},\ \bibinfo {pages} {180509} (\bibinfo {year} {2017})}\BibitemShut
  {NoStop}%
\bibitem [{\citenamefont {Li}\ and\ \citenamefont
  {Benjamin}(2017)}]{LiBenjamin2017}%
  \BibitemOpen
  \bibfield  {author} {\bibinfo {author} {\bibfnamefont {Y.}~\bibnamefont
  {Li}}\ and\ \bibinfo {author} {\bibfnamefont {S.~C.}\ \bibnamefont
  {Benjamin}},\ }\bibfield  {title} {\enquote {\bibinfo {title} {Efficient
  variational quantum simulator incorporating active error minimization},}\
  }\href {http://dx.doi.org/10.1103/PhysRevX.7.021050} {\bibfield  {journal}
  {\bibinfo  {journal} {\emph {Phys. Rev. X}}\ }\textbf {\bibinfo {volume}
  {7}},\ \bibinfo {pages} {021050} (\bibinfo {year} {2017})}\BibitemShut
  {NoStop}%
\bibitem [{\citenamefont {{van den Berg}}\ \emph {et~al.}(2023)\citenamefont
  {{van den Berg}}, \citenamefont {Minev}, \citenamefont {Kandala},\ and\
  \citenamefont {Temme}}]{Berg2023}%
  \BibitemOpen
  \bibfield  {author} {\bibinfo {author} {\bibfnamefont {E.}~\bibnamefont {{van
  den Berg}}}, \bibinfo {author} {\bibfnamefont {Z.~K.}\ \bibnamefont {Minev}},
  \bibinfo {author} {\bibfnamefont {A.}~\bibnamefont {Kandala}}, \ and\
  \bibinfo {author} {\bibfnamefont {K.}~\bibnamefont {Temme}},\ }\bibfield
  {title} {\enquote {\bibinfo {title} {Probabilistic error cancellation with
  sparse {P}auli{\textendash}{L}indblad models on noisy quantum processors},}\
  }\href {http://dx.doi.org/10.1038/s41567-023-02042-2} {\bibfield  {journal}
  {\bibinfo  {journal} {\emph {Nat. Phys.}}\ }\textbf {\bibinfo {volume}
  {19}},\ \bibinfo {pages} {1116} (\bibinfo {year} {2023})}\BibitemShut
  {NoStop}%
\bibitem [{\citenamefont {Spagnoli}\ \emph {et~al.}(2025)\citenamefont
  {Spagnoli}, \citenamefont {Goss}, \citenamefont {Roggero}, \citenamefont
  {Rrapaj}, \citenamefont {Cervia}, \citenamefont {Patwardhan}, \citenamefont
  {Naik}, \citenamefont {Balantekin}, \citenamefont {Younis}, \citenamefont
  {Santiago}, \citenamefont {Siddiqi},\ and\ \citenamefont
  {Aldaihan}}]{Spagnoli2025}%
  \BibitemOpen
  \bibfield  {author} {\bibinfo {author} {\bibfnamefont {L.}~\bibnamefont
  {Spagnoli}}, \bibinfo {author} {\bibfnamefont {N.}~\bibnamefont {Goss}},
  \bibinfo {author} {\bibfnamefont {A.}~\bibnamefont {Roggero}}, \bibinfo
  {author} {\bibfnamefont {E.}~\bibnamefont {Rrapaj}}, \bibinfo {author}
  {\bibfnamefont {M.~J.}\ \bibnamefont {Cervia}}, \bibinfo {author}
  {\bibfnamefont {A.~V.}\ \bibnamefont {Patwardhan}}, \bibinfo {author}
  {\bibfnamefont {R.~K.}\ \bibnamefont {Naik}}, \bibinfo {author}
  {\bibfnamefont {A.~B.}\ \bibnamefont {Balantekin}}, \bibinfo {author}
  {\bibfnamefont {E.}~\bibnamefont {Younis}}, \bibinfo {author} {\bibfnamefont
  {D.~I.}\ \bibnamefont {Santiago}}, \bibinfo {author} {\bibfnamefont
  {I.}~\bibnamefont {Siddiqi}}, \ and\ \bibinfo {author} {\bibfnamefont
  {S.}~\bibnamefont {Aldaihan}},\ }\bibfield  {title} {\enquote {\bibinfo
  {title} {Collective neutrino oscillations in three flavors on qubit and
  qutrit processors},}\ }\href {http://dx.doi.org/10.1103/gjr1-lf8s} {\bibfield
   {journal} {\bibinfo  {journal} {\emph {Phys. Rev. D}}\ }\textbf {\bibinfo
  {volume} {111}},\ \bibinfo {pages} {103054} (\bibinfo {year}
  {2025})}\BibitemShut {NoStop}%
\bibitem [{\citenamefont {Urbanek}\ \emph {et~al.}(2021)\citenamefont
  {Urbanek}, \citenamefont {Nachman}, \citenamefont {Pascuzzi}, \citenamefont
  {He}, \citenamefont {Bauer},\ and\ \citenamefont
  {de~Jong}}]{Urbaneketal2021}%
  \BibitemOpen
  \bibfield  {author} {\bibinfo {author} {\bibfnamefont {M.}~\bibnamefont
  {Urbanek}}, \bibinfo {author} {\bibfnamefont {B.}~\bibnamefont {Nachman}},
  \bibinfo {author} {\bibfnamefont {V.~R.}\ \bibnamefont {Pascuzzi}}, \bibinfo
  {author} {\bibfnamefont {A.}~\bibnamefont {He}}, \bibinfo {author}
  {\bibfnamefont {C.~W.}\ \bibnamefont {Bauer}}, \ and\ \bibinfo {author}
  {\bibfnamefont {W.~A.}\ \bibnamefont {de~Jong}},\ }\bibfield  {title}
  {\enquote {\bibinfo {title} {Mitigating depolarizing noise on quantum
  computers with noise-estimation circuits},}\ }\href
  {http://dx.doi.org/10.1103/PhysRevLett.127.270502} {\bibfield  {journal}
  {\bibinfo  {journal} {\emph {Phys. Rev. Lett.}}\ }\textbf {\bibinfo {volume}
  {127}},\ \bibinfo {pages} {270502} (\bibinfo {year} {2021})}\BibitemShut
  {NoStop}%
\bibitem [{\citenamefont {Shinjo}\ \emph {et~al.}(2026)\citenamefont {Shinjo},
  \citenamefont {Seki}, \citenamefont {Shirakawa}, \citenamefont {Sun},\ and\
  \citenamefont {Yunoki}}]{Shinjo2026}%
  \BibitemOpen
  \bibfield  {author} {\bibinfo {author} {\bibfnamefont {K.}~\bibnamefont
  {Shinjo}}, \bibinfo {author} {\bibfnamefont {K.}~\bibnamefont {Seki}},
  \bibinfo {author} {\bibfnamefont {T.}~\bibnamefont {Shirakawa}}, \bibinfo
  {author} {\bibfnamefont {R.-Y.}\ \bibnamefont {Sun}}, \ and\ \bibinfo
  {author} {\bibfnamefont {S.}~\bibnamefont {Yunoki}},\ }\bibfield  {title}
  {\enquote {\bibinfo {title} {Unveiling clean two-dimensional discrete time
  crystals on a digital quantum computer},}\ }\href
  {http://dx.doi.org/10.1038/s41534-026-01193-3} {\bibfield  {journal}
  {\bibinfo  {journal} {\emph {npj Quantum Inf.}}\ }\textbf {\bibinfo {volume}
  {12}},\ \bibinfo {pages} {41} (\bibinfo {year} {2026})}\BibitemShut {NoStop}%
\bibitem [{\citenamefont {Quantinuum}(2024)}]{QuantinuumRoadmap}%
  \BibitemOpen
  \bibfield  {author} {\bibinfo {author} {\bibnamefont {Quantinuum}},\ }\href
  {https://www.quantinuum.com/press-releases/quantinuum-unveils-accelerated-roadmap-to-achieve-universal-fault-tolerant-quantum-computing-by-2030}
  {\enquote {\bibinfo {title} {Quantinuum roadmap},}\ }\bibinfo {howpublished}
  {https://www.quantinuum.com/press-releases/quantinuum-unveils-accelerated-roadmap-to-achieve-universal-fault-tolerant-quantum-computing-by-2030}
  (\bibinfo {year} {2024}),\ \bibinfo {note} {accessed on April 2,
  2026}\BibitemShut {NoStop}%
\bibitem [{\citenamefont {Ostmeyer}(2023)}]{Ostmeyer2023}%
  \BibitemOpen
  \bibfield  {author} {\bibinfo {author} {\bibfnamefont {J.}~\bibnamefont
  {Ostmeyer}},\ }\bibfield  {title} {\enquote {\bibinfo {title} {Optimised
  {T}rotter decompositions for classical and quantum computing},}\ }\href
  {http://dx.doi.org/10.1088/1751-8121/acde7a} {\bibfield  {journal} {\bibinfo
  {journal} {\emph {J. Phys. A: Math. Theor.}}\ }\textbf {\bibinfo {volume}
  {56}},\ \bibinfo {pages} {285303} (\bibinfo {year} {2023})}\BibitemShut
  {NoStop}%
\bibitem [{\citenamefont {Rost}\ \emph {et~al.}(2020)\citenamefont {Rost},
  \citenamefont {Jones}, \citenamefont {Vyushkova}, \citenamefont {Ali},
  \citenamefont {Cullip}, \citenamefont {Vyushkov},\ and\ \citenamefont
  {Nabrzyski}}]{Rost2020}%
  \BibitemOpen
  \bibfield  {author} {\bibinfo {author} {\bibfnamefont {B.}~\bibnamefont
  {Rost}}, \bibinfo {author} {\bibfnamefont {B.}~\bibnamefont {Jones}},
  \bibinfo {author} {\bibfnamefont {M.}~\bibnamefont {Vyushkova}}, \bibinfo
  {author} {\bibfnamefont {A.}~\bibnamefont {Ali}}, \bibinfo {author}
  {\bibfnamefont {C.}~\bibnamefont {Cullip}}, \bibinfo {author} {\bibfnamefont
  {A.}~\bibnamefont {Vyushkov}}, \ and\ \bibinfo {author} {\bibfnamefont
  {J.}~\bibnamefont {Nabrzyski}},\ }\href
  {http://dx.doi.org/10.48550/ARXIV.2001.00794} {\enquote {\bibinfo {title}
  {Simulation of thermal relaxation in spin chemistry systems on a quantum
  computer using inherent qubit decoherence},}\ } (\bibinfo {year} {2020}),\
  \bibinfo {note} {arXiv:2001.00794 [quant-ph]}\BibitemShut {NoStop}%
\bibitem [{\citenamefont {Lepp\"{a}kangas}\ \emph {et~al.}(2023)\citenamefont
  {Lepp\"{a}kangas}, \citenamefont {Vogt}, \citenamefont {Fratus},
  \citenamefont {Bark}, \citenamefont {Vaitkus}, \citenamefont {Stadler},
  \citenamefont {Reiner}, \citenamefont {Zanker},\ and\ \citenamefont
  {Marthaler}}]{Leppaekangas2023}%
  \BibitemOpen
  \bibfield  {author} {\bibinfo {author} {\bibfnamefont {J.}~\bibnamefont
  {Lepp\"{a}kangas}}, \bibinfo {author} {\bibfnamefont {N.}~\bibnamefont
  {Vogt}}, \bibinfo {author} {\bibfnamefont {K.~R.}\ \bibnamefont {Fratus}},
  \bibinfo {author} {\bibfnamefont {K.}~\bibnamefont {Bark}}, \bibinfo {author}
  {\bibfnamefont {J.~A.}\ \bibnamefont {Vaitkus}}, \bibinfo {author}
  {\bibfnamefont {P.}~\bibnamefont {Stadler}}, \bibinfo {author} {\bibfnamefont
  {J.-M.}\ \bibnamefont {Reiner}}, \bibinfo {author} {\bibfnamefont
  {S.}~\bibnamefont {Zanker}}, \ and\ \bibinfo {author} {\bibfnamefont
  {M.}~\bibnamefont {Marthaler}},\ }\bibfield  {title} {\enquote {\bibinfo
  {title} {Quantum algorithm for solving open-system dynamics on quantum
  computers using noise},}\ }\href
  {http://dx.doi.org/10.1103/physreva.108.062424} {\bibfield  {journal}
  {\bibinfo  {journal} {\emph {Phys. Rev. A}}\ }\textbf {\bibinfo {volume}
  {108}},\ \bibinfo {pages} {062424} (\bibinfo {year} {2023})}\BibitemShut
  {NoStop}%
\bibitem [{\citenamefont {Owrutsky}\ \emph {et~al.}(1994)\citenamefont
  {Owrutsky}, \citenamefont {Raftery},\ and\ \citenamefont
  {Hochstrasser}}]{Owrutsky1994}%
  \BibitemOpen
  \bibfield  {author} {\bibinfo {author} {\bibfnamefont {J.~C.}\ \bibnamefont
  {Owrutsky}}, \bibinfo {author} {\bibfnamefont {D.}~\bibnamefont {Raftery}}, \
  and\ \bibinfo {author} {\bibfnamefont {R.~M.}\ \bibnamefont {Hochstrasser}},\
  }\bibfield  {title} {\enquote {\bibinfo {title} {Vibrational relaxation
  dynamics in solutions},}\ }\href
  {http://dx.doi.org/10.1146/annurev.pc.45.100194.002511} {\bibfield  {journal}
  {\bibinfo  {journal} {\emph {Annu. Rev. Phys. Chem.}}\ }\textbf {\bibinfo
  {volume} {45}},\ \bibinfo {pages} {519} (\bibinfo {year} {1994})}\BibitemShut
  {NoStop}%
\bibitem [{\citenamefont {Egorov}\ and\ \citenamefont
  {Berne}(1997)}]{Egorov1997}%
  \BibitemOpen
  \bibfield  {author} {\bibinfo {author} {\bibfnamefont {S.~A.}\ \bibnamefont
  {Egorov}}\ and\ \bibinfo {author} {\bibfnamefont {B.~J.}\ \bibnamefont
  {Berne}},\ }\bibfield  {title} {\enquote {\bibinfo {title} {Vibrational
  energy relaxation in the condensed phases: {Q}uantum vs classical bath for
  multiphonon processes},}\ }\href {http://dx.doi.org/10.1063/1.474273}
  {\bibfield  {journal} {\bibinfo  {journal} {\emph {J. Chem. Phys.}}\ }\textbf
  {\bibinfo {volume} {107}},\ \bibinfo {pages} {6050} (\bibinfo {year}
  {1997})}\BibitemShut {NoStop}%
\bibitem [{\citenamefont {Mukhopadhyay}\ \emph {et~al.}(2023)\citenamefont
  {Mukhopadhyay}, \citenamefont {Wiebe},\ and\ \citenamefont
  {Zhang}}]{Mukhopadhyay2023}%
  \BibitemOpen
  \bibfield  {author} {\bibinfo {author} {\bibfnamefont {P.}~\bibnamefont
  {Mukhopadhyay}}, \bibinfo {author} {\bibfnamefont {N.}~\bibnamefont {Wiebe}},
  \ and\ \bibinfo {author} {\bibfnamefont {H.~T.}\ \bibnamefont {Zhang}},\
  }\bibfield  {title} {\enquote {\bibinfo {title} {Synthesizing efficient
  circuits for {H}amiltonian simulation},}\ }\href
  {http://dx.doi.org/10.1038/s41534-023-00697-6} {\bibfield  {journal}
  {\bibinfo  {journal} {\emph {npj Quantum Inf.}}\ }\textbf {\bibinfo {volume}
  {9}},\ \bibinfo {pages} {31} (\bibinfo {year} {2023})}\BibitemShut {NoStop}%
\bibitem [{\citenamefont {Di}\ and\ \citenamefont {Wei}(2013)}]{Di2013}%
  \BibitemOpen
  \bibfield  {author} {\bibinfo {author} {\bibfnamefont {Y.-M.}\ \bibnamefont
  {Di}}\ and\ \bibinfo {author} {\bibfnamefont {H.-R.}\ \bibnamefont {Wei}},\
  }\bibfield  {title} {\enquote {\bibinfo {title} {Synthesis of multivalued
  quantum logic circuits by elementary gates},}\ }\href
  {http://dx.doi.org/10.1103/physreva.87.012325} {\bibfield  {journal}
  {\bibinfo  {journal} {\emph {Phys. Rev. A}}\ }\textbf {\bibinfo {volume}
  {87}},\ \bibinfo {pages} {012325} (\bibinfo {year} {2013})}\BibitemShut
  {NoStop}%
\bibitem [{\citenamefont {Di}\ and\ \citenamefont {Wei}(2015)}]{Di2015}%
  \BibitemOpen
  \bibfield  {author} {\bibinfo {author} {\bibfnamefont {Y.-M.}\ \bibnamefont
  {Di}}\ and\ \bibinfo {author} {\bibfnamefont {H.-R.}\ \bibnamefont {Wei}},\
  }\bibfield  {title} {\enquote {\bibinfo {title} {Optimal synthesis of
  multivalued quantum circuits},}\ }\href
  {http://dx.doi.org/10.1103/physreva.92.062317} {\bibfield  {journal}
  {\bibinfo  {journal} {\emph {Phys. Rev. A}}\ }\textbf {\bibinfo {volume}
  {92}},\ \bibinfo {pages} {062317} (\bibinfo {year} {2015})}\BibitemShut
  {NoStop}%
\bibitem [{\citenamefont {Trisciani}\ \emph {et~al.}(2025)\citenamefont
  {Trisciani}, \citenamefont {Cattaneo},\ and\ \citenamefont
  {Zimbor\'{a}s}}]{Trisciani2025}%
  \BibitemOpen
  \bibfield  {author} {\bibinfo {author} {\bibfnamefont {D.}~\bibnamefont
  {Trisciani}}, \bibinfo {author} {\bibfnamefont {M.}~\bibnamefont {Cattaneo}},
  \ and\ \bibinfo {author} {\bibfnamefont {Z.}~\bibnamefont {Zimbor\'{a}s}},\
  }\href {http://dx.doi.org/10.48550/ARXIV.2507.09781} {\enquote {\bibinfo
  {title} {Decomposition of multi-qutrit gates generated by {W}eyl-{H}eisenberg
  strings},}\ } (\bibinfo {year} {2025}),\ \bibinfo {note} {arXiv:2507.09781
  [quant-ph]}\BibitemShut {NoStop}%
\bibitem [{\citenamefont {Furtenbacher}\ \emph {et~al.}(2020)\citenamefont
  {Furtenbacher}, \citenamefont {T\'obi\'as}, \citenamefont {Tennyson},
  \citenamefont {Polyansky},\ and\ \citenamefont
  {Cs\'asz\'ar}}]{Furtenbacher2020a}%
  \BibitemOpen
  \bibfield  {author} {\bibinfo {author} {\bibfnamefont {T.}~\bibnamefont
  {Furtenbacher}}, \bibinfo {author} {\bibfnamefont {R.}~\bibnamefont
  {T\'obi\'as}}, \bibinfo {author} {\bibfnamefont {J.}~\bibnamefont
  {Tennyson}}, \bibinfo {author} {\bibfnamefont {O.~L.}\ \bibnamefont
  {Polyansky}}, \ and\ \bibinfo {author} {\bibfnamefont {A.~G.}\ \bibnamefont
  {Cs\'asz\'ar}},\ }\bibfield  {title} {\enquote {\bibinfo {title} {{W2020}:
  {A} database of validated rovibrational experimental transitions and
  empirical energy levels of {H$_2$${}^{16}$O}},}\ }\href
  {http://dx.doi.org/10.1063/5.0008253} {\bibfield  {journal} {\bibinfo
  {journal} {\emph {J. Phys. Chem. Ref. Data}}\ }\textbf {\bibinfo {volume}
  {49}},\ \bibinfo {pages} {033101} (\bibinfo {year} {2020})}\BibitemShut
  {NoStop}%
\end{thebibliography}

%

\end{document}